\documentclass[twocolumn]{aastex62}

\usepackage{amsmath}
\usepackage[inline]{enumitem}

\usepackage{natbib}
\usepackage{array,multirow}

\received{2018 March 25}
\accepted{2018 May 8}
\submitjournal{AJ}

\shorttitle{Reassessing Exoplanet Light Curves}
\shortauthors{Adams \& Laughlin}

\begin{document}

\title{Reassessing Exoplanet Light Curves with a Thermal Model}

\author[0000-0002-7139-3695]{Arthur D. Adams}
\affiliation{Department of Astronomy, Yale University, New Haven, CT 06520}
\author[0000-0002-3253-2621]{Gregory Laughlin}
\affiliation{Department of Astronomy, Yale University, New Haven, CT 06520}

\begin{abstract}
We present a uniform assessment of existing near-infrared Spitzer Space Telescope observations of planet-bearing stars. Using a simple four-parameter blackbody thermal model, we analyze stars for which photometry in at least one of Spitzer's IRAC bands has been obtained over either the entirety or a significant fraction of the planetary orbit. Systems in this category comprise ten well-studied systems with Hot Jupiters on circular or near-circular orbits (HAT-P-7, HD 149026, HD 189733, HD 209458, WASP-12, WASP-14, WASP-18, WASP-19, WASP-33, and WASP-43), as well as three stars harboring planets on significantly eccentric orbits (GJ 436, HAT-P-2, and HD 80606). We find that our simple model, in almost all cases, accurately reproduces the minimum and maximum planetary emission, as well as the phase offsets of these extrema with respect to transits/secondary eclipses. For one notable exception, WASP-12 b, adding an additional parameter to account for its tidal distortion is not sufficient to reproduce its photometric features. Full-orbit photometry is available in multiple wavelengths for 10 planets. We find that the returned parameter values for independent fits to each band are largely in agreement. However, disagreements in night-side temperature suggest distinct atmospheric layers, each with their own characteristic minimum temperature. In addition, a diversity in albedos suggests variation in opacity of the photospheres. While previous works have pointed out trends in photometric features based on system properties, we cannot conclusively identify analogous trends for physical model parameters. To make the connection between full-phase data and physical models more robust, a higher signal-to-noise must come from both increased resolution and a careful treatment of instrumental systematics.
\end{abstract}


\section{Introduction}\label{sec:introduction} 
While planets on short orbital periods cannot be directly imaged, their orbital phases can often be inferred to high precision from transits and Doppler velocity measurements, and so the geometric conditions governing their time-dependent illumination phase functions can be known precisely. For transiting planets in particular, a variety of observational strategies -- including transmission and reflection spectroscopy -- can be brought to the task of characterizing the planetary atmospheres. For the most well-studied planets, there is a broad consensus that measured emission spectra permit inferences of the molecular compositions and pressure-temperature profiles of their atmospheres \citep{mad09, mad11}. Data from observations have been used, for example, to support arguments for disequilibrium chemistry \citep{ste10,mor17}, thermal inversions \citep{knu08,nug17}, and varying C/O ratios \citep{mad11,arc18}.

A primary result of observations over the last decade has been the detection of molecules, especially H$_2$O and CO, in exoplanet atmospheres. This work has been done primarily in the near-infrared, via ground- and space-based spectroscopy. The first claimed detection of CO was made in \citet{sne10} for the well-studied planet HD 209458 b, using the CRISES spectrograph on the Very Large Telescope (VLT). While this particular detection has not yet been reproduced \citep{sch15a}, a detection of CO in HD 189733 b was published in 2013 \citep{rod13}. HD 209458 b has also showed spectral evidence of H$_2$O \citep{bea10}, and since 2010, several other planets, including GJ 1214 b \citep{ber12}, WASP-12, 17, and 19 b \citep{man13}, WASP-121 b \citep{eva17}, and HD 189733 b \citep{cro14} have also shown spectral evidence of H$_2$O in their transmission spectra. Beyond CO and H$_2$O, there is tentative evidence for nitrogen-containing molecules such as NH$_3$ and HCN \citep{mac17}.

Transmission spectra have revealed molecular signatures, but in many cases have also suggested significant clouds/hazes. Flatter-than-expected spectra have been observed in planets such as GJ 1214 b \citep{ber12} and GJ 3470 b \citep{cro13}. Clouds/hazes have been invoked to explain flat spectra, including lower-than-expected signals of certain molecular species, due to the opacity-increasing effect of clouds/hazes \citep{hui12,eva13,man13,cro14,sch15b}.

Within the framework of models of atmospheric composition and circulation, the effective albedo and heat recirculation/redistribution efficiency are the most robustly empirically constrained parameters. Their constraint relies on a combination of measurements taken during both transit and secondary eclipse \citep{cow11b}. The prediction that the day-night temperature contrast, a consequence of recirculation efficiency, should increase with increasing equilibrium temperature is supported by calculations of the recirculation efficiencies for a number of Hot Jupiters \citep{won16}. This trend is currently explained as a consequence primarily of atmospheric radiative heating/cooling \citep{kom16}.

The expectation of thermal inversions in the atmospheres of at least some giant exoplanets is motivated both by the presence of an inversion layer in Jupiter's atmosphere, as well as the relatively much stronger instellation for close-in planets that can drive such an inversion. \citet{knu08} used eclipse observations of HD 209458 b in multiple Spitzer IRAC bands to suggest that atmospheric models with thermal inversion better explain the observed depths than models without inversion. These conclusions were consistent with the data, but there was an acknowledgement that a truly robust determination of inversion layers would require more precise observations \citep{mad10}. The evidence for inversion in HD 209458 b in particular remains controversial \citep[see e.g., ][]{dia14,sch15a,lin16}. Analyses of photometry from other planets also suggest that the data are consistent with weak thermal inversions \citep{mad11,oro14}. The expectation persists that inversion should exist in at least some subset of highly-irradiated planets \citep{for08,spi09,par15}, and recent studies of the highly-irradiated planets WASP-18 b \citep{arc18} and WASP-33 b \citep{hay15,nug17} indicate their data are consistent with inversion layers.

In stark contrast to the analyses that have presented evidence for atmospheric chemistry, inversions, and structural profiles, authors such as \citet{han14} have been argued that, in the majority of cases, the data are no more consistent within statistical uncertainty with spectral retrieval models than they are with a simple blackbody radiative model. This range in optimism of interpretation is quite striking, and has important implications for how the forthcoming observations of short-period planets with JWST are to be interpreted. The current disconnect that permits a startling range of interpretation can be ascribed in part to historical accident. The Spitzer Mission was designed and constructed prior to the discovery of transiting extrasolar planets \citep{wer04}, and so its suite of instruments was not necessarily optimized for monitoring short-period planets. Indeed, given that the Spitzer Space Telescope was not specifically tuned for exoplanet observations, it is remarkable that some of the most exciting scientific results from the mission have come in connection with exoplanet-related observations.

Careful attention has been paid to the instrumental systematics in Spitzer data. The two major types of systematic effects are
\begin{enumerate*}[label={(\roman*)}]
\item the detector ``ramp'', a measurable brightening in time which has been observed in IRAC photometry, most notably at 8.0 $\mu$m \citep{dem06,gri07,knu08,knu09b,des09,ago10,tod10}, and
\item the ``pointing oscillation'' \citep{knu08,dem12,tod14}.
\end{enumerate*} Similar systematics appear to exist in the mid-infrared Spitzer MIPS instrument \citep{cro12}, complicating constraints on light curve parameters. The properties of the instrumental effects on pixel-to-pixel and intra-pixel variability are now well-documented \citep{car12,bei14}, and various analyses have employed novel reduction techniques designed to both address known systematics \citep[e.g.][]{dia14,tod14} and check the consistency of results derived from the photometry \citep[e.g.][]{ing16}. These considerations will certainly persist as we move to the next generation of space-based missions.

With Spitzer now in its end stages, and with JWST not yet on sky, there is a window of opportunity for a fresh assessment of the strengths of the various observational interpretations that have been offered. In this paper, we report the results of a uniform assessment of the existing secondary eclipse measurements and full- or extended-phase light curve photometry of close-in giant planets. We argue that a simple, but physically grounded 4-parameter model, which includes the planetary rotation rate, the time scale of atmospheric radiative response, the global planetary energy surface flux, and the planetary albedo can be used to characterize the observed planets at a level of detail that consistently matches the quality of the underlying data.

The plan of this paper is as follows. In \S \ref{sec:photometry}, we review the current catalog of secondary eclipse measurements that have been carried out from ground and space, and collect and discuss a normalized set of Spitzer's full phase photometric measurements. This background is motivated by the wealth of individual analyses available in the field, and lays a foundation for the overarching approach we take in re-analyzing the Spitzer light curves.  The methodology of our analysis, including model physics, is covered in \S \ref{sec:model}. We present the analysis in \S \ref{sec:results}, and discuss how these results tie in with future observation planning in \S \ref{sec:discussion}.

\section{Photometry}\label{sec:photometry}
Broadband photometry offers a low-resolution but high measurement signal-to-noise probe of planetary thermal emission. For a selection of planets orbiting relatively bright parent stars, photometry in near-infrared bands has been obtained throughout the orbit; such data are referred to as light curve observations. Full- or extended-phase photometry can be obtained through either fortuitous acquisition or planned observations. For example, the \textit{Kepler} satellite returned data for a handful of planets that are hot enough or reflective enough to produce significant optical emission \citep{fai15,shp17,mil17}. The optical light curves of these planets were discerned after the primary transits were detected. In other cases, the \emph{Spitzer Space Telescope} has made specifically targeted long-duration observations (sometimes lasting more than a week) of individual transiting planet bearing stars \citep[e.g.][]{lew13}.

The Spitzer mission, during both its cryogenic and ``warm'' phases, has obtained spectra, eclipse, transit, or full-phase observations of over 100 planet-bearing stars \citep{hane14}. Spitzer's IRAC detector \citep{wer04} has operated in four wavelength channels centered at 3.6, 4.5, 5.8, and 8.0 $\mu$m. During the warm phase, only the 3.6 and 4.5 $\mu$m channels have been available. The Infrared Spectrograph \citep{hou04} has provided spectra as well as photometry at 16 $\mu$m in the peak-up imaging mode, and the Multiband Imaging Photometer (MIPS) instrument \citep{rie04} provides photometry in the mid- to far infrared, particularly at 24 $\mu$m. However, all light curves analyzed in this work (listed in \S \ref{sec:photometry:phasecurves}) were reduced and rebinned from data from the IRAC channels alone.

\subsection{Secondary Eclipse Measurements}\label{sec:photometry:eclipses}
Full- or extended-phase light curve observations only exist for a small number of planets; a much larger sample of planets have had their secondary eclipse depths measured. Secondary eclipses permit an assessment of the day-side temperatures and meteorological conditions. During the cryogenic phase of its mission, Spitzer observed 15 exoplanets in secondary eclipse in a selection of its 3.6, 4.5, 5.8, and 8 $\mu$m channels. Following depletion of its liquid helium, a substantial number of additional secondary eclipses at 3.6 and 4.5 $\mu$m have been detected, and at present, a total of 32 planets have eclipse depths measured. In total there are over 100 individual measurements made at bandpasses ranging from the Kepler optical band \citep[e.g.][]{ang15} to ground-based measurements in J (1.22 $\mu$m), H (1.63 $\mu$m), and K (2.19 $\mu$m) \citep[e.g.][]{cro11,and13,che14}.

A handful of planets have been measured (albeit asynchronously) in five or more bands, including  J, H, and K, and many planets have been observed in three or more bands. A set of secondary eclipse measurements for a given planet at a range of different bandpasses amount to a low-resolution day-side planetary emission spectrum, and there has been a substantial effort to interpret the observational results with theoretical models. For example, \citet{for08} suggest that strongly irradiated atmospheres (corresponding roughly to the half of the observed planets that receive the largest orbit-averaged fluxes) have thermal inversions. Inversions are believed to arise from the presence of hardy molecules such as TiO or VO which absorb and re-radiate starlight at low pressures high in the atmosphere; this prediction has been supported by more recent models such as in \citet{par15}. A related, long-running class of models was presented by \citet{bur08}, which, in addition to varying the presence of a generic high-altitude gray absorber, also incorporate a heat sink added at depth to facilitate redistribution of heat from the planetary day-side to the night-side. \citet{knu10} presented evidence for empirical correlations which suggest that chromospheric activity inhibits the formation of such thermal inversions, perhaps by destroying the inversion-producing molecules through an elevated flux of high-energy photons. \citet{mad12} as well as \citet{mad10} proposed that C/O ratios constitute a key dimension along which planets can exhibit planet-to-planet variation in secondary eclipse depths.

In the last several years, developments in the reduction techniques for Spitzer photometry have spurred re-analyses of existing data \citep[e.g.][]{lew13,tod13,tod14,won14,won15,won16}. With sufficient photometric signal-to-noise, finer structure in the eclipse light curve can reveal features of the spatial intensity of the planet's day-side \citep{rau07}. This manifests itself in variations of the ingress and egress, which can be interpreted using spherical harmonics in brightness on the planet's visible disk \citep{maj12,dew12}. In addition to single-eclipse profile variations, multiple observations of eclipses for a given planet over several years constrain not only the orbital period and ephemerides, but have led to measurements of orbital periastron precession \citep[e.g.][]{won14}.

A handful of planets on modestly to highly eccentric orbits have been observed during their eclipses, including GJ 436 b \citep{dem07,ste10}, HAT-P-2 b \citep{lew13}, XO-3 b \citep{won14}, and HD 80606 b \citep{lau09,dew16}. Since the atmospheres of eccentric planets undergo time-dependent instellation, in theory, the dynamics may be significantly different than those on circular orbits. For HAT-P-2 b and HD 80606 b, the two most eccentric cases, the orientation of their orbits with respect to our line of sight is such that periastron and secondary eclipse occur within intervals that are relatively short compared with their orbital periods, allowing valuable assessment of the intense day-side heating that occurs near periastron.

\begin{figure}[htb!]
\begin{center}
\includegraphics[width=8.5cm]{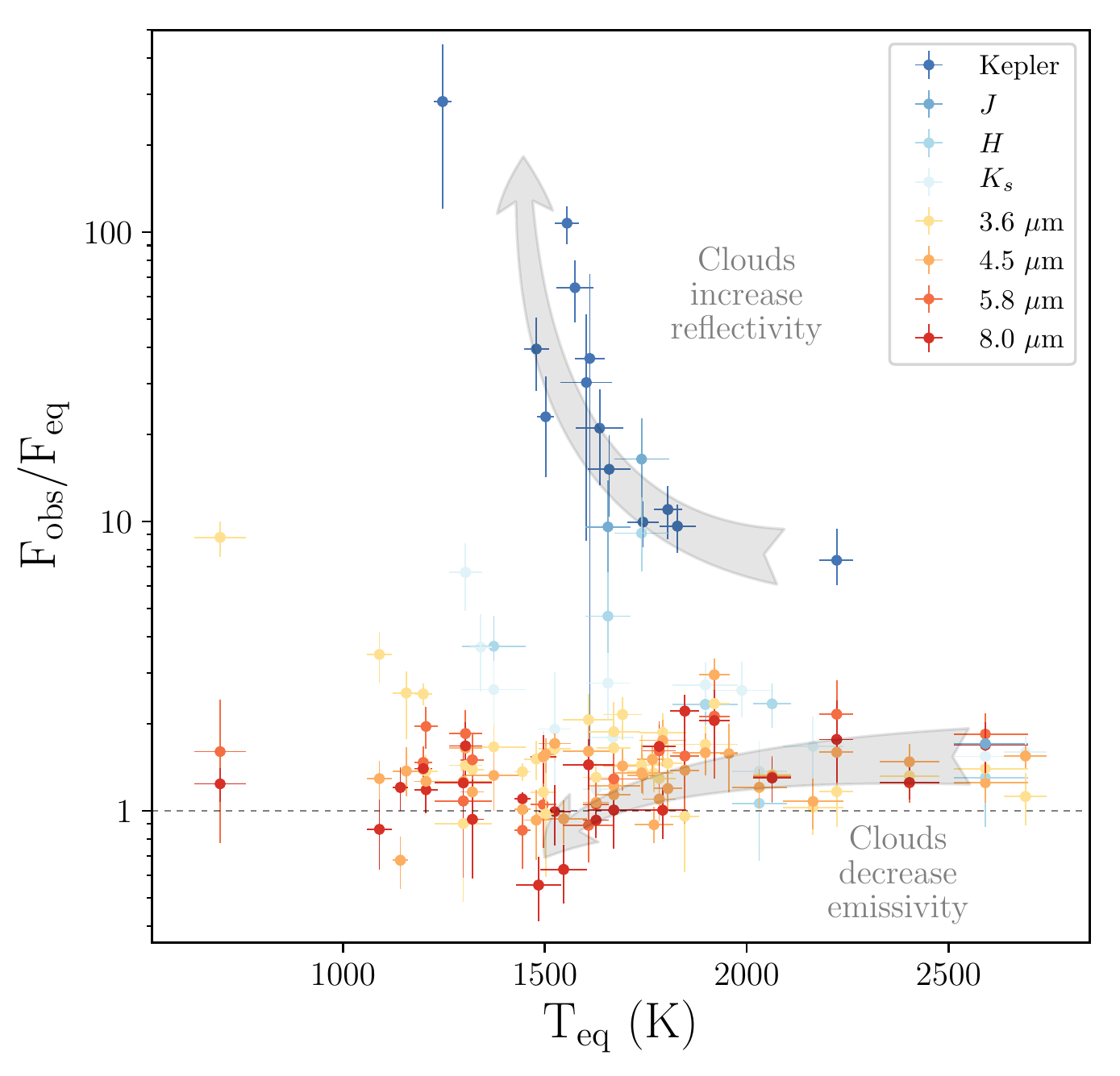}
\caption{Collected secondary eclipse measurements of 53 planets, spanning one optical and seven infrared photometric bands. We plot the ratio, $F_{\rm obs}/F_{\rm eq}$ of the observed flux, $F_{\rm obs}$, of the planet at secondary eclipse to the flux, $F_{\rm eq}$ that would result if the planet were a globally uniform blackbody radiating at the orbit-averaged equilibrium temperature $\bar{T}_{\rm eq}$. Values for $F_{\rm obs}/F_{\rm eq}$ generally exceed unity. This trend likely arises from inefficient transfer of heat from the day side to the night side, but can be attribued to a variety of additional contributions discussed in \S \ref{sec:photometry:eclipses}.}
\label{fig:eclipse_flux}
\end{center}
\end{figure}

\begin{figure}[htb!]
\begin{center}
\includegraphics[width=8.5cm]{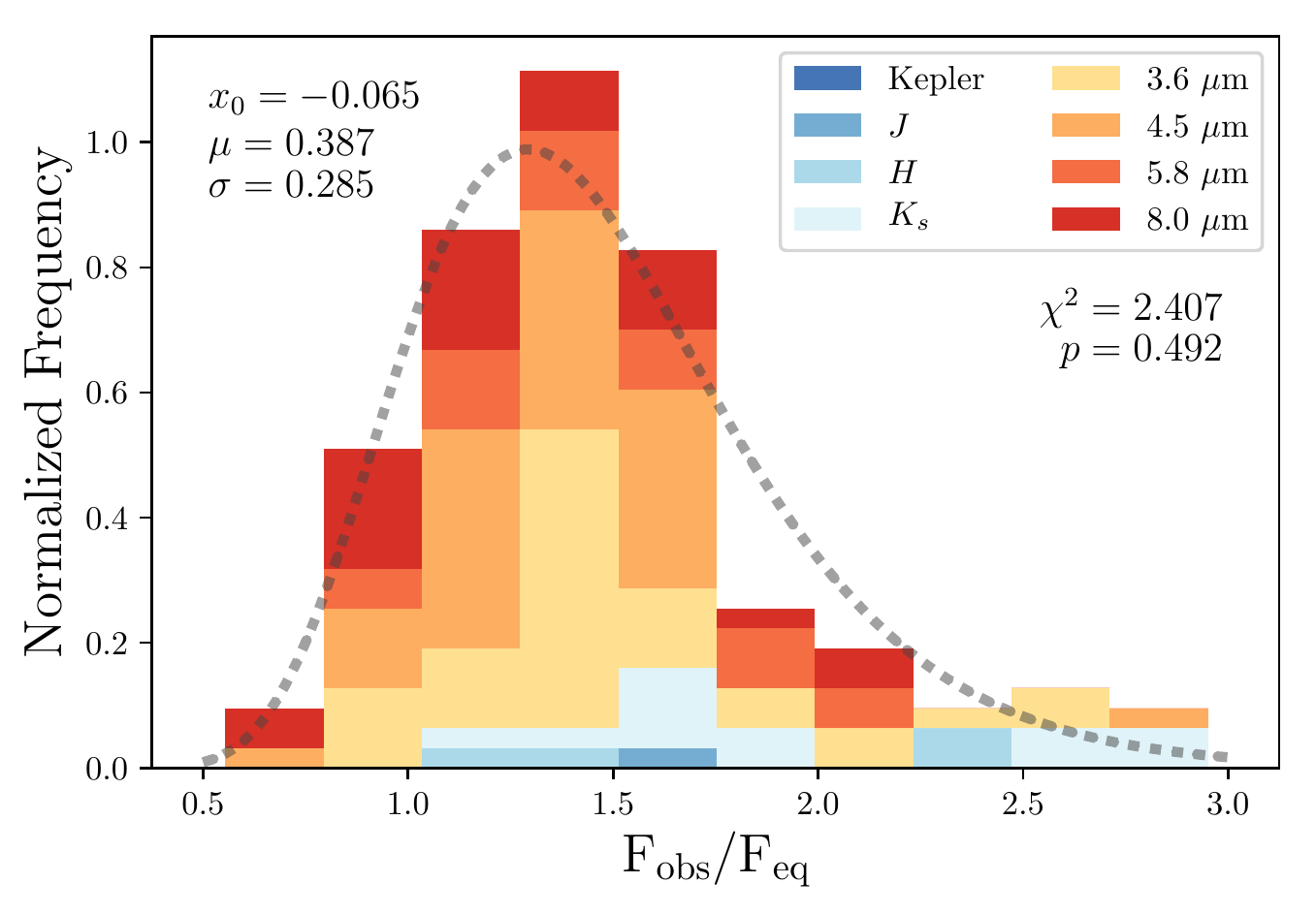}
\caption{The stacked, normalized distribution of eclipse flux ratios in the Spitzer IRAC bands (warm colors) as well as $JHK_s$ (cool colors) appears to follow a log-normal distribution (Equation \ref{eq:lognormal}). Note that we have truncated the histogram at a flux ratio of 3, but have included them in Figure \ref{fig:eclipse_flux}.}
\label{fig:eclipse_flux_lognorm}
\end{center}
\end{figure}

Figures \ref{fig:eclipse_flux}--\ref{fig:eclipse_flux_lognorm} show distributions of the observed eclipse depth in units of the expected depth if the planet were uniformly ($4\pi$) radiating as a blackbody at its equilibrium temperature, $\bar{T}_{\rm eq} = T_{\mathrm{eff}} \sqrt{R_\star / \left(2 a\right)}$, where $T_{\mathrm{eff}}$ and $R_\star$ are the effective temperature and radius of the host star, and $a$ the orbital semi-major axis. A ratio considerably above unity suggests the planet is radiating more strongly than would be expected from a blackbody at the planetary orbit-averaged equilibrium temperature. Elevated flux ratios can be due to any number of factors spanning a variety of atmospheric and meteorological conditions.

We note the general trend in Figure \ref{fig:eclipse_flux} of increasing observed-to-thermal ratios with decreasing wavelength. For the bluest bands shown, the highest observed fluxes stem from reflected light. On the redder end, 13 eclipsing planets, or about 1/4 of the sample, have sub-thermal fluxes (ratios below unity). All of the sub-thermal measurements are from Spitzer IRAC bands, and span equilibrium temperatures from 1142--1862 K. The minimum value of 0.63 occurs for WASP-17 b at 8.0 $\mu$m, with $\bar{T}_{\rm eq} = 1547$ K. This points to a prediction \citep{gao18} which identifies a possible cloud transition at roughly 1500 K. With this is mind, we propose a context for the observed eclipse depth ratio distribution where the reddest bands exhibit a minimum near this temperature, while in the optical the flux ratio reaches a peak. If clouds play an important role, the emissivity in the infrared may be suppressed, while the contribution to the flux in reflected light will be enhanced.

Measurements of secondary eclipse depths for exoplanets are commonly reported in connection with comparisons to model spectra from model atmospheres of irradiated planets \citep[e.g.][]{mah13,bas13}. The atmospheric models used for comparison often have a substantial degree of sophistication and are informed by multiple free parameters and physical assumptions. In most studies, some of the atmospheric parameters, such as the presence or absence of a high-altitude inversion-producing absorber, or the global average efficiency of day- to night-side heat redistributions are varied, whereas others, such as the assumption of hydrostatic equilibrium and global energy balance, are assumed settled.

Invariably, the number of implicit and explicit parameter choices substantially exceed the number of measurements, and make it difficult to evaluate the degree to which a given, highly detailed planetary atmospheric model exhibits explanatory power. The Central Limit Theorem states that any quantity that is formed from a sum of $n$ completely independent random variables will approach a normal (Gaussian) distribution as $n\rightarrow \infty$. By extension, any quantity that is the product of a large number of random variables will be distributed approximately log-normally. To examine whether variations in eclipse fluxes exhibit this behavior, we fit the distributions of flux ratios with a log-normal probability distribution
\begin{equation}\label{eq:lognormal}
f\!\left( x \right) = \frac{1}{\sqrt{2 \pi}\,\sigma \left(x-x_0\right)} \exp{\left\{ -\frac{1}{2} \left[ \frac{\log{\left(x-x_0\right)}-\mu}{\sigma} \right]^2 \right\}}
\end{equation}
with free parameters $x_0$, $\mu$, and $\sigma$. Under the assumption the distribution is purely a product of strictly positive random variables, $x_0$ should theoretically be zero. 

As can be seen in Figure \ref{fig:eclipse_flux_lognorm}, the stacked nomalized distribution of eclipse flux ratios can be fit by a log-normal distribution. We choose here to focus on the eclipse depths within the range where thermal emission should be most dominant. Considering the ten histogram bins below a flux ratio of 3, the associated chi-square statistic for the fit is $\chi^2=2.41$, with an associated p-value of $p=0.49$ for the null hypothesis that the data are consistent with log-normal. This lends some credence to the hypothesis that the observed infrared fluxes are affected by a significant number of physical parameters that vary strongly from one planet to the next. In other words, the observed aggregation of planets points towards a substantial diversity of worlds; any number of proposed physical processes from the literature could contribute to this distribution.

\begin{figure*}[htb!]
\begin{center}
\includegraphics[width=17cm]{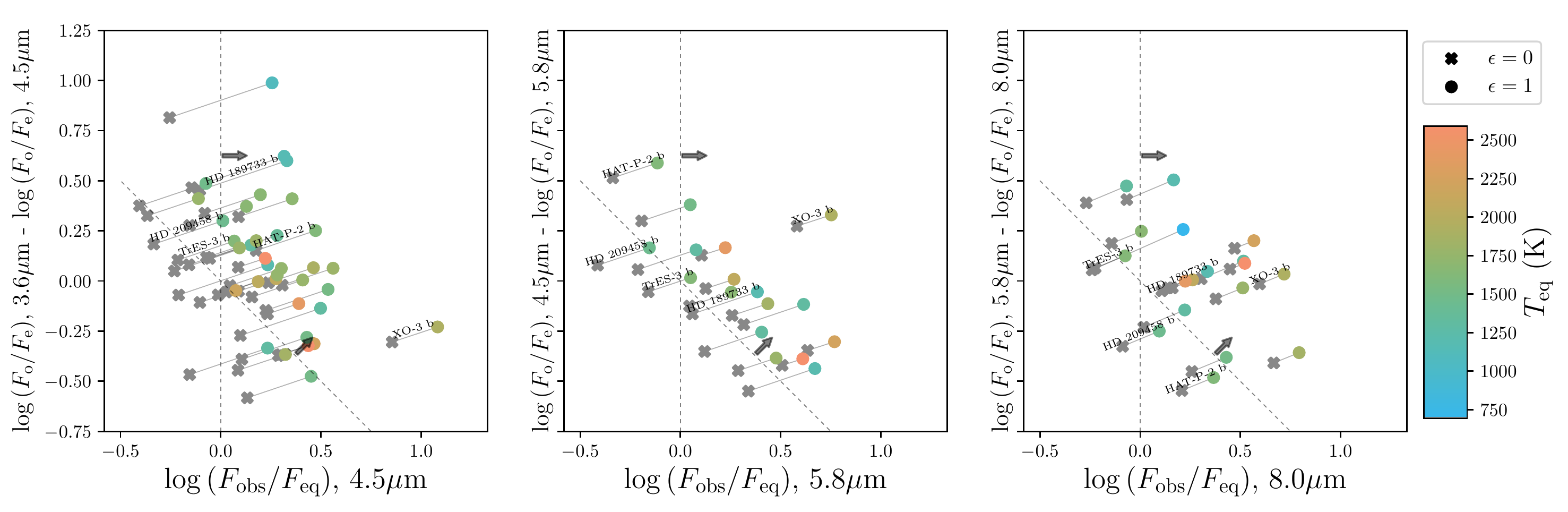}
\caption{Eclipse color-magnitude diagrams are constructed by plotting the logarithm of the eclipse flux ratios in a given band (``magnitude'') versus the difference in logarithmic flux ratios between two bands (``color'', written as the difference of wavelengths as shorthand). We present both the assumed case of perfect heat redistribution (uniform temperature, denoted $\epsilon = 1$) and the opposite limiting case (i.e. no circulation to the night-side, denoted $\epsilon = 0$), which amounts to multiplying the temperature by a factor of $\sqrt{2}$ \citep[see e.g.][]{cow11b}. The vertical dashed line shows where the redder band is thermal (i.e. log of the flux ratio is zero), and the diagonal dashed line shows the equivalent for the bluer band. The arrows point to the region of the diagram where both bands are super-thermal.}
\label{fig:eclipse_flux_color}
\end{center}
\end{figure*}

The idea of comparing eclipse depths in two bands relative to the equilibrium depth was explored in \citet{bas13}. The day-side emission of close-in planets have been incorporated into color-magnitude diagrams that explore whether Jupiter-size planets \begin{enumerate*}[label={(\roman*)}]
\item continue the trend of low-mass stars and brown dwarfs,
\item behave similarly to blackbodies, or
\item neither
\end{enumerate*} \citep{tri14}. In principle, we can combine elements of both approaches by using flux ratios (observed flux relative to the expectation of thermal radiation) in place of absolute magnitudes in a color-magnitude diagram. We construct flux ``colors'' by taking the difference in flux ratios of any two bands, and choose to plot the logarithms of the flux ratios as an analogy to traditional magnitudes in Figure \ref{fig:eclipse_flux_color}. Here we plot two sets of points, one for each extreme of heat redistribution efficiency. The orbit-averaged equilibrium temperature assumes perfect redistribution; for the opposite case that all the energy remains on the planet's day-side, we adopt the equilibrium temperature definition from \citet{cow11b}, which is higher than the isothermal case by a factor of $\sqrt{2}$\footnote{Since the thermal flux ratios depend on the blackbody spectrum at a given temperature, the corresponding scaling of the flux ratios seen in Figure \ref{fig:eclipse_flux_color} is non-linear.}. Our results support the observation from \citet{bas13} that all observed planets have super-thermal fluxes in at least one band, assuming perfect $4\pi$ heat redistribution. In each color, there additionally appears to be a trend of decreasing color metrics with increasing flux ratio in the redder band. While there is no clear single reason that would produce this trend, it is consistent with a combination of physical processes which depend on wavelength.

\subsection{Light Curve Measurements}\label{sec:photometry:phasecurves}
Full phase photometry (or significant partial phase photometry) exists for the 13 planets that we discuss in this paper. These data permit empirical testing of evolutionary predictions of radiative models over time and over a range of longitudinal views. Moreover, for all of the planets that we consider, save GJ 436~b and HD 209458~b, Spitzer light curves in multiple bands are available, inviting analysis of the consistency (or the lack of consistency) of the best-fit physical parameters across different wavelengths.

The history of published phase photometry from Spitzer now spans over a decade, starting with the report of observations of a significant fraction of the orbit of HD 189733~b in 8.0 $\mu$m \citep{knu07a}. Since phase variations can in principle offer valuable evidence of atmospheric properties, publications of light curves often include analyses of the phase variations within the structure of radiative and dynamical models. In general, the physical realism of these models has developed over time, as more data and multiple bands become available. Models have ranged from analytical treatments that use a limited number of parameters incorporating thermal emission, ellipsoidal variations, and in some cases reflected light \citep[e.g.][]{knu07a,lau09,cow12,dew16,von16}, to fully 3-dimensional circulation models \citep[e.g.][]{sho02,sho09,sho15,lew10,lew14,lew17,dob13,amu16,kom16}. To set the context and to chart the development of ideas, we review the planet-specific histories of published full- or extended-phase light curves, and any models or predictions of properties of the planetary atmospheres used to analyze the data. The order reflects the chronology of the earliest available phase photometry for each planet.
\begin{description}
\item[HD 189733 b] One of the earliest transiting Hot Jupiters, HD 189733 b was the first exoplanet to have been mapped thermally \citep{knu07b}. Also within the publication of this map was an observation of a significant fraction of the planetary orbit in IRAC's 8.0 $\mu$m band. This early example of exoplanetary phase photometry includes both transit and eclipse, and allowed the authors to calculate a full range of brightness temperatures. The light curve also demonstrated a non-zero phase offset\footnote{Phase offsets often refer to the offset in phase of the time of maximum flux with respect to the time of secondary eclipse. If the planet heats and cools instantaneously, one expects the highest flux at the ``full'' phase of observation, i.e. the time of eclipse. A sufficiently nonzero offset may indicate winds, a non-instantaneous radiative timescale, a combination of both, or possibly another set of factors.}, which the authors show could be attributed to an eastward hotspot. \citet{knu09a} complemented their 8.0 $\mu$m light curve with a partial light curve in 24 $\mu$m. The phase offset and brightness temperatures are very similar in the two bands, with the latter suggesting a vertical temperature homogenization (since the redder photosphere is expected to lie at a deeper isobar). Expectations for the night-side methane abundances and equilibrium chemistry, however, are inconsistent with both this conclusion and a high-altitude cloud cover \citep{knu09a}. The authors conclude that the presence of efficient day-night heat transport is clear, but note that more complex phenomena would require better 1-D and 3-D modeling.

HD 189733 b presents a very high signal-to-noise target, and nearly-full phase coverage  in 3.6 and 4.5 $\mu$m was presented by \citet{knu12}. The phase offsets for both the minima and maxima in these bands are consistent with a coherent longitudinal wind, which was first suggested on the basis of the 8.0 $\mu$m data. On the instrumental side, \citet{knu12} revisit the 8.0 $\mu$m phase data, taking a more developed account of known instrumental systematic effects, concluding that the apparent flux minimum in the 8.0 $\mu$m light curve could well be attributed to the detector ``ramp'' effect, rather than a physical phase offset.

Beyond this inference, the larger phase amplitude for 3.6 $\mu$m compared with 4.5 $\mu$m counters the expectation from 1-D models that redder wavelengths probe deeper atmospheric layers with larger heat capacity and smaller thermal variation. The addition of the bluer IRAC bands permits a more coherent analysis of the carbon chemistry (primarily CO and CH$_4$). From differences in transit depths between 3.6 and 4.5 $\mu$m, \citet{knu12} infer a possible excess in CO absorption. The corresponding full-phase light curves have a smaller amplitude of phase variations than what is predicted from 3-D circulation models which assume equilibrium chemistry. From this, \citet{knu12} point to a possible disequilibrium: a vertical mixing of atmospheric layers on the planet's night side, which could create an excess CO absorption signature.

\item[HD 80606 b] While its closest approach distance to its host star ($\sim 0.03$ AU) would qualify it as a Hot Jupiter, this planet's extreme eccentricity puts it on a 111-day orbit. \citet{lau09} presented roughly 30 hours worth of photometry at 8.0 $\mu$m, encompassing the planet's periastron passage. Given the orientation of the planet's orbit relative to our line of sight, the secondary eclipse occurs just a few hours prior to periastron passage. \citet{lau09} fitted the data using a 2-D hydrodynamical model with three parameters (two for pressure of the photosphere, one for albedo) that are fixed at physically-motivated values. In the analysis they also assume a pseudo-synchronous rotation period of roughly 40 hours. They find a radiative time scale of 4.5 hours at 8.0 $\mu$m, which is significantly shorter than the equivalent time scale on Earth. Given the assumed rotation rate and inferred radiative time scale, there is an expected significant decrease in flux following periastron which is not present in the data. From this the authors conclude there could either be efficient day-night heat advection, or a rotation period considerably different from the pseudo-synchronous value that emerges from visco-elastic tidal theory \citep{hut81}.

More recently, \citet{dew16} presented a new reduction of both the existing 8.0 $\mu$m photometry and a newer 4.5 $\mu$m light curve which also includes periastron and eclipse, and spans about 80 hours. They apply a simple radiative model with four parameters, similar to the one we employ for our current analysis, and find that HD 80606 b appears to rotate at a rate markedly slower than the theoretical pseudo-synchronous rate. The inferred radiative timescale of 4 hours, moreover, is much shorter than the inferred rotation period.

Most recently, \citet{lew17} have presented their own analysis of the 4.5 and 8.0 $\mu$m photometry. Their interpetation rests on a fully 3-D circulation model; it constrains the rotation rate and invokes cloud/aerosol dynamics to explain the time development of the light curves. Specifically, they investigate models with the planet spinning at the pseudo-synchronous rate, as well as at one-half and at twice the rate. The slowest rotation -- twice the pseudo-synchronous period -- most closely reproduces the amplitude and the timing of the phase variations occurring near periastron, although the phase offset departs slightly from full agreement as the period increases. Additionally, in connection with the longer-duration 4.5 $\mu$m light curve model, they find that admitting he presence of certain cloud species (most notably MgSiO$_3$) reduces the flux as the planet moves away from periastron, permitting approach to the near-zero levels seen in the data. 

\item[HD 149026 b] At a mass of 0.37 $M_J$ and a radius of 0.81 $R_J$, HD 149026 b was the first close-in gas giant observed to not have a significantly inflated radius. Partial-phase photometry in 8.0 $\mu$m was published in \citet{knu09b}, covering just under half an orbit. The light curve comprises 14 data points binned in 2.2-hour intervals, including one point obtained during eclipse. From this they employ a 2-hemisphere model \citep{cow08} which admits longitudinal variations in flux. Given the large uncertainties in the available data, they were not able to retrieve a phase offset statistically significant from zero.

\citet{zha18} present their own recent reduction of phase photometry in 3.6 and 4.5 $\mu$m from 2011. Their work uses calculated phase offsets to highlight a correlation between offset and the planet's orbit-averaged equilibrium temperature. The best-fit phase offsets from the 3.6 and 4.5 $\mu$m photometry are both significantly different from zero and in complete disagreement with each other. Given this disagreement, the authors point to potential systematic errors in the 3.6 $\mu$m data, whose offset is not supported by thermal models that assume synchronous rotation. These problems are not amenable to correction by any known algorithm at the pixel-to-pixel level, and even with a separate treatment for the systematics in the 3.6 $\mu$m data, the authors conclude that any physical inferences stemming from phase offsets or variation amplitudes are unreliable.

\item[WASP-12 b] WASP-12 b is thought to be one of the most tidally distorted planets known; its measured radius is roughly 1/2 the radius of its Roche lobe. \citet{cow12} present full-phase light curves in both 3.6 and 4.5 $\mu$m. The authors employ sinusoids to model both the thermal and ellipsoidal variations, the latter of which connects to the expectation that the tidal distortion of the planet is significant enough to affect the shape of the phase curve. The measured transit depth at 3.6 $\mu$m is deeper than at 4.5 $\mu$m, implying a larger effective radius for the planet (from the observed cross-section defined by the day-night terminator). Since this is opposite the expectation of the thermal models considered in the paper, the authors suggest the presence of a haze with a large scale height. Comparing the eclipse depths with 1-D radiative models, the authors find that their observations are consistent with an enhanced C/O ratio, but the uncertainties in the depths limit this to a marginal detection.

The ellipsoidal variations appear much more significant for the 4.5 $\mu$m data than for 3.6 $\mu$m. If of physical origin, this implies the 4.5 $\mu$m photosphere occurs much higher in the planet's tidally distorted atmosphere. Under this interpretation, the outermost, Roche-filling layers of the atmosphere would be optically thin. The authors note, however, that uncharacterized detector systematics could explain much of the difference in the strength of the ellipsoidal variations between the two bands.

\item[WASP-18 b] This iron-density Hot Jupiter has a mass comparable with the boundary between planet and brown dwarf (10.38 $M_J$), and orbits its star in just under an Earth day. \citet{max13} present full-phase light curves in both 3.6 and 4.5 $\mu$m. They employ a thermal model to explain modest phase variations, and note that in both bands, there appears to be uncharacterized systematic noise at the level of $10^{-4}$ in the flux ratio. They note that the amplitude of the WASP-12 b ellipsoidal variations at 4.5 $\mu$m presented in \citet{cow12} are about an order of magnitude larger than the noise floor imposed by the systematics. Similar variations are not present in the 4.5 $\mu$m data for WASP-18 b.

\item[HAT-P-2 b] A highly massive Jupiter-sized planet at 9.09 $M_J$ and 1.16 $R_J$, implying a density twice that of Earth, HAT-P-2 b is currently the planet with the most comprehensive set of near-infrared photometry. Full phase data are available in both 3.6 and 4.5 $\mu$m, as well as partial coverage of the orbit (including both transit and eclipse) in 8.0 $\mu$m \citep{lew13}. Considering just eclipse depths, \citet{lew13} find a best match to models which include high-altitude absorbers; however, an additional eclipse depth at 5.8 $\mu$m disagrees with such an interpretation. The light curves are compared with 1-D radiative transfer models which assume instantaneous radiative timescales. The shape and phase offsets suggest a westward cool spot following transit, which is the opposite behavior to what one would expect from a super-rotating wind. Further analysis of the peak observed in 8.0 $\mu$m, which is higher than even the flux predicted with zero Bond albedo, suggest that chemistry and/or vertical temperature variations must play a role in shaping the phase variations. The authors fit a sinusoidal model similar to \citet{cow08}, but, accounting for the eccentric orbit, report the expected flux as a function of true anomaly rather than time. The consequence of such a model is that it implies the day-side brightness profile is constant, which is not expected from a physical standpoint for planets on significantly eccentric orbits. The conclusion is that such a model fits the phase variations largely due to geometric effects from our line of sight.

\item[HD 209458 b] Two of its notable milestones include both being the first planet detected via transit, as well as the first planet with a detected atmosphere. The only full-phase light curve to date is in 4.5 $\mu$m, published in \citet{zel14}. The authors find the day-side flux is comparable with the results of the model predictions of \citet{sho09}. However, the models overestimate the night-side flux; this disagreement is explained as a consequence of assuming equilibrium chemistry. In particular, a high abundance of CH$_4$ on the night-side could mitigate the difference between models and data. Beyond this, some systematics remain in the reduced data around secondary eclipse and near quadrature; \citet{zel14} find it likely neither to be Spitzer pointing errors nor transient stellar variability. Given a similar systematic effect for HD 189733 b in 3.6 and 4.5 $\mu$m \citep{knu12}, the authors conclude this is likely the result of some residual, uncharacterized Spitzer systematics.

\item[GJ 436 b] GJ 436 b's significant eccentricity $e=0.16$, and orbit which is almost orthogonal to its host star's spin, present intriguing clues to the past evolution of this system. \citet{lan14} present an analysis of a full-phase light curve in 8.0 $\mu$m in a paper which uses occultation depths spanning 3.6--24 $\mu$m to confirm previous inferences about the planet's high atmospheric metallicity. The authors compare the results from 3-D circulation models from \citet{lew10}, which account for atmospheric metallicity and pseudo-synchronous rotation, with their own model which assumes isotropic thermal emission. The \citet{lew10} models need a metallicity much higher than solar to approach the thermal flux variations observed in the full-phase photometry. While their measured transit and eclipse depths at 3.6, 4.5, and 8.0 $\mu$m are shallower than preceding works, their conclusions about atmospheric chemistry are not affected qualitatively.

\item[WASP-43 b] WASP-43 b has the shortest orbital distance in our sample of planets with phase photometry, with $a=0.015$ AU. \citet{kat15} compare broadband light curves from the Wide Field Camera WFC3 of the Hubble Space Telescope, as well as light curves from near-infrared bands ranging from 1.14--1.63 $\mu$m, with 3-D atmospheric circulation models. The models predict a strong equatorial jet and are able to reproduce the peak day-side emission with a metallicity 5 times that of the solar metallicity. However, their model could not reproduce a low enough night-side flux to match the observed minimum. The authors suggest that either the night-side is brighter at lower effective temperatures, or that high-altitude clouds could play a role in limiting the observed flux.

More recently, \citet{ste17} present a sophisticated analysis of Spitzer 3.6 and 4.5 $\mu$m full-phase light curves. In their analysis, the metallicity of the atmosphere is consistent with solar, and they constrain the abundance of H$_2$O vapor on the day-side. In order to match the observed night-side fluxes, the authors include effects from clouds/hazes, though their results suggest the cloud cover would need to be inhomogeneous. Furthermore, the authors note that their measured night-side fluxes at 3.6 $\mu$m were dependent on date of observation, which means the emission levels must be partially degenerate with some position-dependent Spitzer systematic.

\item[WASP-14 b] WASP-14 b's high mass of 7.59 $M_J$ at a radius of 1.24 $R_J$ makes it one of the few Hot Jupiters with a density rivaling that of the Earth's. \citet{won15} present full-phase light curves in both 3.6 and 4.5 $\mu$m. While the 3-D circulation models they adapt from the work of \citet{sho09} capture the day-side fluxes in both bands as intermediate to the models with and without TiO/VO absorbers, the night-side fluxes they predict disagree with the observed minima; the 3.6 $\mu$m minimum is over-predicted while the 4.5 $\mu$m is under-predicted. The authors point to an enhanced C/O ratio, or perhaps the presence of high-altitude silicate clouds, to explain both disagreements. Their dynamical model, the 3-D SPARC model \citet{sho09}, assumes the planet on its modestly eccentric orbit is rotating pseudo-synchronously, which is only slightly different from assuming spin-orbit synchronization.

\item[HAT-P-7 b] The calculated spin-orbit misalignment of the HAT-P-7 system is almost exactly 180$^\circ$, implying an almost perfectly retrograde orbit. \citet{won16} present full-phase light curves in both 3.6 and 4.5 $\mu$m. Here their analysis of the eclipse depths lead to supporting a day-side thermal inversion with inefficient day-night heat redistribution. 1-D and 3-D models cannot explain the low night-side temperature in 3.6 $\mu$m. The authors suggest this is evidence for a high C/O ratio. HAT-P-7 b is also placed in the context of a trend of planets with relatively inefficient heat redistribution and high instellation. Even with a detailed correction for the Spitzer ramp as well as intrapixel and stellar variability, the authors note there still exists uncorrected noise in the 3.6 $\mu$m photometry.

In addition to Spitzer, \citet{arm16} present a phase curve from \emph{Kepler} photometry, making HAT-P-7 b the first planet with phase photometry in both the optical and near-IR. Their recent analysis suggests atmospheric variability may be due to variable wind speeds causing variable cloud cover on the day-side.

\item[WASP-19 b] WASP-19 b has the shortest orbital period of the planets we consider, at just under 19 hours. \citet{won16} present full-phase light curves in both 3.6 and 4.5 $\mu$m. Unlike HAT-P-7 b, the authors find no evidence for a thermal inversion in the atmosphere; additionally, the day-night heat redistribution is more efficient. 1-D and 3-D model comparisons cannot explain the low night-side temperature in either band. The authors suggest this discrepancy could be resolved by including high-altitude silicate clouds. Their analysis places WASP-19 b within the trend of planets with increasingly inefficient redistribution with instellation.

\item[WASP-33 b] Orbiting an A-type $\delta$ Scuti variable star, WASP-33 b is the hottest planet in our sample, with $T_{\mathrm{eq}} = 2723$ K. \citet{zha18} present their own recent reduction of full-phase photometry in 3.6 and 4.5 $\mu$m from 2012. As with HD 149026 b, the phase offset of WASP-33 b is used to suggest that observed phase offsets appear to correlate with equilibrium temperature. However, the uncharacterized systematics affecting the 3.6 $\mu$m light curve for HD 149026 b are not seen in the WASP-33 b data.

\end{description}

The analyses of Spitzer occultations and light curves has developed from initial measurements of phase offsets and brightness temperatures in the 8.0 $\mu$m band, to inferences of thermal inversions, carbon chemistry, and optical absorbers with the addition of warm Spitzer (3.6, 4.5 $\mu$m) photometry. Despite these advances in multi-wavelength photometry, no data are spared from persistent systematic effects in the Spitzer instrumental response, some of which appear to remain uncharacterized even in recent analyses. Additionally, 3-D modeling has not yet demonstrated a consistent ability to capture broad features of the available light curves, for example the amplitude of phase variations (e.g. HD 209458 b). The distribution of eclipse depths relative to the thermal expectation (\S \ref{sec:photometry:eclipses}) suggests we are encountering contributions from a potentially large number of physical processes, an interpretation which could very well extend to extended phase photometry. With this in mind, we put forth a straightforward physically rudimentary model in our current analysis (\S \ref{sec:model}), to explore the limitation of sophistication of physical interpretation which might be imposed by the quality of Spitzer observations of exoplanets.

\section{Modeling the Orbit Geometry and Thermal Evolution}\label{sec:model}
Our analysis begins with published properties of exoplanet systems (Table \ref{table:planet_properties}). The orbital geometries of all the systems studied are well constrained, with fractional uncertainities at a maximum of $10^{-5}$ in period and $10^{-3}$ in eccentricity.

\newcommand\tn{\tablenotemark{*}}
\newcommand\ttn{\tablenotemark{**}}

\begin{deluxetable*}{ccccccccc}
\centering
\tabletypesize{\scriptsize}
\tablecaption{Orbital and Stellar Properties for Exoplanets with Spitzer Light Curves}
\tablewidth{0pt}
\tablehead{
\colhead{Name} & \colhead{$P$ (days)} & \colhead{$e$} & \colhead{$\omega\left(^\circ\right)$} & \colhead{Ref.} & \colhead{$M_\star$ $\left(M_\odot\right)$} & \colhead{$R_\star$ $\left(R_\odot\right)$} & \colhead{$T_{\textrm{eff}}$ (K)} & \colhead{Ref.}}

\startdata
GJ 436 b & $2.64389803^{+0.00000027}_{-0.00000025}$\ & $0.1616^{+0.0041}_{-0.0032}$ & $327.2^{+1.8}_{-2.2}$ & 1 & $0.556^{+0.071}_{-0.065}$ & $0.455 \pm 0.018$ & $3416\pm54$ & 2 \\    
HAT-P-2 b & $5.6334729\pm6.1\times10^{-6}$ & $0.5171\pm0.0033$ & $185.22\pm0.95$ & 3 & $1.36\pm0.04$ & $1.69^{+0.09}_{-0.08}$ & $6290\pm60$ & 3 \\
HAT-P-7 b & $2.2047372 \pm 1.1\times10^{-6}$ & $0.0016^{+0.0034}_{-0.0010}$ & $165^{+93}_{-66}$ & 4 & $1.47^{+0.08}_{-0.05}$ & $1.84\pm0.17$  & $6441\pm69$ & 5, 6 \\
HD 80606 b & $111.43740\pm7.2\times10^{-4}$ & $0.93286\pm5.5\times10^{-4}$ & $300.83 \pm 0.15$ & 7 & $1.007\pm0.024$ & $1.01 \pm 0.05$ & $5574\pm50$ & 8, 9 \\
HD 149026 b & $2.8758911 \pm 2.5\times10^{-6}$ & $\sim 0$ & N/A & 10 & $1.345 \pm 0.020$ & $1.541^{+0.046}_{-0.042}$ & $6160 \pm 50$ & 10 \\
HD 189733 b & $2.218575200 \pm 7.7\times10^{-8}$ & $\sim 0$ & N/A & 11 & $0.846 \pm 0.049$ & $0.805 \pm 0.016$ & $4875 \pm 43$ & 12, 13 \\
HD 209458 b & $3.52474859 \pm 3.8\times10^{-7}$ & $\sim 0$ & N/A & 14 & $1.131^{+0.026}_{-0.024}$ & $1.203\pm0.061$ & $6092 \pm 103$ & 13, 15 \\
WASP-12 b & $1.09142119 \pm 2.1\times10^{-7}$ & $0.0447\pm0.0043$ & $272.7^{+2.4}_{-1.3}$ & 16, 17 & $1.280 \pm 0.05$ & $1.630 \pm 0.08$ & $6300^{+200}_{-100}$ & 18, 19 \\
WASP-14 b & $2.24376524 \pm 4.4\times10^{-7}$ & $\sim 0$ & N/A & 20 & $1.392 \pm 0.040$ & $1.004 \pm 0.016$ & $5568 \pm 71$ & 21 \\
WASP-19 b & $0.788838989 \pm 4.0\times10^{-8}$ & $0.0020^{+0.0140}_{-0.0020}$ & $259^{+13}_{-170}$ & 4 & $0.904\pm0.040$ & $1.004 \pm 0.016$ & $5568 \pm 71$ & 22, 6\\
WASP-33 b & $1.2198669 \pm 1.2\times10^{-6}$ & $\sim 0$ & N/A & 23 & $1.495\pm0.031$ & $1.444 \pm 0.034$ & $7430 \pm 100$ & 23\\
WASP-43 b & $0.81347753 \pm 7.1\times10^{-7}$ & $0.0035^{+0.0060}_{-0.0025}$ & $328^{+115}_{-34}$ & 24 & $1.036 \pm 0.019$ & $0.667^{+0.011}_{-0.010}$ & $4520 \pm 120$ & 24\\
\enddata

\tablerefs{(1) \citet{mac14}; (2) \citet{von12}; (3) \citet{pal10}, \citet{lew13}; (4) \citet{won16}; (5) \citet{pal08} ($M_\star$, $R_\star$); (6) \citet{tor12} ($T_{\mathrm{eff}}$); (7) \citet{win09a}; (8) \citet{hebr10} ($M_\star$, $R_\star$); (9) \citet{mou09} ($T_{\mathrm{eff}}$); (10) \citet{car09}; (11) \citet{bal15}; (12) \citet{dek13} ($M_\star$); (13) \citet{boy15} ($R_\star$, $T_{\mathrm{eff}}$); (14) \citet{knu07a}; (15) \citet{tak07} ($M_\star$); (16) \citet{tur16} ($P$, $e$); (17) \citet{knu14} ($\omega$); (18) \citet{eno10}; (19) \citet{heb09}; (20) \citet{won15}; (21) \citet{jos09}; (22) \citet{tre13} ($M_\star$, $R_\star$); (23) \citet{col10}; (24) \citet{gil12}.}

\label{table:planet_properties}
\end{deluxetable*}

We model the planets using a latitude/longitude grid with cells of dimension $5^\circ\times 5^\circ$. Each cell is treated as a blackbody, with the thermal energy set by a combination of time-dependent instellation and a parametrized baseline temperature which attempts to capture, to first-order, heating from the planet interior (see \S \ref{sec:model:parameters}). Advection is not treated explicitly; instead the model uses a bulk rotation parameter that can represent either a differential average rotation of radiating surface material or a coherent solid-body rotation.

The time evolution of grid cell temperatures is combined with the known geometry of the orbit (with respect to Earth) and instrumental bandpasses to generate model full-orbit light curves for a given set of parameter values, which are adjusted to fit the data.

\subsection{Model Parameters}\label{sec:model:parameters}
Our fits to the \emph{Spitzer Space Telescope} data employ a simple 4-parameter blackbody radiative model. Each parameter captures a broad physical property of the planet.

\begin{description}
\item[Rotation Period $\left( P_{\textrm{rot}} \right)$] Coherent rotation of the planet is parametrized in units of either the orbital period (for circular orbits) or the theoretical pseudo-synchronous rate (for eccentric orbits)\footnote{The pseudo-synchronous rate approaches the orbital period as $e\rightarrow0$.}. As discussed below, this parameter may also equivalently represent a coherent velocity of an atmospheric layer probed by a specific wavelength.

\item[Radiative Timescale $\left( \tau_{\textrm{rad}} \right)$] The radiative timescale determines how quickly parcels of material change temperature in response to instellation. In general, the rate of change depends both on this parameter as well as the equilibrium temperature. The equilibrium temperature is determined by time-dependent properties including the star-planet separation and local stellar altitude (Equation \ref{eq:Teq}). Therefore, for the purposes of our analysis, we choose to report our results in terms of the corresponding radiative timescale $\tau_{eq}$ at a reference temperature given by the orbit-averaged equilibrium temperature $\bar{T_{\mathrm{eq}}}$.

\item[Minimum (``Night-Side'') Temperature $\left( T_0 \right)$] All energy sources other than instellation contributing to heating the atmosphere are folded into the minimum temperature. Examples of sources include tidal heating \citep{lev07,agu14} and, for the hottest planets, energy from the dissociation and recombination of hydrogen \citep{bel18}. However, our model cannot account for time-dependent variations in these processes.

\item[Albedo $\left( A \right)$] The scaling of the fraction of incident stellar flux absorbed allows us to account for a mean global reflectivity due to cloud cover. This parameter represents a characteristic Bond albedo.
\end{description}

This formulation allows us to model the largest-scale features observed in the Spitzer light curves. Most prominently, as shown in detail for planets such as HD 189733 b \citep{knu07b, knu09a, knu12} and HD 209458 b \citep{zel14}, the observed times of minimum/maximum flux occur noticeably prior to transit/secondary eclipse, respectively. These phase offsets, or ``eastward hotspots'', have been demonstrated in global atmospheric simulations; a summary of such simulations is detailed both in \citet{zel14} and \citet{dem17}. Our model's ability to capture these offsets in extrema is provided by a combination of the rotation parameter and the radiative timescale. In the limit $\tau_{\mathrm{eq}} \ll P_{\textrm{rot}}$, the offset vanishes, as cells respond thermally much more quickly than they are advected longitudinally into the view of the observer.

\subsection{Thermal Evolution}\label{sec:model:thermal}
Our planetary model is initialized with a uniform surface temperature, $T_0$. For eccentric orbits, the planet is initialized at apastron. For all planets except HD 80606 b, we choose to divide the orbit into 200 equal time intervals; HD 80606 b has a much longer orbital period and so we use 5,000 divisions. At each successive time step, the temperature of each cell is calculated as

\begin{equation}
    \dot{T}\!\left( \phi, \theta, t  \right) = \frac{T_{\textrm{eq}}\!\left( \phi, \theta, t  \right)}{4 \tau_{\textrm{rad}}} \left\{ 1 - \left[\frac{T\!\left( \phi, \theta, t  \right)}{T_{\textrm{eq}}}\right]^4 \right\}\, ,
\end{equation}
where the cell-specific equilibrium temperature can be calculated at a given point in the orbit via

\begin{equation}\label{eq:Teq}
    T_{\textrm{eq}}^4\!\left( \phi, \theta, t  \right) = \left(1 - A\right) \left[ \frac{L_\star}{4 \pi \sigma r\!\left(t\right)^2} \right] \alpha\!\left( \phi, \theta, t  \right) + T_0^4,
\end{equation}
where the longitude and latitude are $\phi$ and $\theta$, respectively.
Here $r\!\left(t\right) = a \left[ 1 - e \cos E\!\left(t\right)\right]$ is the star-planet separation obtained from Kepler's equation

\begin{equation}\label{eq:Kepler}
M\!\left(t\right) = E\!\left(t\right) - e \sin E \, ,
\end{equation}
where $E\!\left(t\right)$ is the eccentric anomaly and $M\!\left(t\right) = 2\pi\left(t - t_{\mathrm{per}}\right)/P_{\textrm{rot}}$ for periastron passage time $t_{\mathrm{per}}$. The quantity $\alpha$ is the attenuation factor of the instellation due to the apparent altitude of the star at a particular cell on the planet surface, and can be calculated from the longitude, latitude, rotation period, and true anomaly $\nu\!\left(t\right)$ as

\begin{equation}
\alpha\!\left( \phi, \theta, t  \right) = 
\begin{cases}
\cos \left( \phi-\phi_{\mathrm{ss}} \right) \cos\theta, &\left|\phi-\phi_{\mathrm{ss}} \right| \leq \pi/2 \\
0, &\left|\phi-\phi_{\mathrm{ss}} \right| > \pi/2
\end{cases}
\end{equation}
for the substellar longitude $\phi_{\mathrm{ss}} \equiv 2 \pi t/P_{\textrm{rot}} + \nu\!\left(t\right)$.

\subsection{Rotation Rates and the Expectation for Eccentric Planets}\label{sec:model:rotation}
Short-period planets on circular orbits are expected to be spin-synchonized. In these cases, our parameterized rotation rate is a proxy for equatorially directed advection. The instellation pattern on a synchronous planet is unchanging, providing a steady forcing to the atmosphere. In such cases, one expects the atmosphere to exist in a quasi-steady state, where the physical phenomena governing the balance operate on the planetary scale or smaller, and as a consequence, might be difficult or ambiguous to determine.

By contrast, planets on eccentric orbits undergo predictable, time-variable stellar forcing. The periastron passage is associated with an impulsive and well-constrained energy input which induces a dynamical response in the atmosphere. Such planets thus offer tangible clues to the primary meteorological drivers, and their relative prominence is increased because at fixed semimajor axis, the ($\varpi$-averaged) geometric transit probability of an exoplanet increases with orbital eccentricity by a factor of $1/(1-e^2)$.

Tidal evolution is expected to drive spins to the \emph{pseudo-synchronous} rate, where the planet effectively approximates tidal locking around periastron. Assuming a viscoelastic planetary rheology, \citet{hut81} calculates the pseudo-synchronous rotation frequency as a function of eccentricity using a tidal argument, with the result

\begin{equation}
\Omega_{\mathrm{PSR}} = \sqrt{\frac{a^3}{G M_\star}} \left[ \frac{1 + \frac{15}{2} e^2 + \frac{45}{8} e^4 + \frac{5}{16} e^6}{\left( 1 + 3 e^2 + \frac{3}{8} e^4 \right) \left(1 - e^2 \right)^{3/2}} \right]\, ,
\end{equation}

\noindent where $a$ is the semimajor axis, $M_\star$ the stellar mass, and $e$ the orbital eccentricity. In the limit $\left( e \rightarrow 0 \right)$, the spin frequency matches the orbital frequency.

No strong empirical constraints exist for the rotation rates of any eccentric exoplanets. \citet{dew16} have recently calculated the rotation rate of the extremely eccentric HD 80606 b as $93^{+85}_{-35}$ hours, which is inconsistent with the expected pseudo-synchronous period of $39.9$ hours. While an apparent deviation from synchronous rotation in a circular-orbit planet could be explained as arising from a bulk equatorial surface flow, it is not clear whether this phenomenon underlies HD 80606 b's slow apparent spin.

\section{Application to Known Exoplanets}\label{sec:results}
\subsection{Generating Light Curves from the Thermal Model}\label{sec:results:lightcurve}
For a given set of parameters, our model evolves the thermal response of the planet through successive orbits until the temperatures are consistent from one orbit to the next. In practice, only a few orbits are required, during which the elapsed time exceeds any feasible rotation periods or radiative timescales; we have run all models presented here through 5 orbits. The results from the final orbits of each model are used to fit to the data.

To generate model light curves from temperature maps, we note that each cell has a specific radiative intensity given by its blackbody temperature. Then, integrating over the hemisphere of the planet that would be visible to the observer given the known orbital geometry, we can calculate the total observed flux at a given wavelength. Integrating these fluxes over the filter profile of the Spitzer IRAC bands, we generate planet-star contrast ratios (light curves) over a full orbit. We choose to use a blackbody spectrum at the effective temperature for each star, and we neglect any stellar variability.

\subsection{Statistical Methods}\label{sec:results:stats}
We sample parameter space uniformly over physically feasible ranges in each, and calculate Gaussian likelihoods in each to get a sense of the landscape of likelihoods. An annealing Metropolis-Hastings algorithm allows us to use Markov-Chain Monte Carlo techniques to rapidly find an optimal likelihood. In order to quantify the uncertainty in the parameters we run new MCMC chains without annealing, starting at points in parameter space corresponding to the using the 68\% ranges of explored values on either side of the best-fit parameter values to define $1\sigma$ uncertainties. For the uncertainty ranges in the light curves, we pull the 68\% ranges in likelihood for each set of parameters, and plot the range in fluxes for the corresponding light curves. In the following subsection, we discuss the individual cases.

\begin{deluxetable*}{ccccccc}
\tabletypesize{\footnotesize}
\tablewidth{0pt}
\tablecaption{Best-Fit Parameters from Radiative Model}
\tablehead{\colhead{Planet} &
           \colhead{$\lambda$ ($\mu$m)} &
           \multicolumn{2}{c}{$P_{\mathrm{rot}}$} &
           \colhead{$\tau_{\mathrm{eq}}$ (hr)} &
           \colhead{$T_0$ (K)} &
           \colhead{A} \\
           \colhead{} &
           \colhead{} &
           \colhead{$\left(P_{\mathrm{PSR}}\right)$} &
           \colhead{(days)} &
           \colhead{} &
           \colhead{} &
           \colhead{}}
\startdata
GJ 436 b & 8.0 & $0.46^{+0.08}_{-0.16}$ & $1.05^{+0.18}_{-0.37}$ & $4.3^{+17.4}_{-2.4}$ & $616^{+49}_{-78}$ & $0.19^{+0.11}_{-0.11}$ \\
\hline
\multirow{3}{*}{HAT-P-2 b} & 3.6 & $0.78^{+0.01}_{-0.27}$ & $1.48^{+0.02}_{-0.52}$ & $2.2^{+5.5}_{-1.2}$ & $1200^{+128}_{-118}$ & $<0.19$ \\ & 4.5 & $0.83^{+0.50}_{-0.04}$ & $1.57^{+0.94}_{-0.08}$ & $0.3^{+24.2}_{-0.00}$ & $1832^{+61}_{-55}$ & $0.42^{+0.15}_{-0.10}$ \\ & 8.0 & $1.25^{+0.18}_{-0.37}$ & $2.36^{+0.33}_{-0.71}$ & $0.3^{+3.5}_{-0.2}$ & $918^{+133}_{-472}$ & $<0.06$ \\
\hline
\multirow{2}{*}{HAT-P-7 b} & 3.6 & $1.04^{+0.04}_{-0.01}$ & $2.29^{+0.08}_{-0.03}$ & $24.0^{+11.0}_{-16.4}$ & $177^{+425}_{-86}$ & $<0.05$ \\ & 4.5 & $1.19^{+0.08}_{-0.04}$ & $2.62^{+0.19}_{-0.10}$ & $12.1^{+8.8}_{-6.6}$ & $2043^{+52}_{-37}$ & $0.11^{+0.09}_{-0.04}$ \\
\hline
\multirow{2}{*}{HD 80606 b} & 4.5 & $0.74^{+0.18}_{-0.37}$ & $1.24^{+0.30}_{-0.62}$ & $0.6^{+27.9}_{-0.4}$ & $561^{+182}_{-427}$ & $0.67^{+0.06}_{-0.15}$ \\ & 8.0 & $1.23^{+0.01}_{-0.34}$ & $2.05^{+0.02}_{-0.56}$ & $3.9^{+15.8}_{-1.9}$ & $930^{+68}_{-281}$ & $0.85^{+0.05}_{-0.16}$ \\
\hline
\multirow{2}{*}{HD 149026 b} & 3.6 & $1.11^{+0.05}_{-0.05}$ & $3.18^{+0.15}_{-0.15}$ & $97.0^{+19.3}_{-11.9}$ & $1192^{+234}_{-225}$ & $0.01^{+0.10}_{-0.01}$ \\ & 4.5 & $0.51^{+0.28}_{-0.03}$ & $1.47^{+0.80}_{-0.09}$ & $4.1^{+12.3}_{-2.3}$ & $1354^{+101}_{-187}$ & $0.57^{+0.06}_{-0.09}$ \\
\hline
\multirow{3}{*}{HD 189733 b} & 3.6 & $0.68^{+0.05}_{-0.10}$ & $1.52^{+0.11}_{-0.23}$ & $17.2^{+3.6}_{-5.3}$ & $982^{+18}_{-64}$ & $0.17^{+0.03}_{-0.10}$ \\ & 4.5 & $0.73^{+0.32}_{-0.05}$ & $1.62^{+0.70}_{-0.10}$ & $6.7^{+7.2}_{-4.6}$ & $1094^{+23}_{-13}$ & $0.60^{+0.05}_{-0.01}$ \\ & 8.0 & $0.79^{+0.06}_{-0.28}$ & $1.75^{+0.13}_{-0.62}$ & $0.3^{+4.0}_{-0.2}$ & $1181^{+10}_{-24}$ & $0.64^{+0.02}_{-0.04}$ \\
\hline
HD 209458 b & 4.5 & $0.46^{+0.20}_{-0.03}$ & $1.63^{+0.71}_{-0.12}$ & $9.9^{+10.0}_{-0.8}$ & $1036^{+67}_{-30}$ & $0.26^{+0.08}_{-0.03}$ \\
\hline
\multirow{2}{*}{WASP-12 b} & 3.6 & $0.69^{+0.22}_{-0.01}$ & $0.74^{+0.24}_{-0.01}$ & $8.1^{+34.7}_{-0.8}$ & $465^{+1277}_{-52}$ & $<0.02$ \\ & 4.5 & $0.95^{+0.01}_{-0.01}$ & $1.02^{+0.01}_{-0.01}$ & $34.6^{+6.8}_{-6.6}$ & $2259^{+22}_{-21}$ & $<0.03$ \\
\hline
\multirow{2}{*}{WASP-14 b} & 3.6 & $1.03^{+0.05}_{-0.07}$ & $2.22^{+0.12}_{-0.15}$ & $2.8^{+11.8}_{-1.5}$ & $1243^{+63}_{-80}$ & $<0.04$ \\ & 4.5 & $1.12^{+0.16}_{-0.07}$ & $2.41^{+0.34}_{-0.16}$ & $4.5^{+1.9}_{-2.7}$ & $1694^{+50}_{-27}$ & $0.18^{+0.09}_{-0.03}$ \\
\hline
\multirow{2}{*}{WASP-18 b} & 3.6 & $1.05^{+0.07}_{-0.03}$ & $0.98^{+0.07}_{-0.02}$ & $7.8^{+10.6}_{-4.4}$ & $1556^{+85}_{-67}$ & $<0.04$ \\ & 4.5 & $1.05^{+0.09}_{-0.01}$ & $0.99^{+0.09}_{-0.01}$ & $10.2^{+1.6}_{-6.7}$ & $1519^{+47}_{-49}$ & $<0.01$ \\
\hline
\multirow{2}{*}{WASP-19 b} & 3.6 & $0.96^{+0.01}_{-0.02}$ & $0.76^{+0.01}_{-0.02}$ & $9.5^{+5.3}_{-3.7}$ & $1366^{+30}_{-28}$ & $0.09^{+0.02}_{-0.03}$ \\ & 4.5 & $0.95^{+0.01}_{-0.01}$ & $0.75^{+0.01}_{-0.01}$ & $10.8^{+2.4}_{-2.4}$ & $1403^{+21}_{-25}$ & $0.10^{+0.03}_{-0.02}$ \\
\hline
\multirow{2}{*}{WASP-33 b} & 3.6 & $0.96^{+0.08}_{-0.08}$ & $1.17^{+0.10}_{-0.10}$ & $4.7^{+3.9}_{-3.3}$ & $2336^{+42}_{-38}$ & $0.55^{+0.05}_{-0.04}$ \\ & 4.5 & $0.61^{+0.00}_{-0.20}$ & $0.74^{+0.00}_{-0.25}$ & $2.3^{+0.3}_{-0.7}$ & $2171^{+14}_{-126}$ & $0.43^{+0.01}_{-0.12}$ \\
\hline
\multirow{2}{*}{WASP-43 b} & 3.6 & $0.94^{+0.03}_{-0.03}$ & $0.76^{+0.03}_{-0.02}$ & $3.1^{+1.6}_{-1.6}$ & $434^{+48}_{-130}$ & $0.04^{+0.02}_{-0.02}$ \\ & 4.5 & $0.97^{+0.01}_{-0.01}$ & $0.79^{+0.00}_{-0.01}$ & $14.1^{+3.8}_{-3.4}$ & $847^{+30}_{-37}$ & $0.23^{+0.01}_{-0.03}$ \\
\hline
\enddata
\tablecomments{The parameter values from our blackbody model returning the most favorable likelihood from MCMC algorithms. Uncertainties listed are $1\sigma$ ranges of a Metropolis-Hastings algorithm walk around the region of most favorable likelihood in parameter space. Upper limits imply the best-fit values are zero, with a $1\sigma$ uncertainty given by the upper limit.}

\label{table:fits}
\end{deluxetable*}


\startlongtable
\begin{deluxetable*}{m{2.5cm} m{3.75cm} m{3.75cm} m{3.75cm} | m{3.75cm}}
\tabletypesize{\scriptsize}
\tablecaption{Model Light Curves and Orbital Diagrams}
\tablehead{
           \colhead{}       &
           \colhead{}       &
           \colhead{$\lambda$ ($\mu$m)} &
           \colhead{}       &
           \colhead{}    \\
           Planet & 
           \colhead{3.6}    &
           \colhead{4.5}    &
           \colhead{8.0}       &
           \colhead{Orbit} 
           }
           
\startdata
\shortstack{\textbf{\small{GJ 436 b}} \\ \\ $M = 0.07 M_J$ \\ $R = 0.37 R_J$ \\ $T_{\mathrm{eq}} = 686$ K \\ $\lambda =80^{\circ}$} & & & \hspace{-0.790cm}\includegraphics[width=4.540cm]{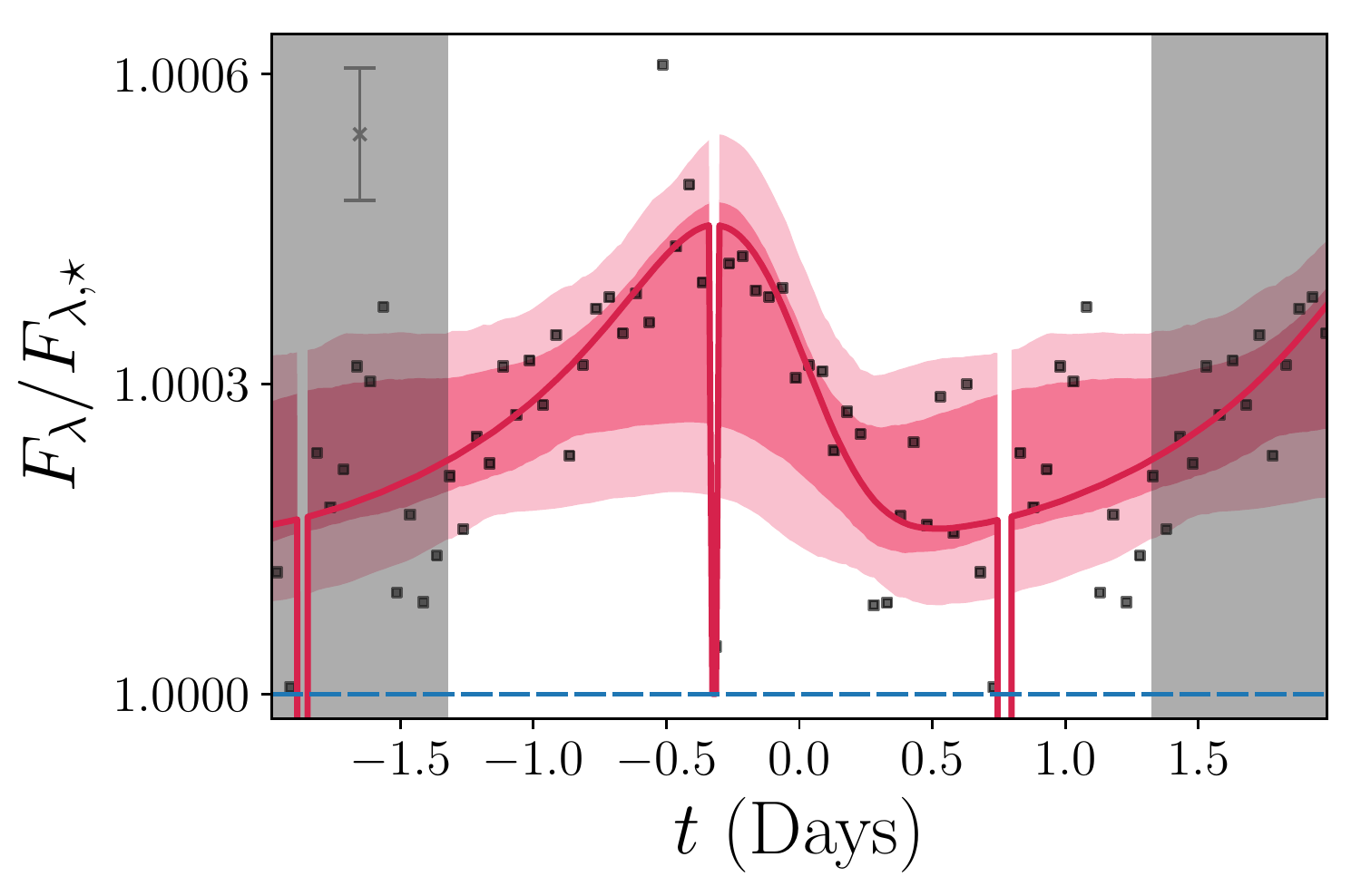} & \includegraphics[height=3.5cm]{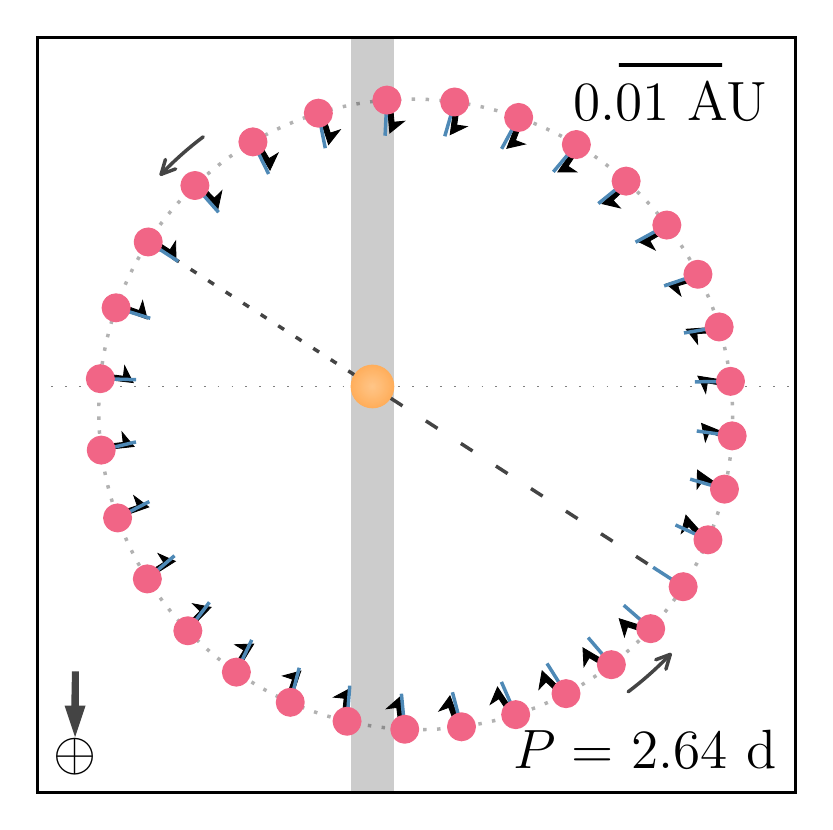}\\
\shortstack{\textbf{\small{HAT-P-2 b}} \\ \\ $M = 9.09 M_J$ \\ $R = 1.16 R_J$ \\ $T_{\mathrm{eq}} = 1540$ K \\ $\lambda =9^{\circ}$} & \hspace{-0.790cm}\includegraphics[width=4.540cm]{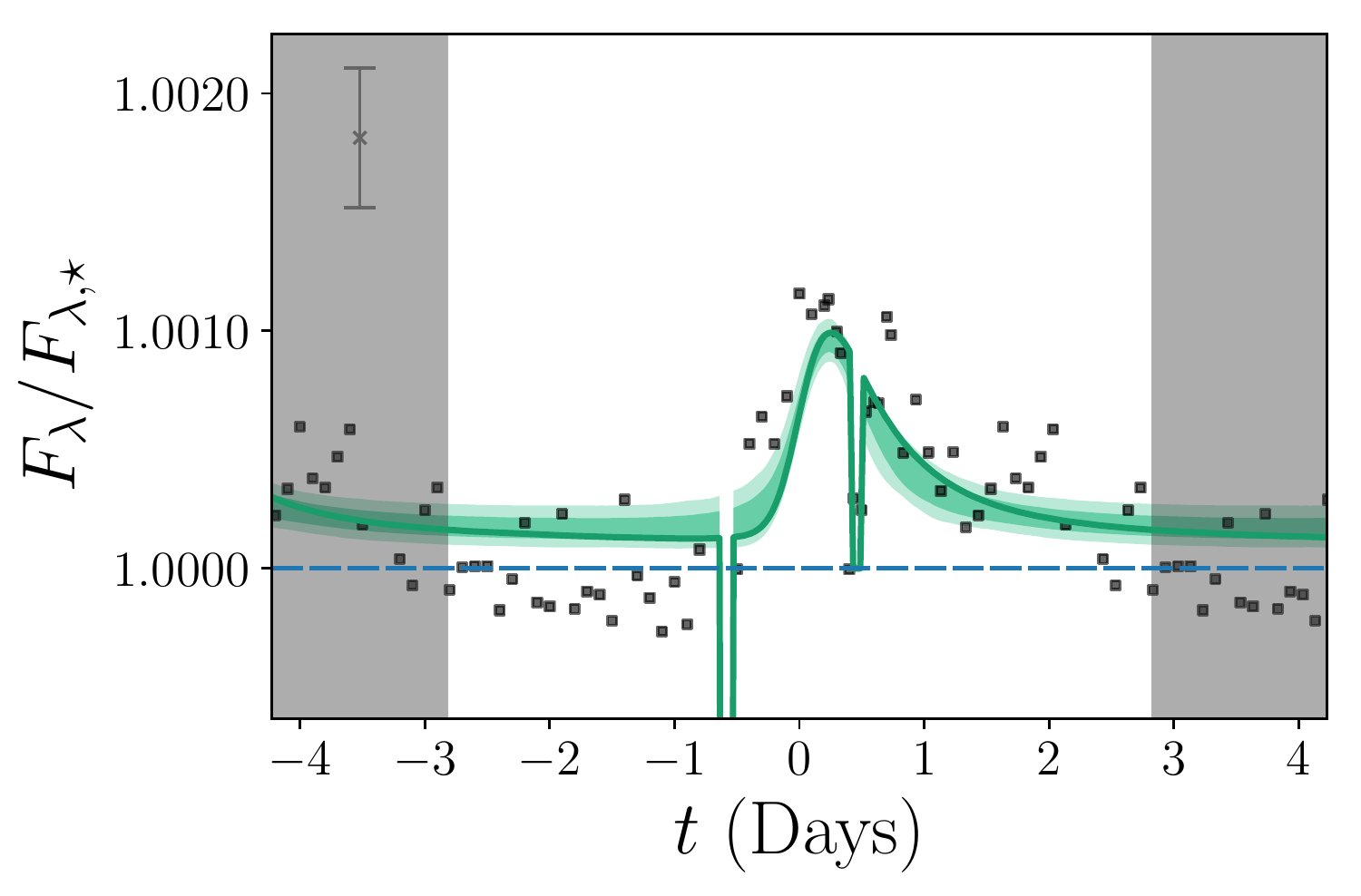} & \includegraphics[width=3.75cm]{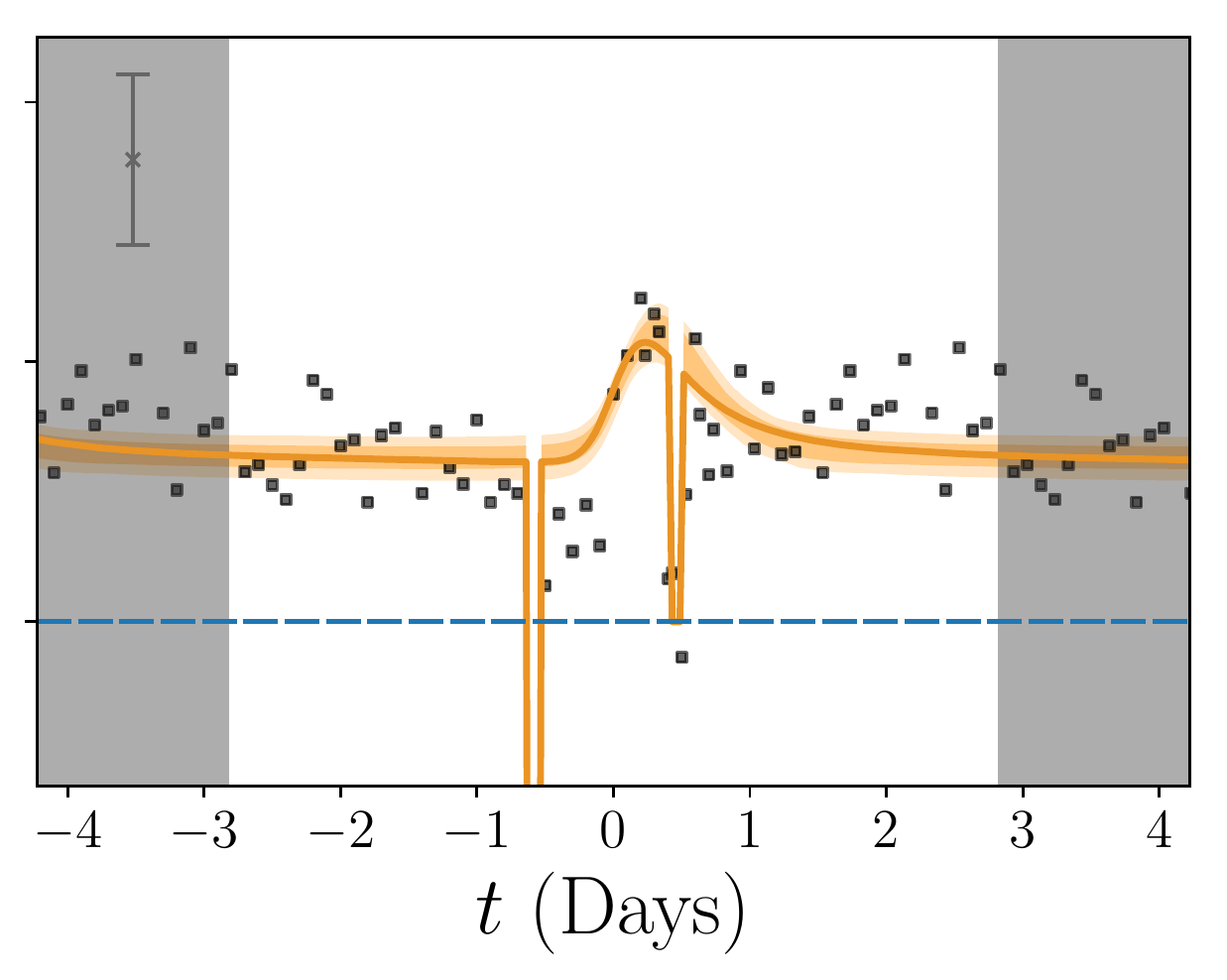} & \includegraphics[width=3.75cm]{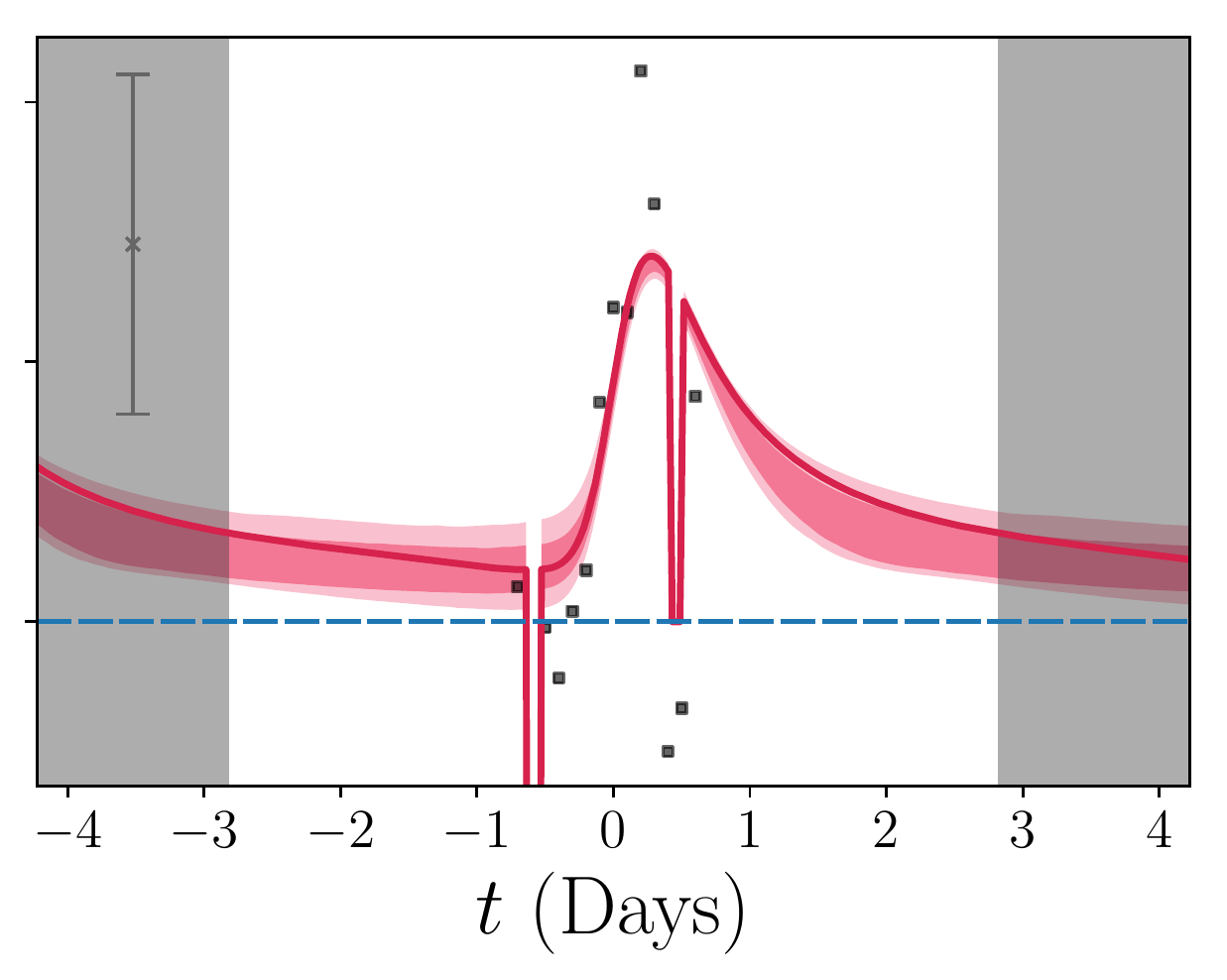} &  \includegraphics[height=3.5cm]{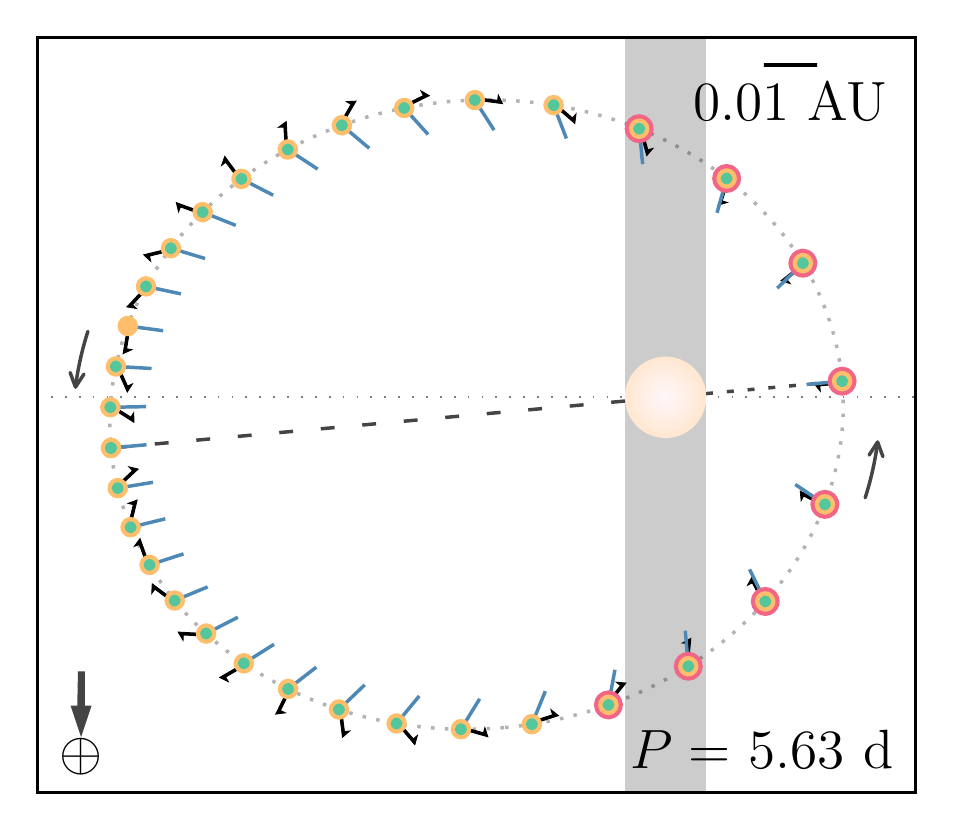}\\
\shortstack{\textbf{\small{HAT-P-7 b}} \\ \\ $M = 1.68 M_J$ \\ $R = 1.49 R_J$ \\ $T_{\mathrm{eq}} = 2700$ K \\ $\lambda = 182.5^{\circ}$} & \hspace{-0.790cm}\includegraphics[width=4.540cm]{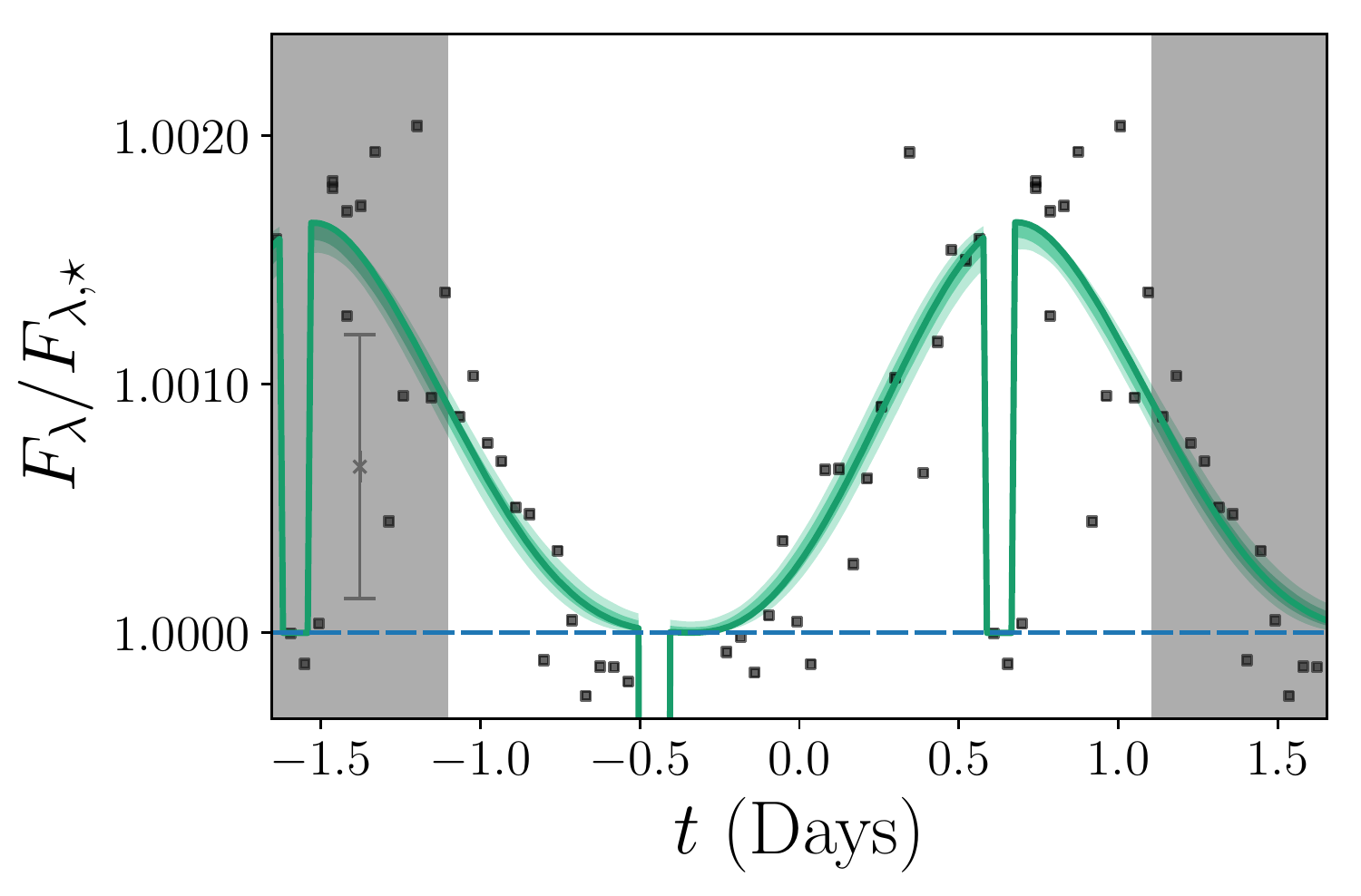} & \includegraphics[width=3.75cm]{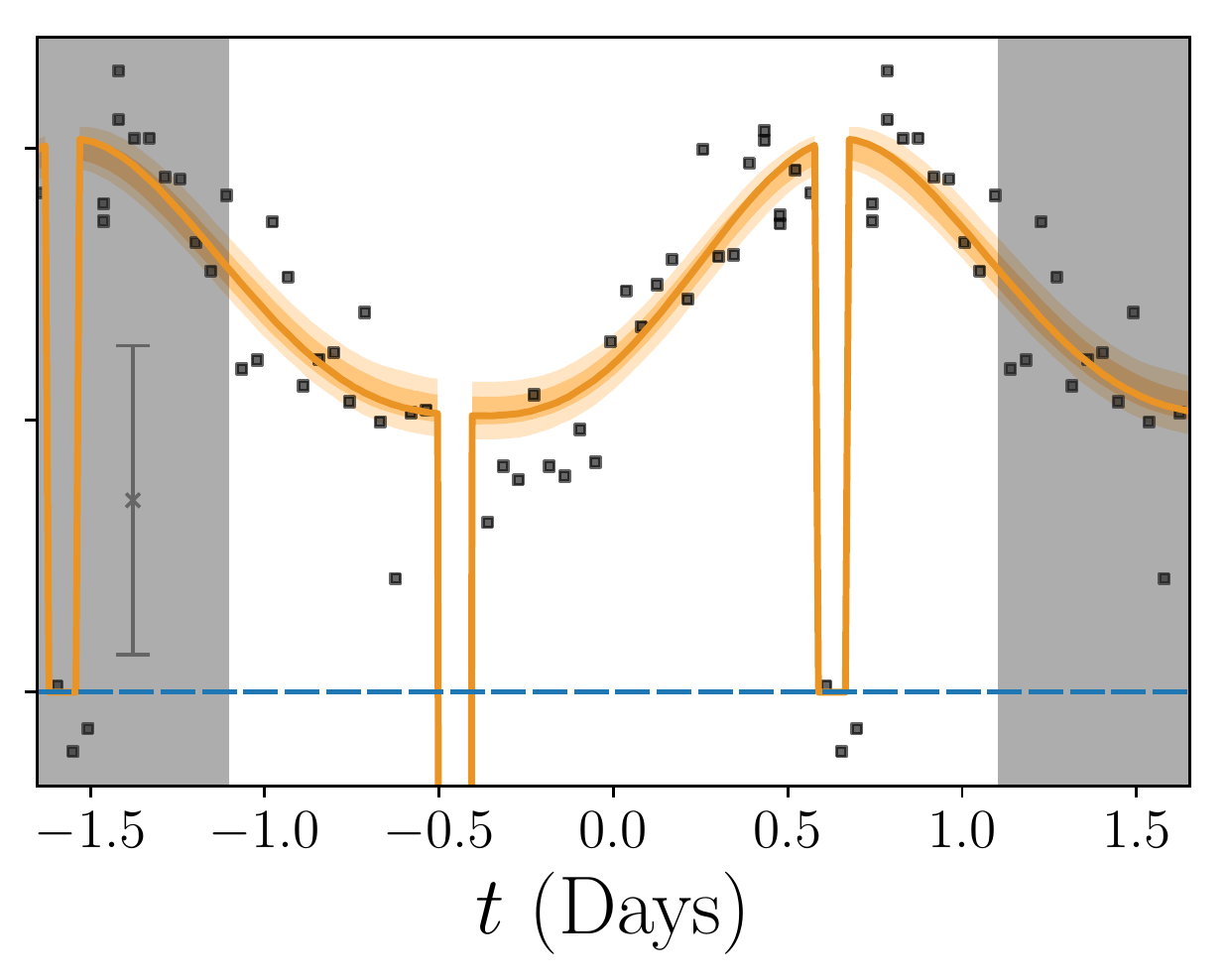} & &  \includegraphics[height=3.5cm]{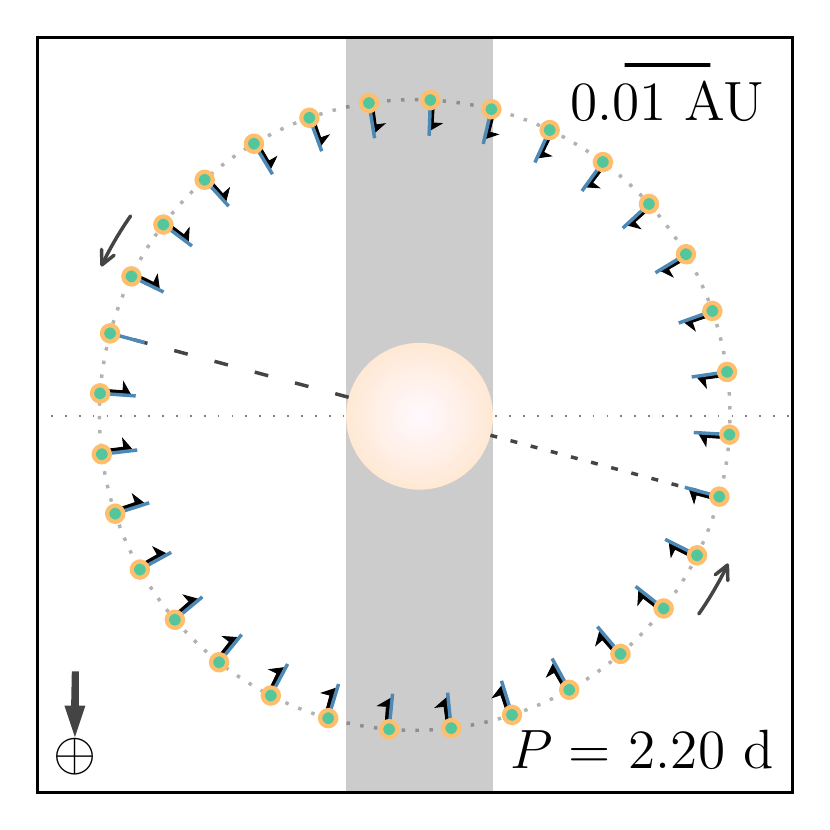}\\
\shortstack{\textbf{\small{HD 80606 b}} \\ \\ $M = 3.94 M_J$ \\ $R = 0.98 R_J$ \\ $T_{\mathrm{eq}} = 405$ K \\ $\lambda = 42^{\circ}$} & & \hspace{-0.790cm}\includegraphics[width=4.540cm]{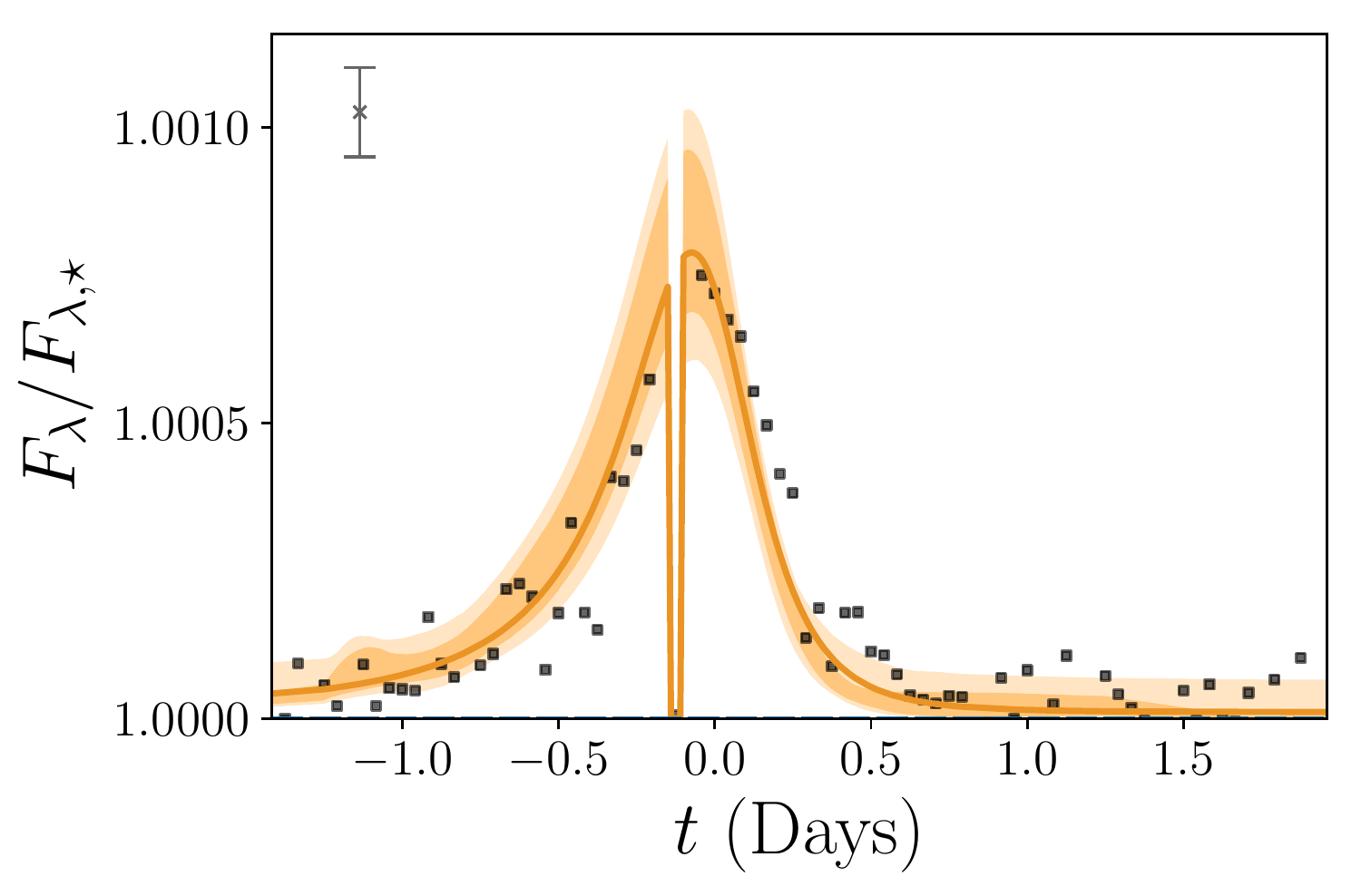} & \includegraphics[width=3.75cm]{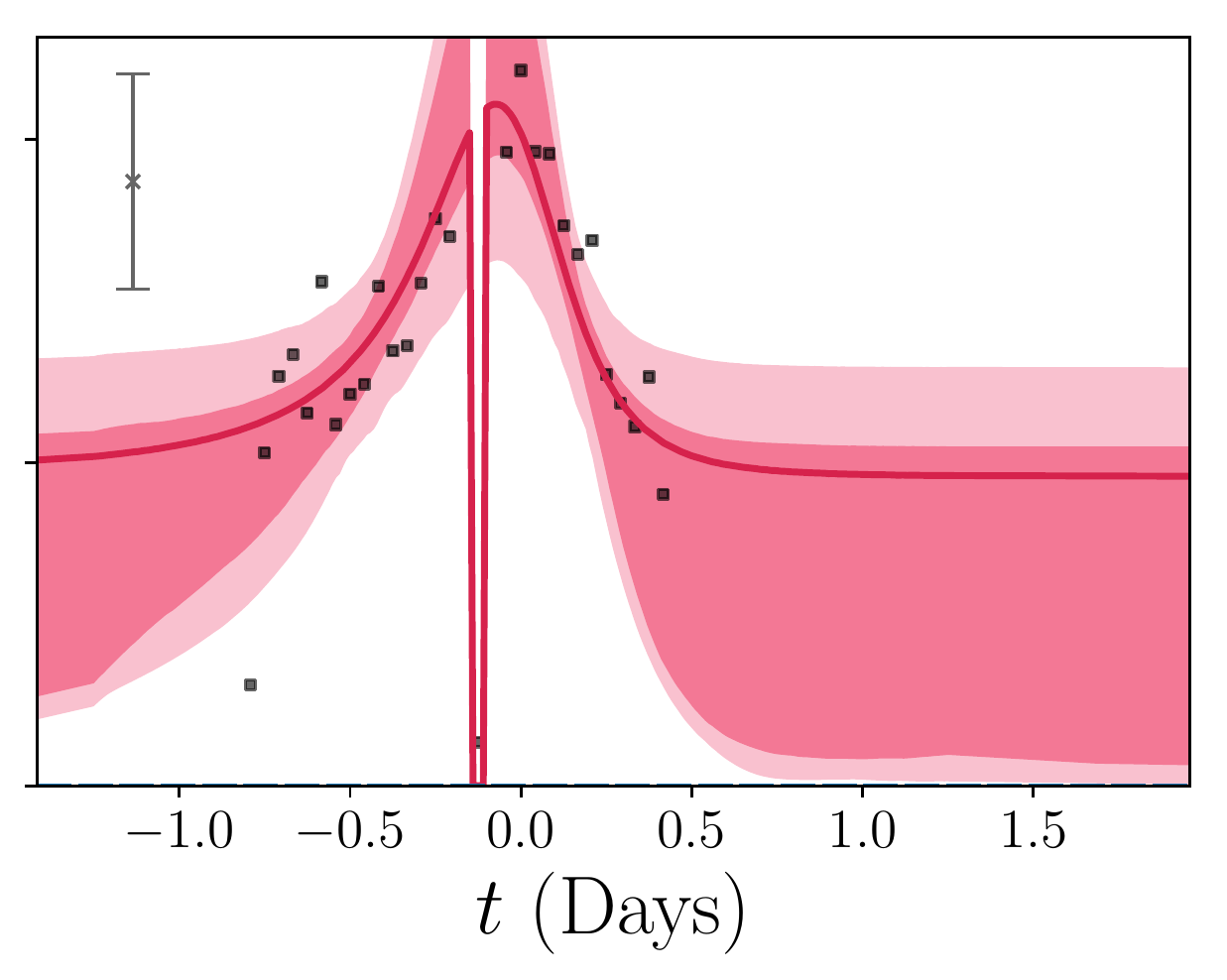} & \includegraphics[height=3.5cm]{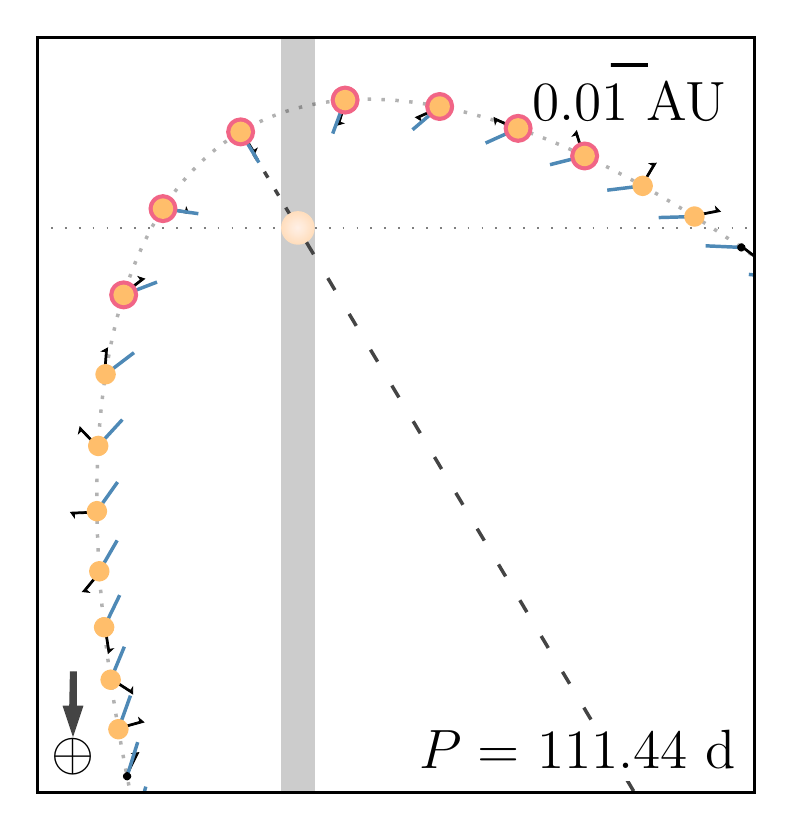}\\
\shortstack{\textbf{\small{HD 149026 b}} \\ \\ $M = 0.37 M_J$ \\ $R = 0.81 R_J$ \\ $T_{\mathrm{eq}} = 1781$ K \\ $\lambda = 12.0^{\circ}$} & \hspace{-0.790cm}\includegraphics[width=4.540cm]{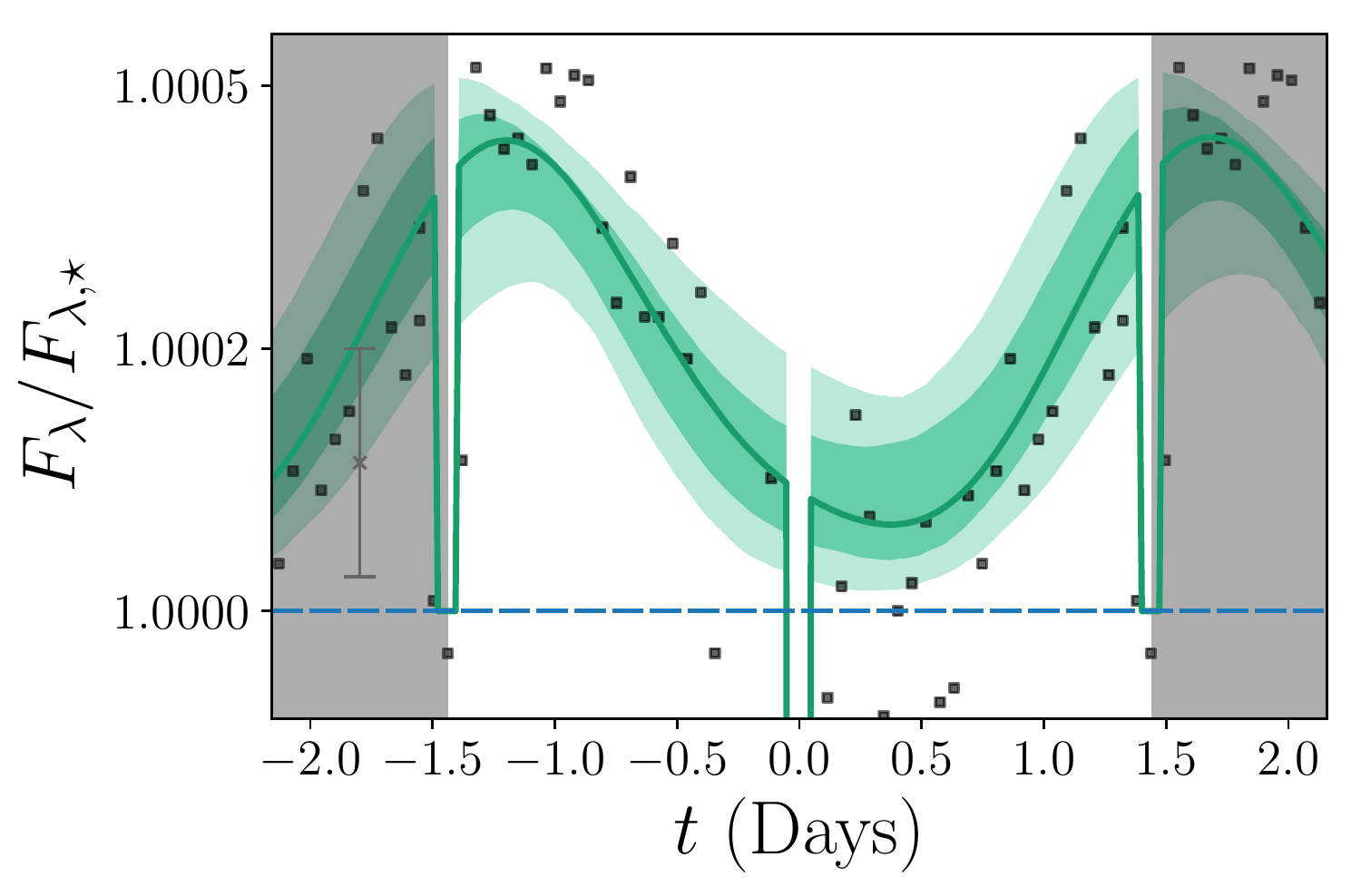} & \includegraphics[width=3.75cm]{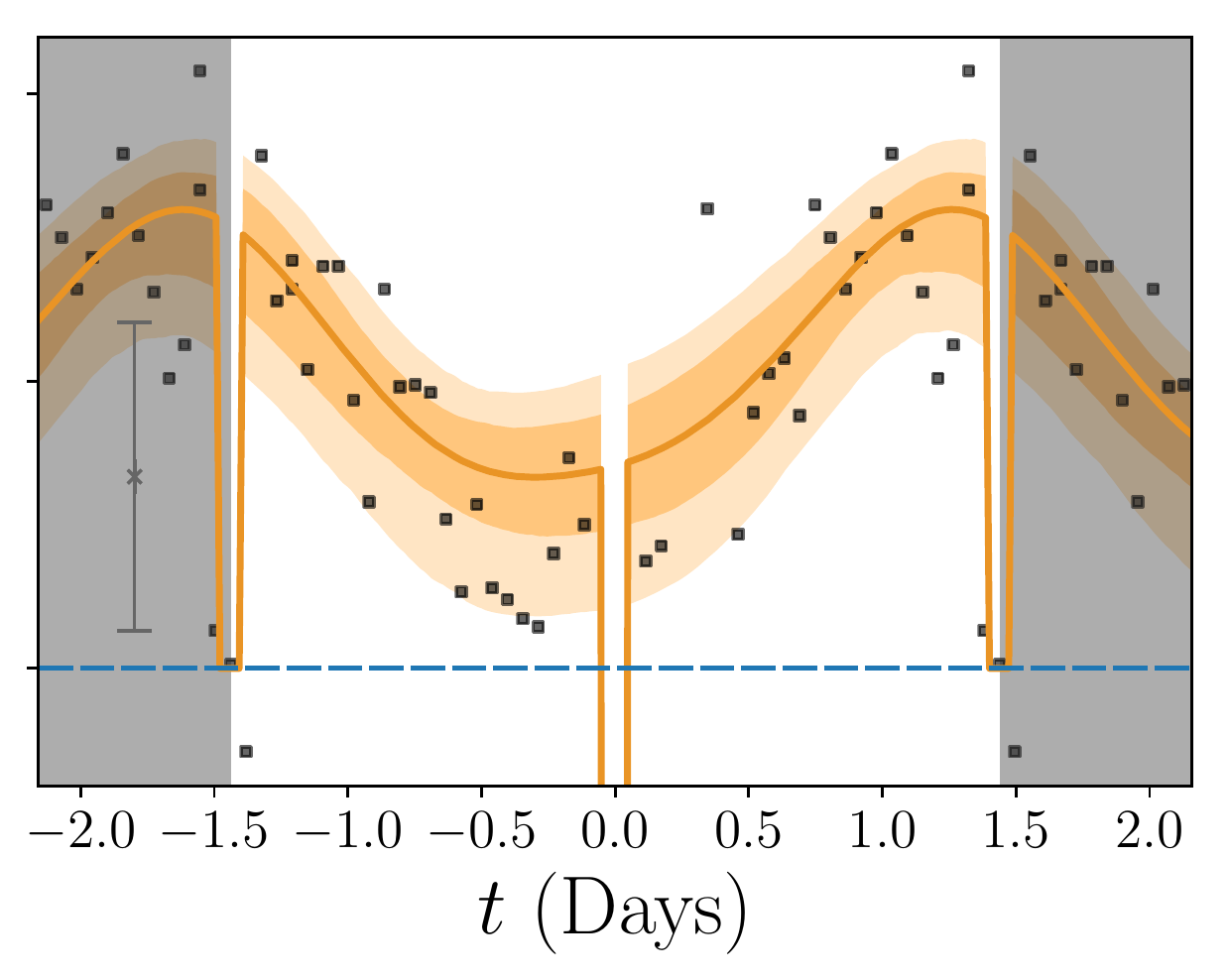} & & \includegraphics[height=3.5cm]{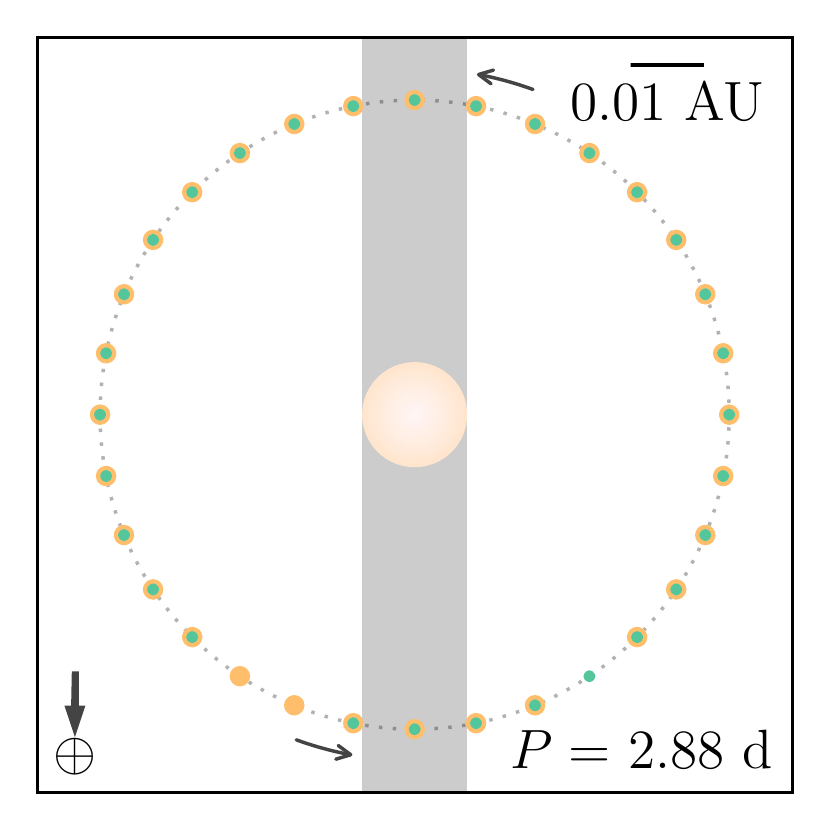}\\
\shortstack{\textbf{\small{HD 189733 b}} \\ \\ $M = 1.16 M_J$ \\ $R = 1.22 R_J$ \\ $T_{\mathrm{eq}} = 1200$ K \\ $\lambda = -0.85^{\circ}$} & \hspace{-0.790cm}\includegraphics[width=4.540cm]{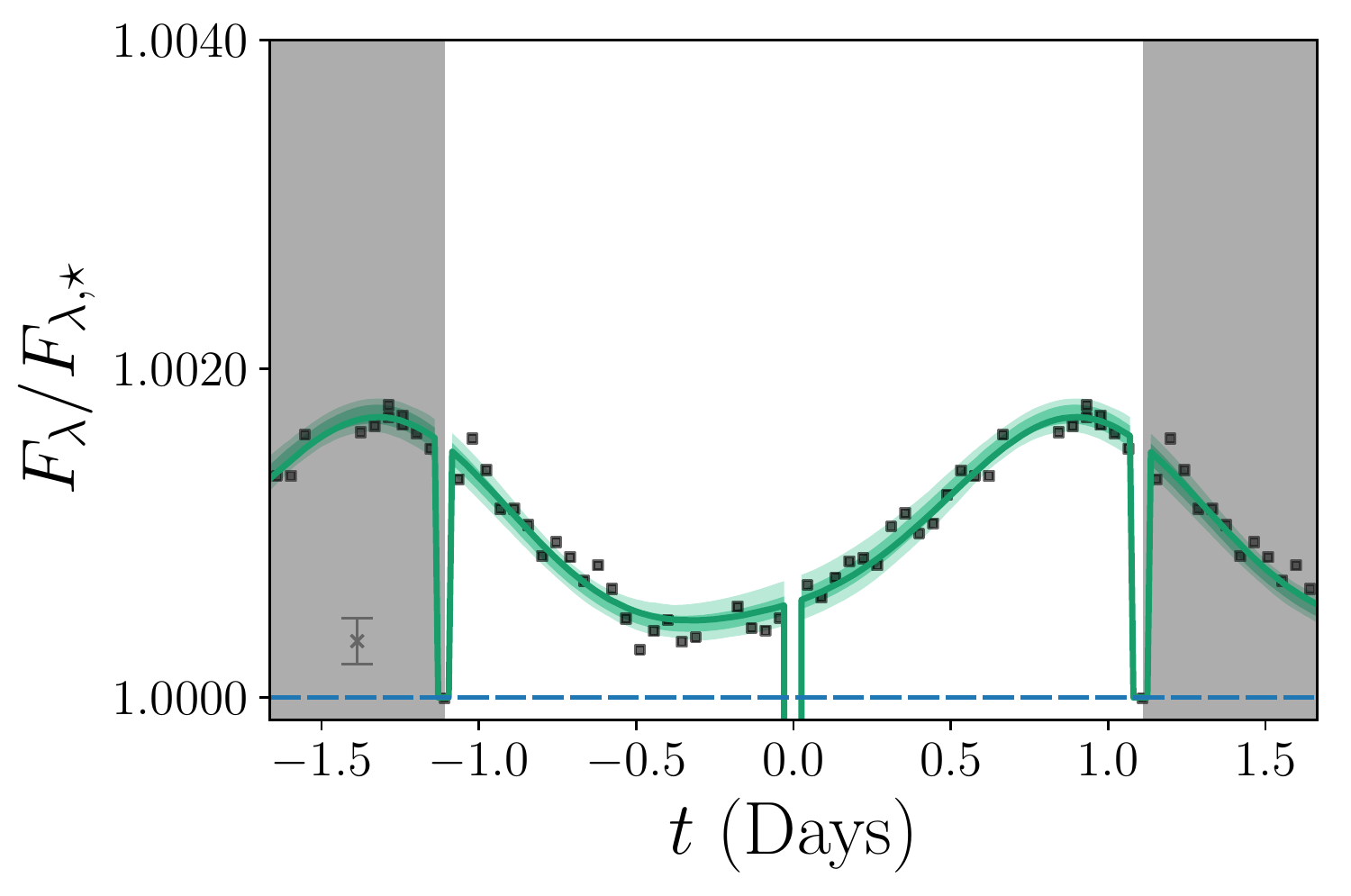} & \includegraphics[width=3.75cm]{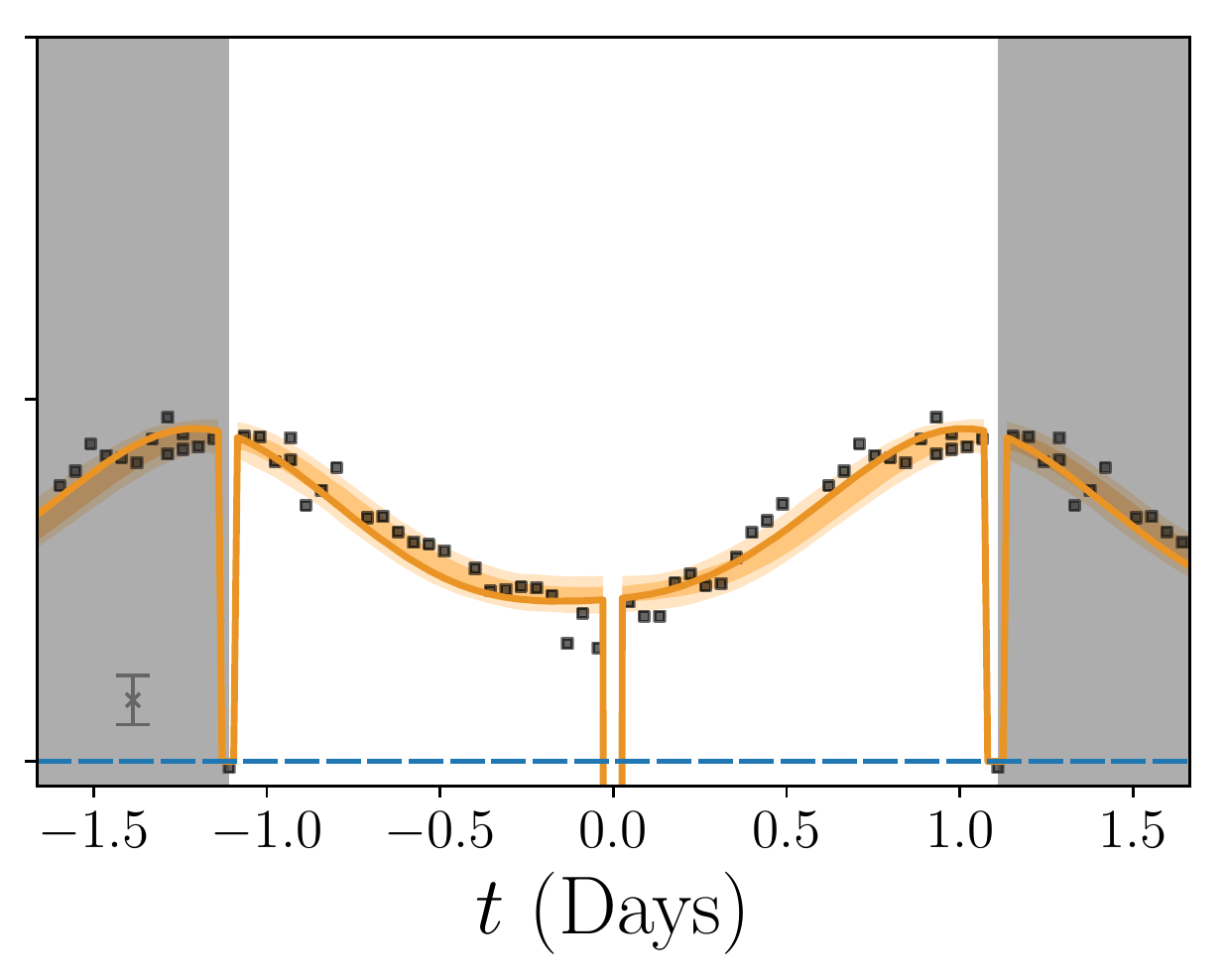} & \includegraphics[width=3.75cm]{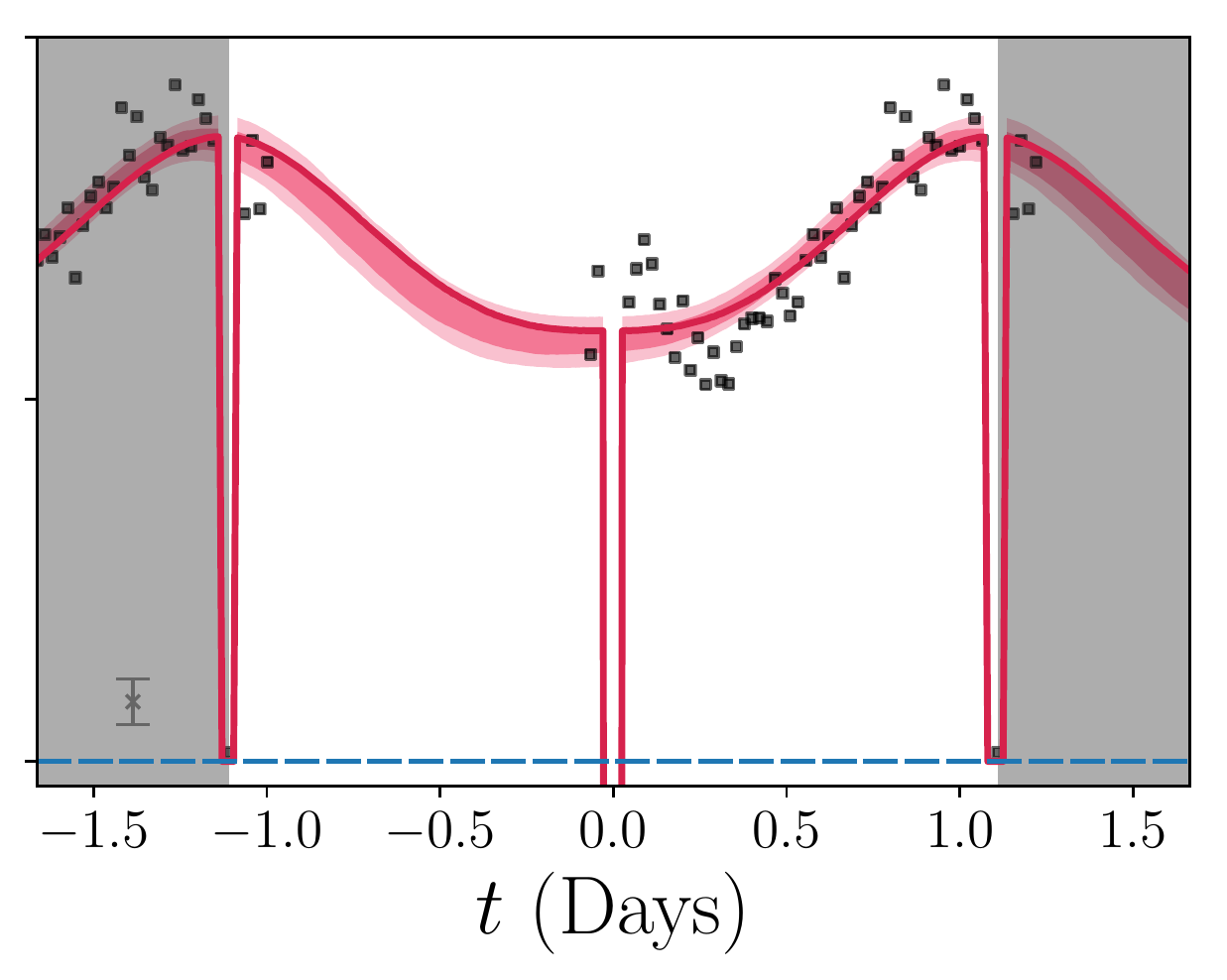} & \includegraphics[height=3.5cm]{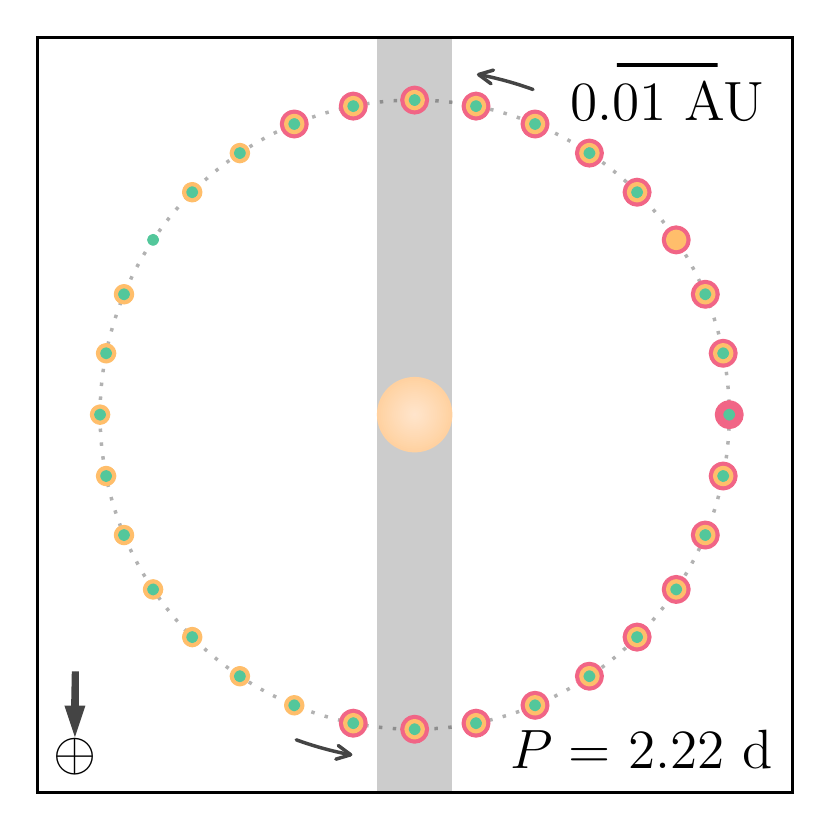}\\
\shortstack{\textbf{\small{HD 209458 b}} \\ \\ $M = 0.71 M_J$ \\ $R = 1.38 R_J$ \\ $T_{\mathrm{eq}} = 1459$ K \\ $\lambda = -4.4^{\circ}$} & & \hspace{-0.790cm}2\includegraphics[width=4.540cm]{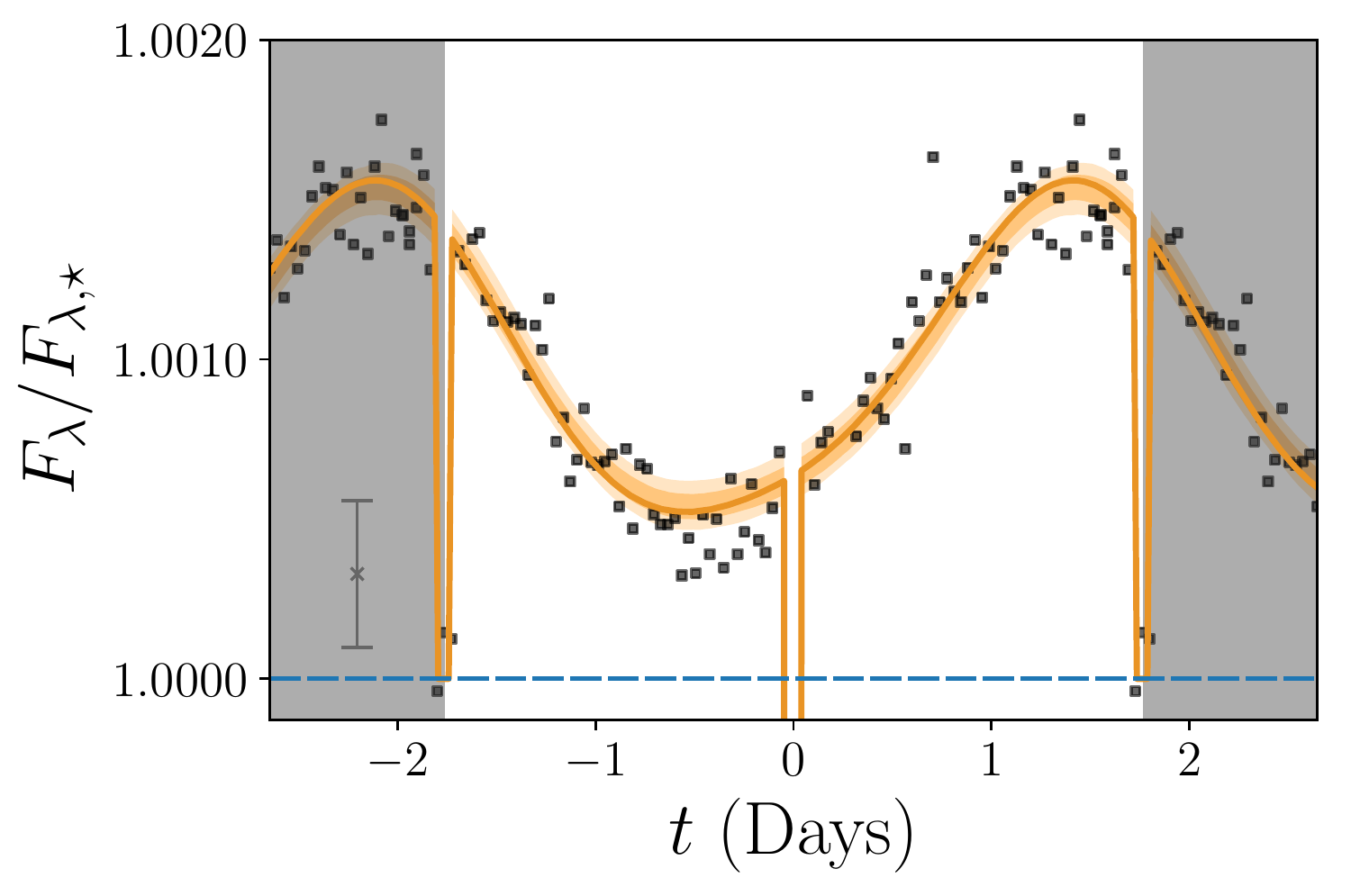} & & \includegraphics[height=3.5cm]{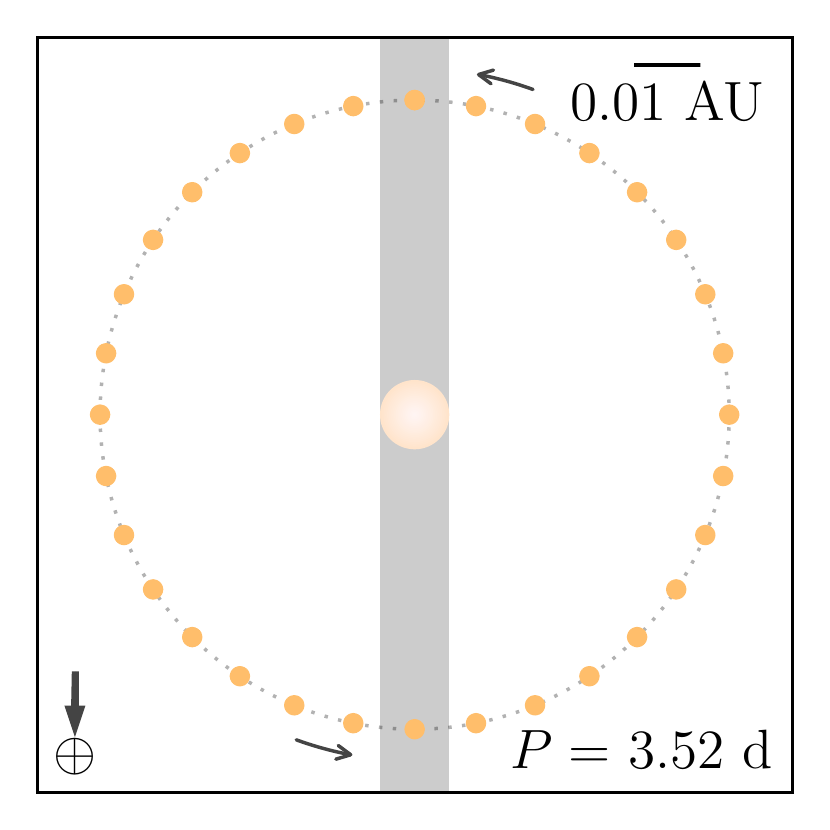}\\
\shortstack{\textbf{\small{WASP-12 b}} \\ \\ $M = 1.43 M_J$ \\ $R = 1.83 R_J$ \\ $T_{\mathrm{eq}} = 2523$ K \\ $\lambda = 59^{\circ}$} & \hspace{-0.790cm}\includegraphics[width=4.540cm]{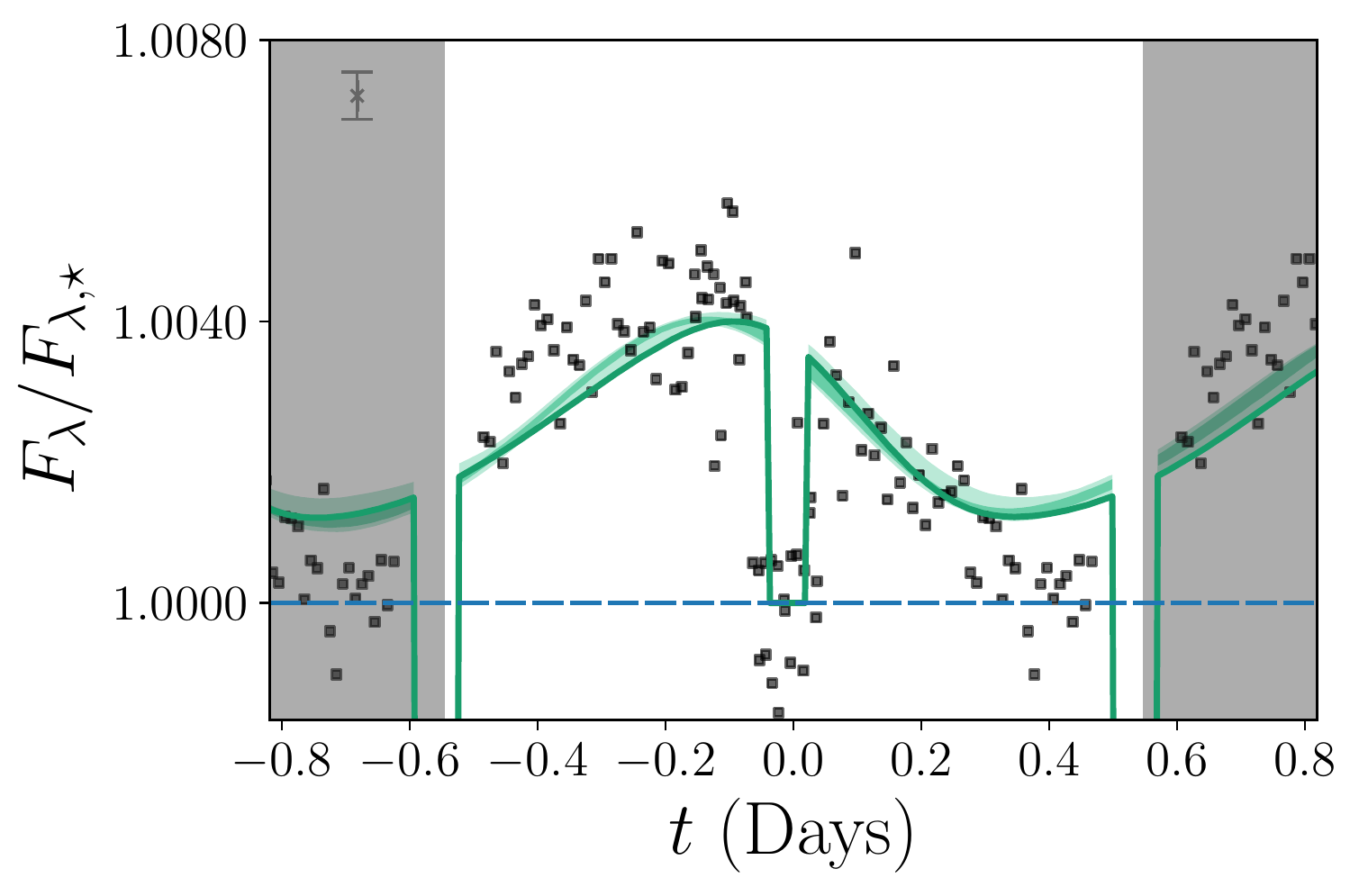} & \includegraphics[width=3.75cm]{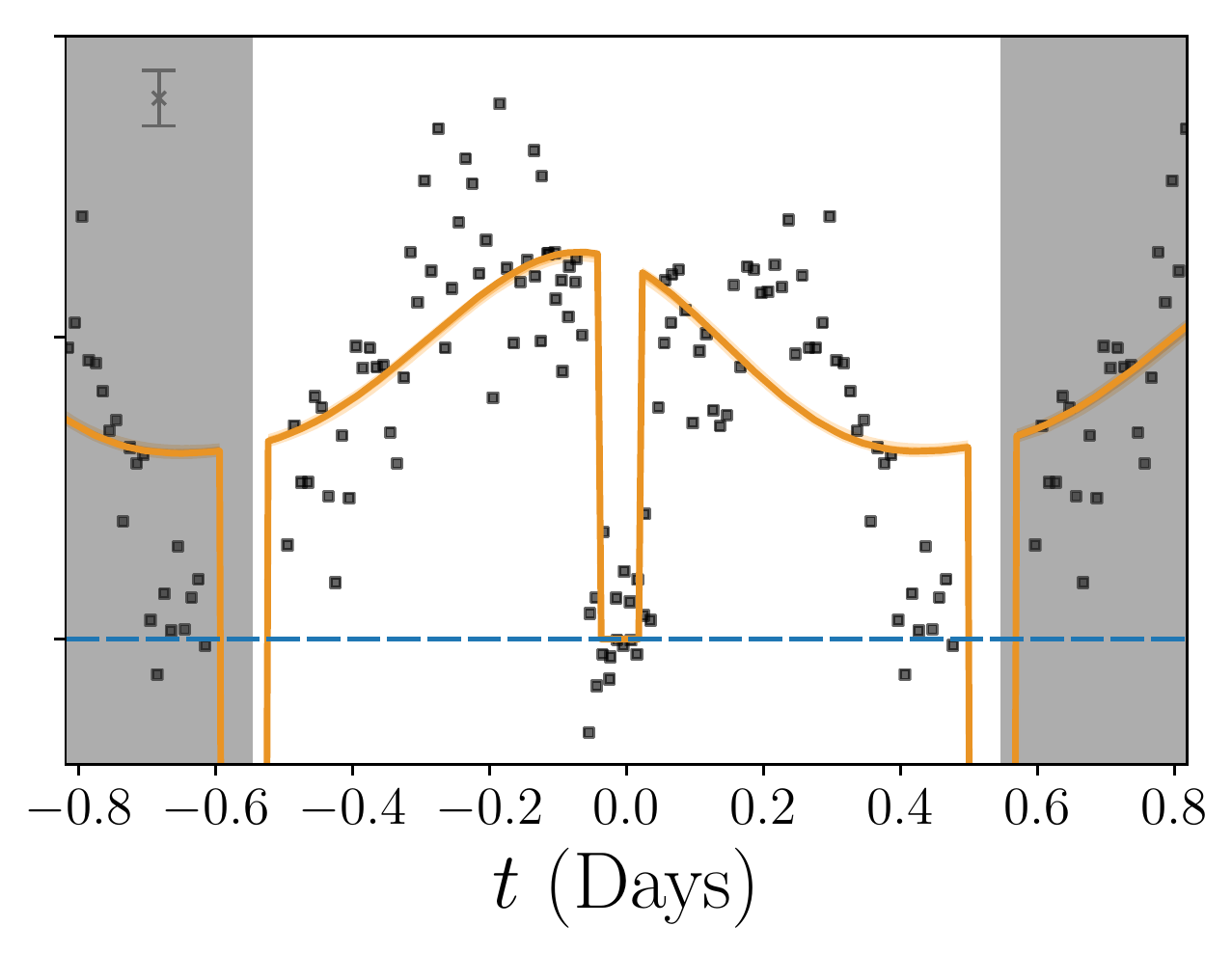} & & \includegraphics[height=3.5cm]{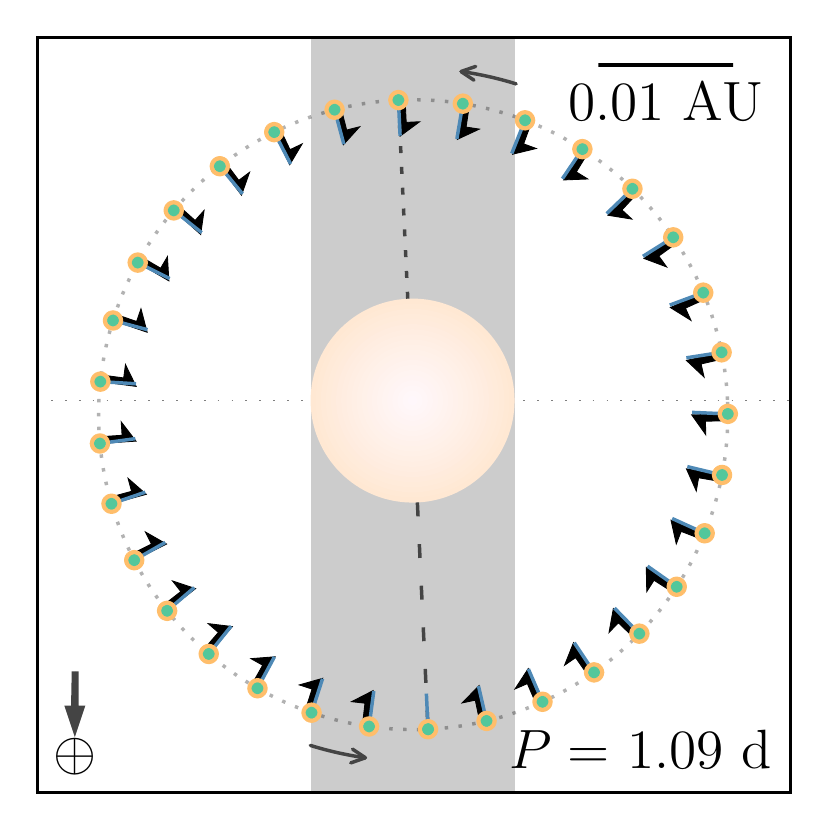}\\
\shortstack{\textbf{\small{WASP-14 b}} \\ \\ $M = 7.59 M_J$ \\ $R = 1.24 R_J$ \\ $T_{\mathrm{eq}} = 1872$ K \\ $\lambda = -33.1^{\circ}$} & \hspace{-0.790cm}\includegraphics[width=4.540cm]{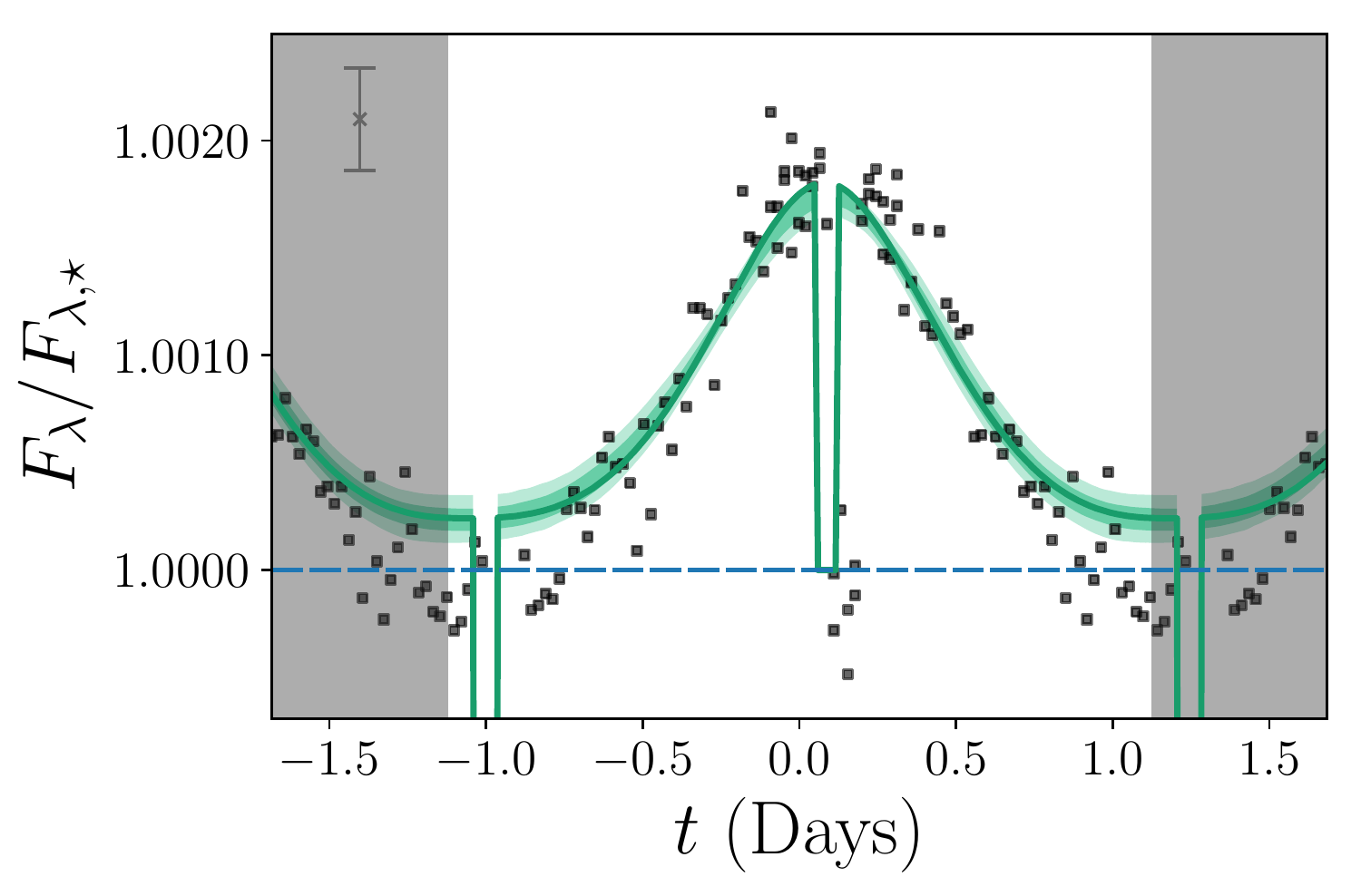} & \includegraphics[width=3.75cm]{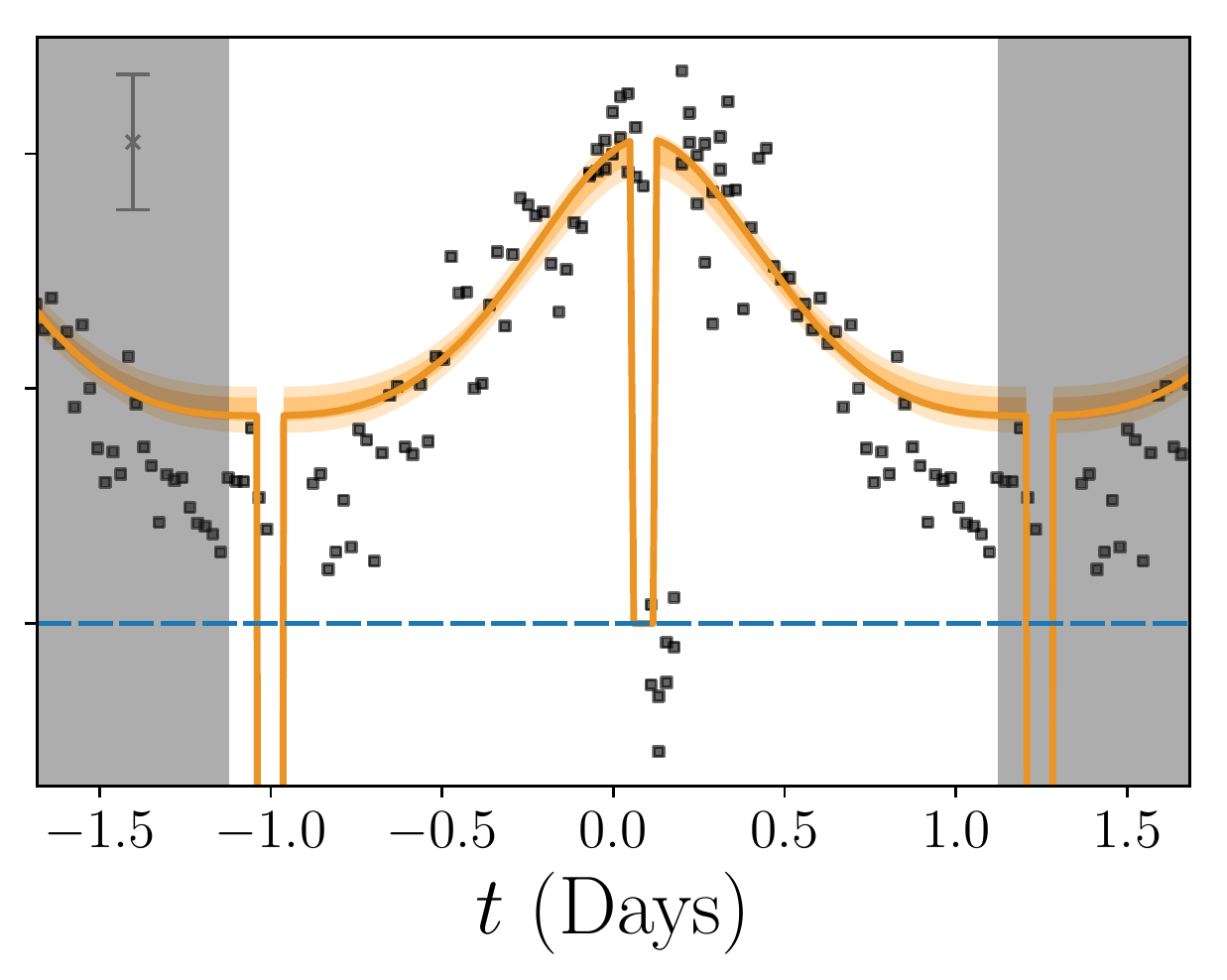} & & \includegraphics[height=3.5cm]{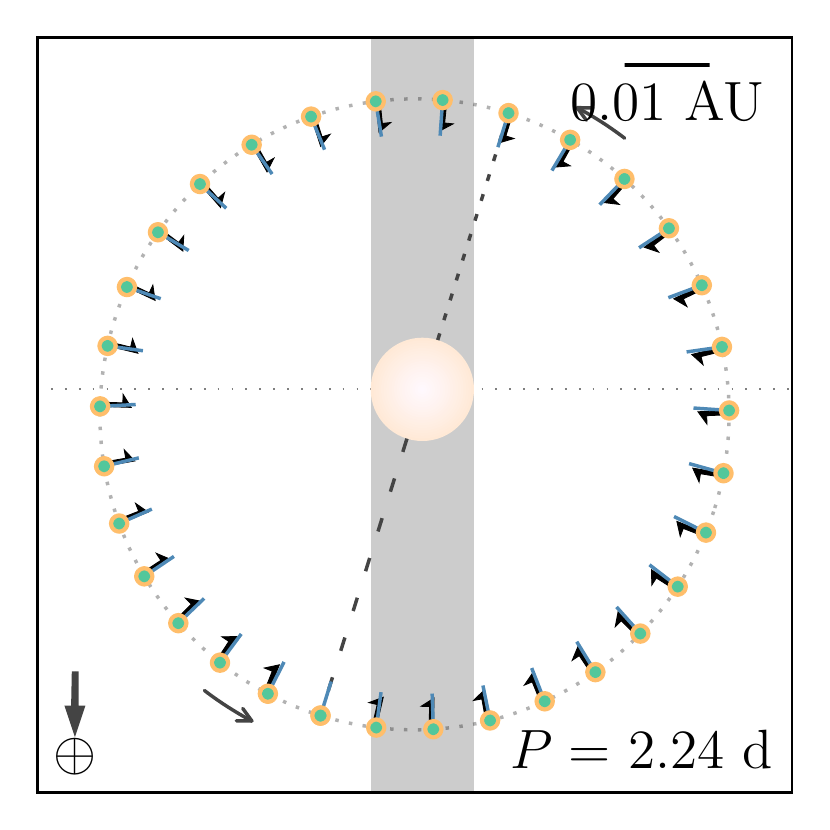}\\
\shortstack{\textbf{\small{WASP-18 b}} \\ \\ $M = 10.38 M_J$ \\ $R = 1.16 R_J$ \\ $T_{\mathrm{eq}} = 2413$ K \\ $\lambda = 4.0^{\circ}$} & \hspace{-0.790cm}\includegraphics[width=4.540cm]{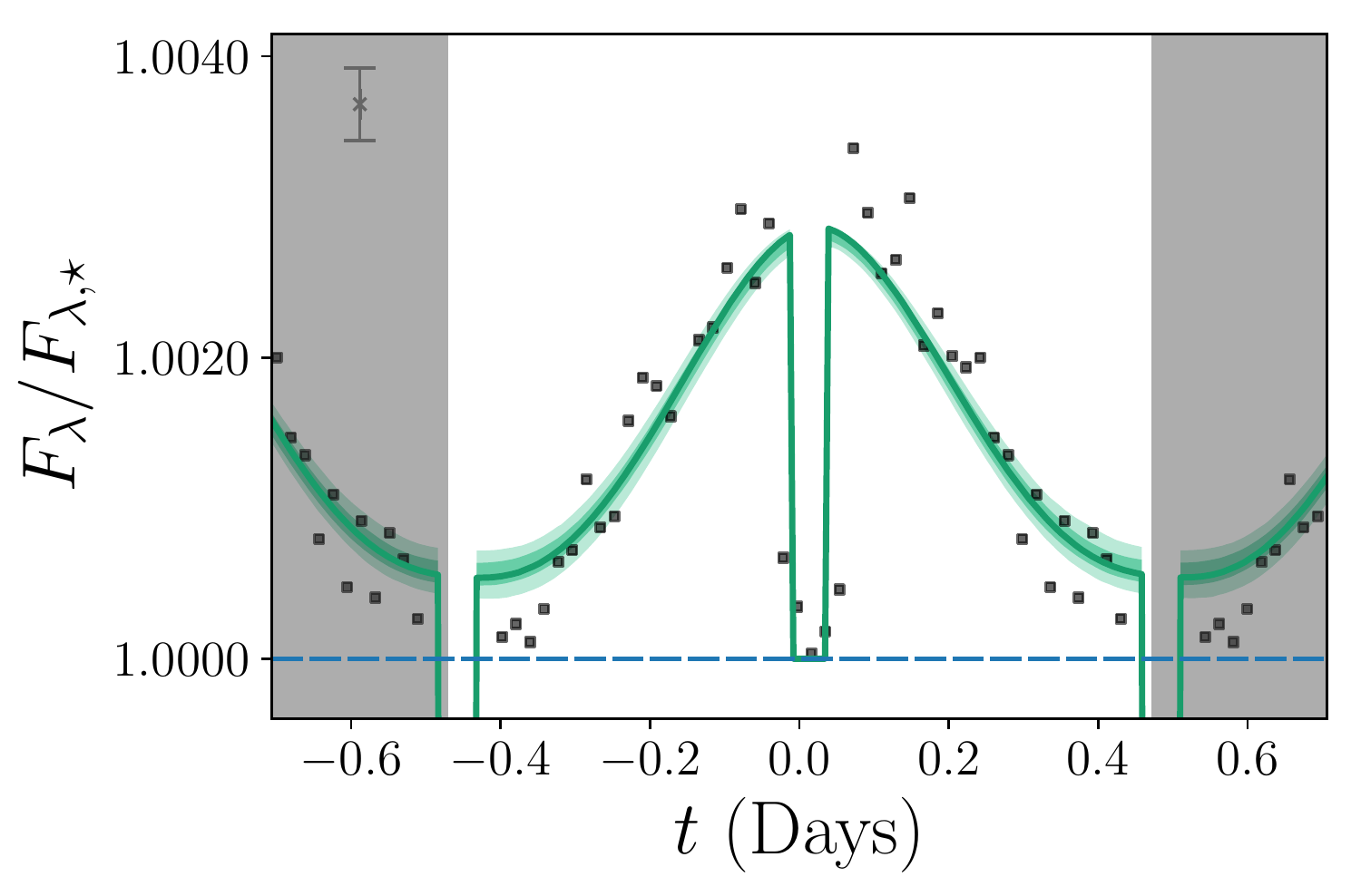} & \includegraphics[width=3.75cm]{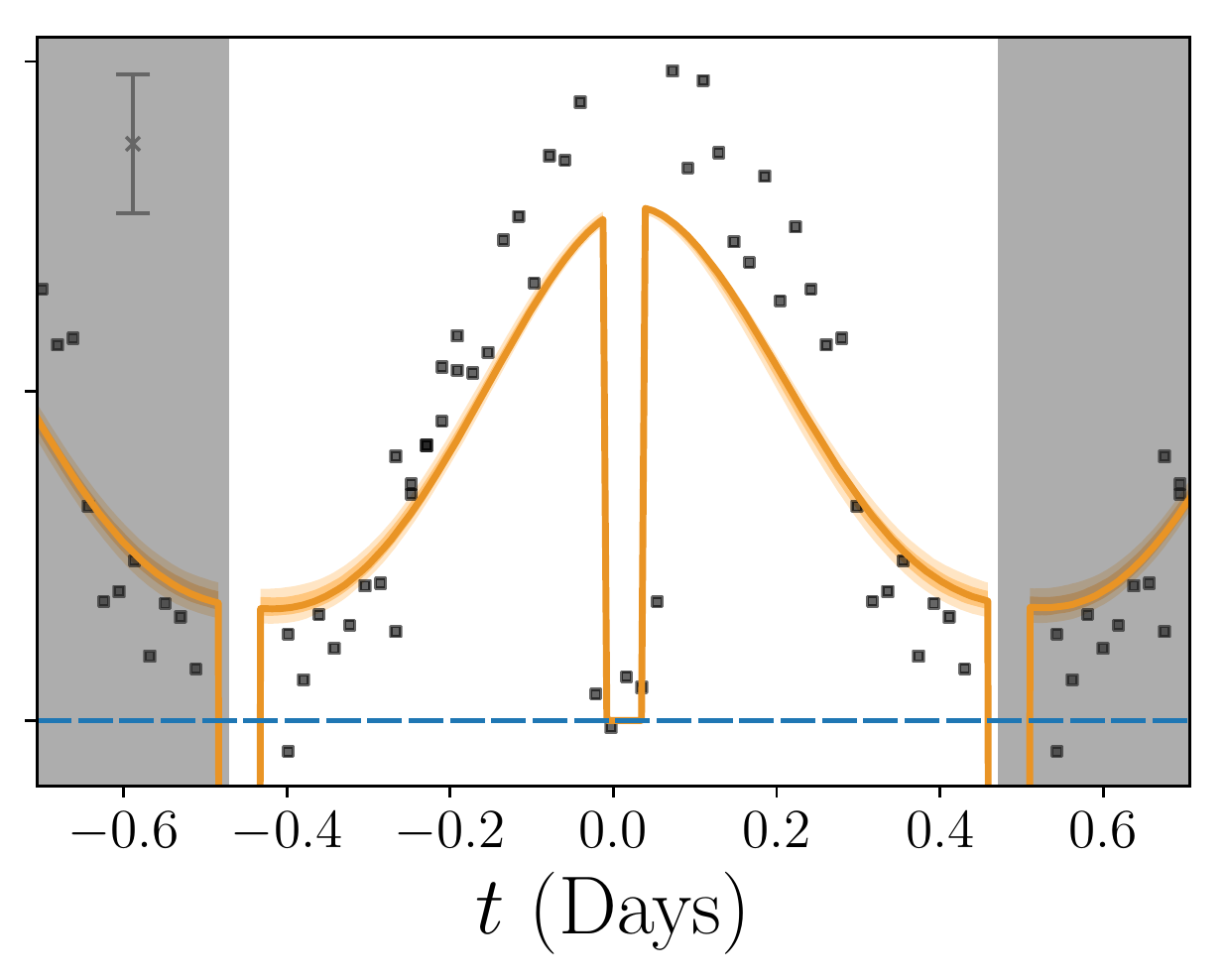} & & \includegraphics[height=3.5cm]{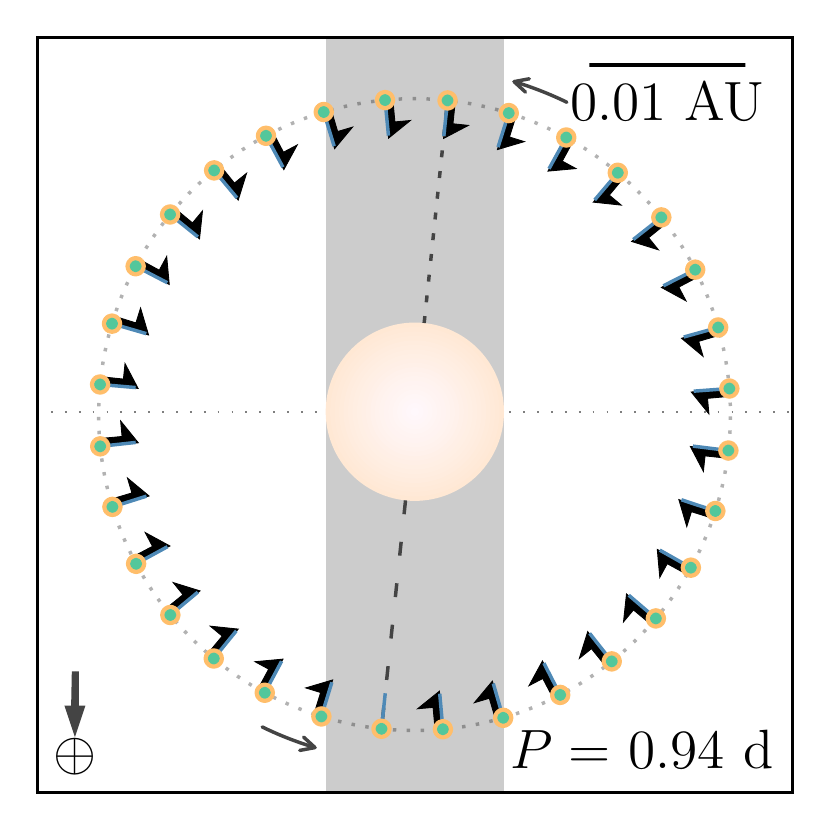}\\
\shortstack{\textbf{\small{WASP-19 b}} \\ \\ $M = 1.07 M_J$ \\ $R = 1.39 R_J$ \\ $T_{\mathrm{eq}} = 2520$ K \\ $\lambda = 4.6^{\circ}$} & \hspace{-0.790cm}\includegraphics[width=4.540cm]{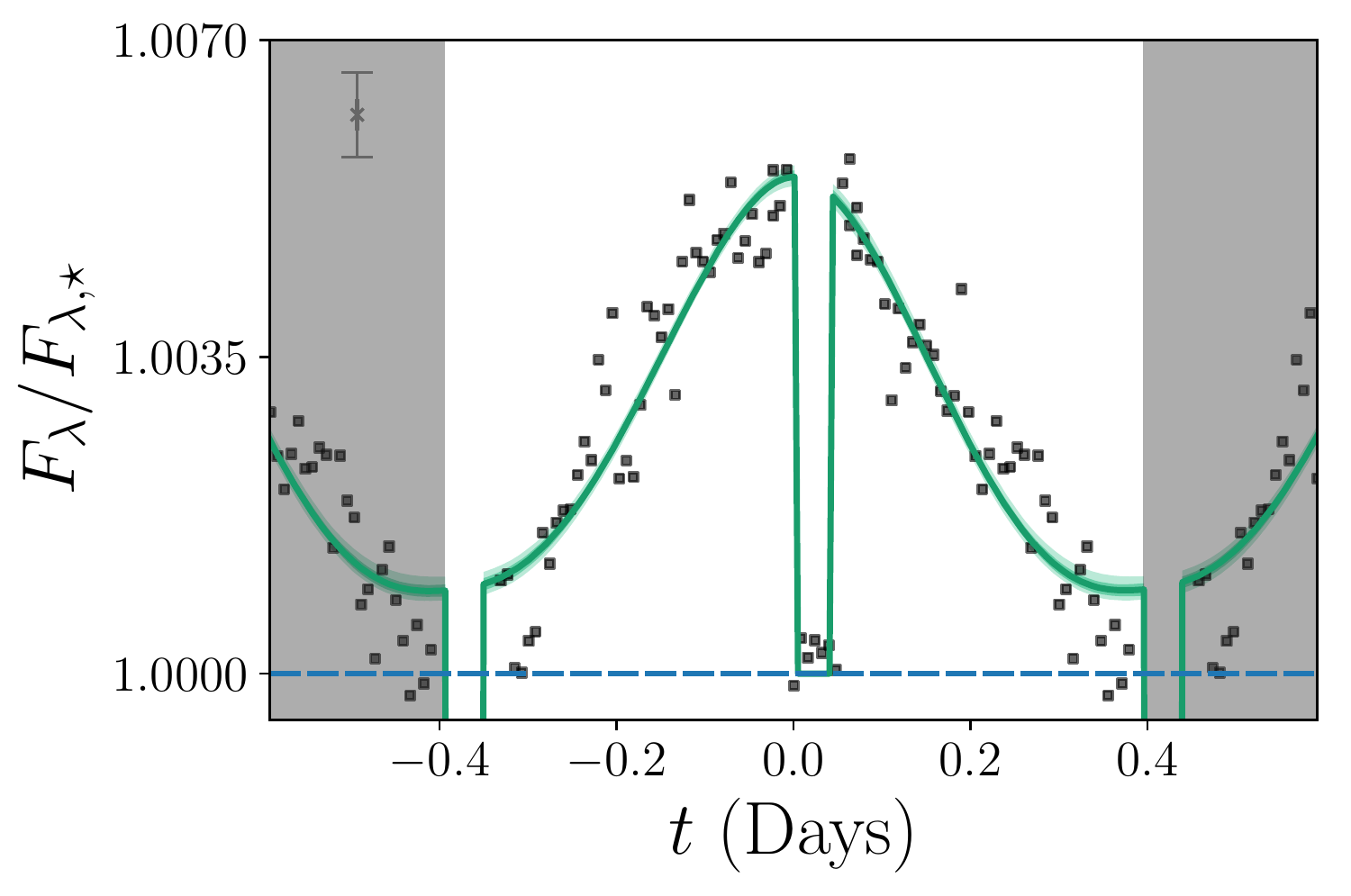} & \includegraphics[width=3.75cm]{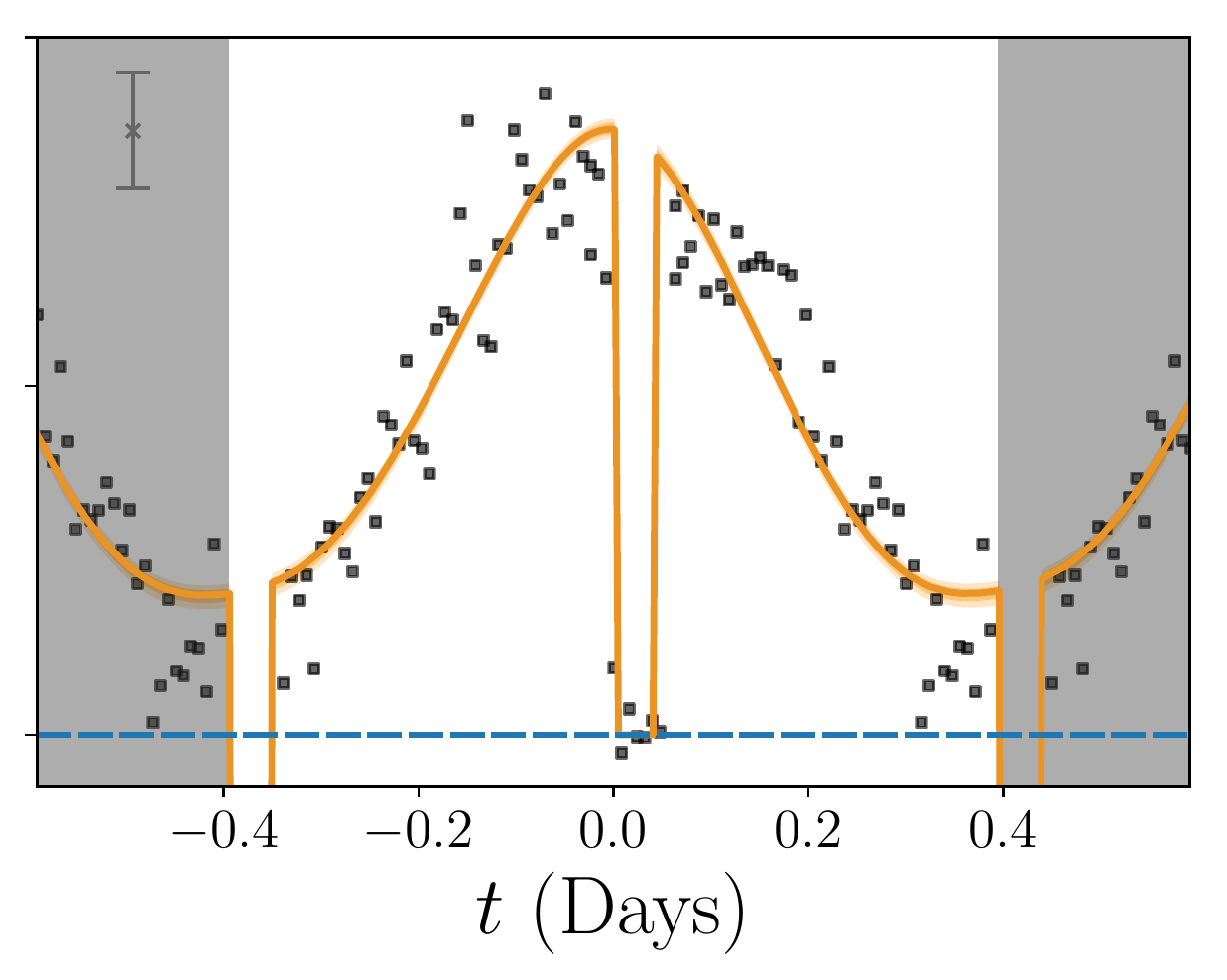} & & \includegraphics[height=3.5cm]{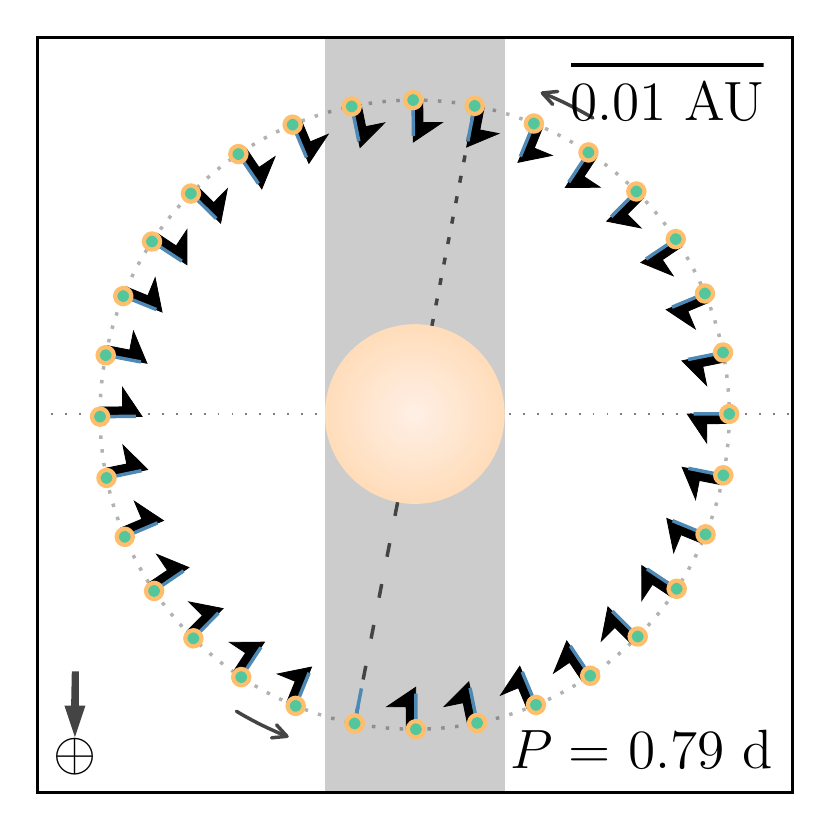}\\
\shortstack{\textbf{\small{WASP-33 b}} \\ \\ $M = 3.28 M_J$ \\ $R = 1.68 R_J$ \\ $T_{\mathrm{eq}} = 2723$ K \\ $\lambda = -108.8^{\circ}$} & \hspace{-0.790cm}\includegraphics[width=4.540cm]{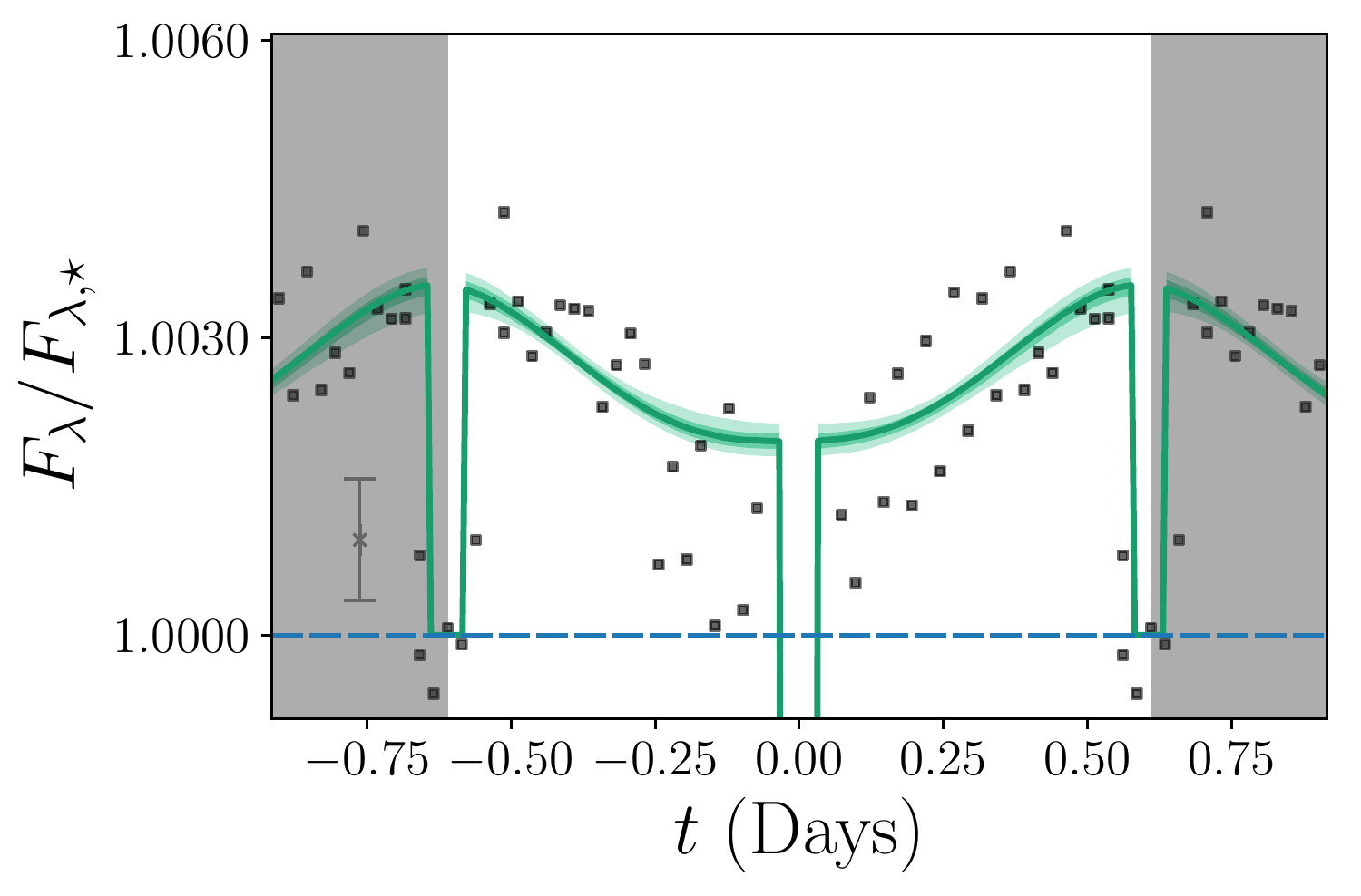} & \includegraphics[width=3.75cm]{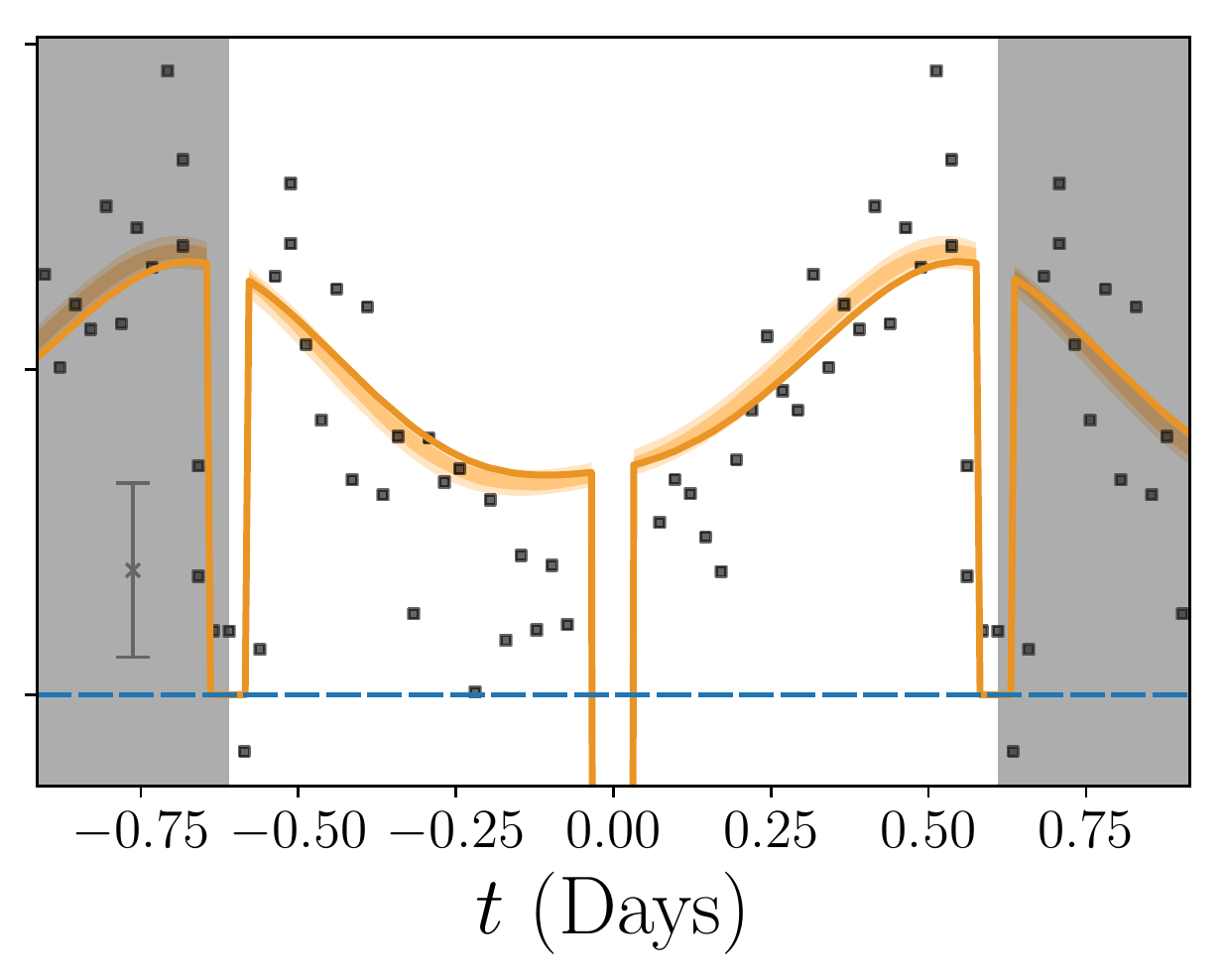} & & \includegraphics[height=3.5cm]{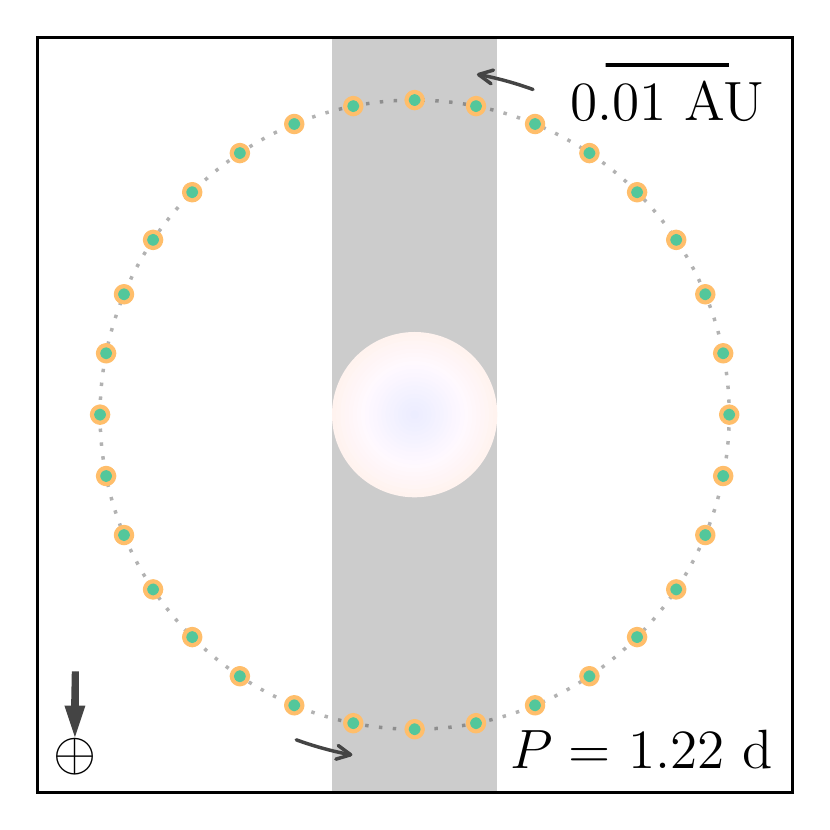}\\
\shortstack{\textbf{\small{WASP-43 b}} \\ \\ $M = 1.78 M_J$ \\ $R = 0.93 R_J$ \\ $T_{\mathrm{eq}} = 1350$ K} & \hspace{-0.790cm}\includegraphics[width=4.540cm]{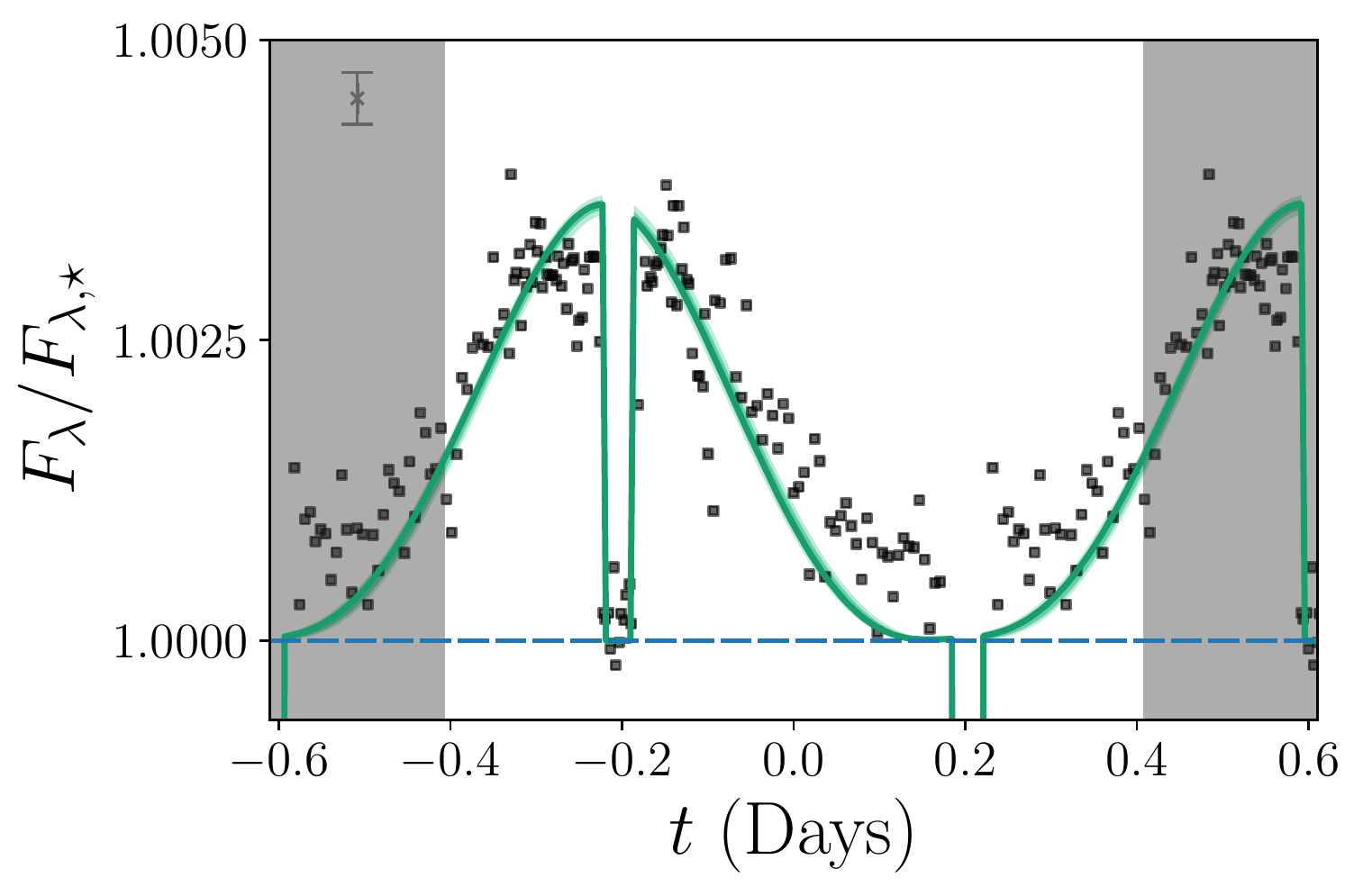} & \includegraphics[width=3.75cm]{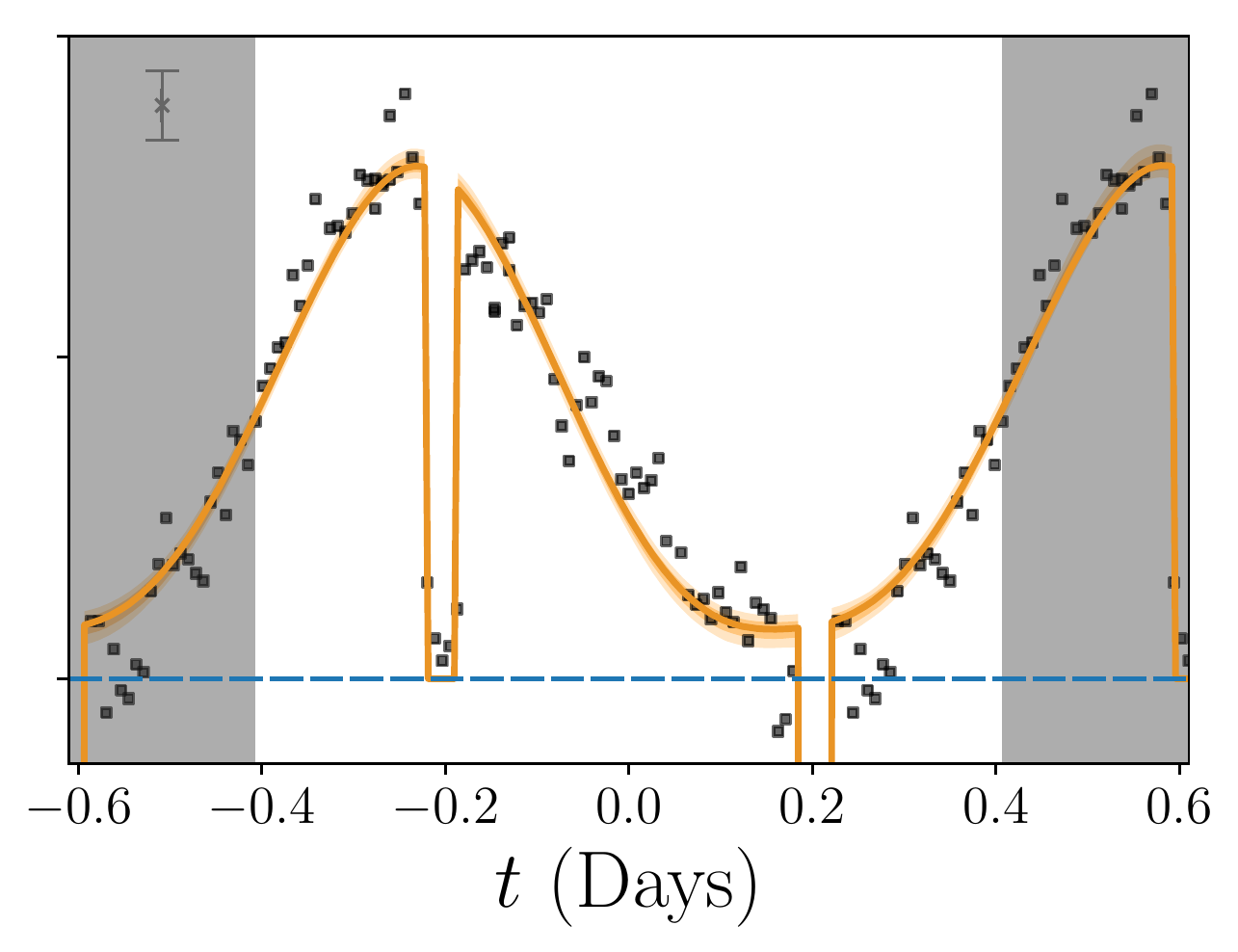} & & \includegraphics[height=3.5cm]{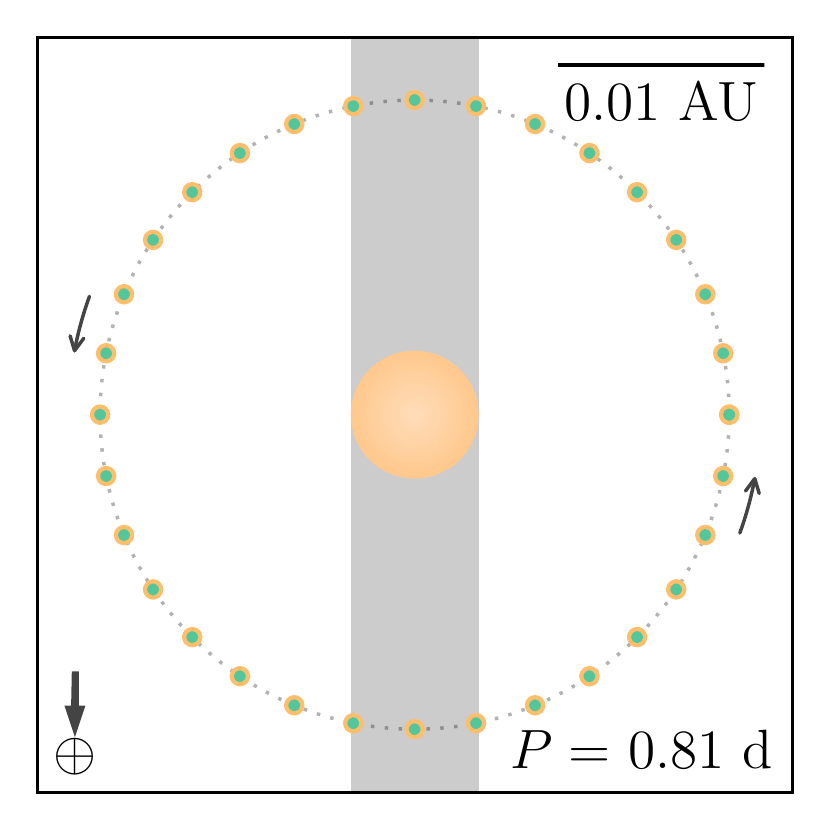}\\
\enddata
\tablerefs{\emph{Planet properties:} \citet{tur16} (GJ 436 b), \citet{pal10} (HAT-P-2 b), \citet{won16} (HAT-P-7 b), \citet{pon09} (HD 80606 b), \citet{car09} (HD 149026), \citet{sou10} (HD 189733 b, HD 209458 b), \citet{sou12} (WASP-12 b), \citet{rae15} (WASP-14 b), \citet{sou12} (WASP-18 b), \citet{won16} (WASP-19 b), \citet{tur16} (WASP-33 b), \citet{hel11b} and \citet{sal15} (WASP-43 b). \emph{Spin-orbit misalignments:} \citet{bou17} (GJ 436 b), \citet{alb12} (HAT-P-2 b), \citet{win09b} (HAT-P-7 b), \citet{hebr10} (HD 80606 b), \citet{alb12} (HD 149026 b), \citet{tri09} (HD 189733 b), \citet{win05} (HD 209458 b), \citet{alb12} (WASP-12 b), \citet{joh09} (WASP-14 b), \citet{tri10} (WASP-18 b), \citet{hel11a} (WASP-19 b), \citet{col10} (WASP-33 b).}

\label{fig:lightcurves}
\end{deluxetable*}

\subsection{Circular Orbit Hot Jupiters}\label{sec:results:circular}
\subsubsection{HAT-P-7 b}\label{sec:results:circular:HAT-P-7b}
The best fits in both bands agree quite well in preferred rotation period. The 3.6 $\mu$m model is marginally consistent with spin synchronization, and slightly longer in the 4.5 $\mu$m model. The models also agree on a radiative timescale within 1$\sigma$, and disagree only slightly on albedo, with the 3.6 $\mu$m model preferring effectively zero albedo and the 4.5 $\mu$m model retaining a modest reflectivity. The night-side temperatures differ significantly within the uncertainties, with the 3.6 $\mu$m model returning a minimum temperature effectively consistent with zero, and the 4.5 $\mu$m model returning a significant minimum temperature exceeding 2000 K.

\begin{figure*}[htb!]
\begin{center}
\textbf{HAT-P-7 b}\par\medskip
\begin{tabular}{cc}
\includegraphics[height=5cm]{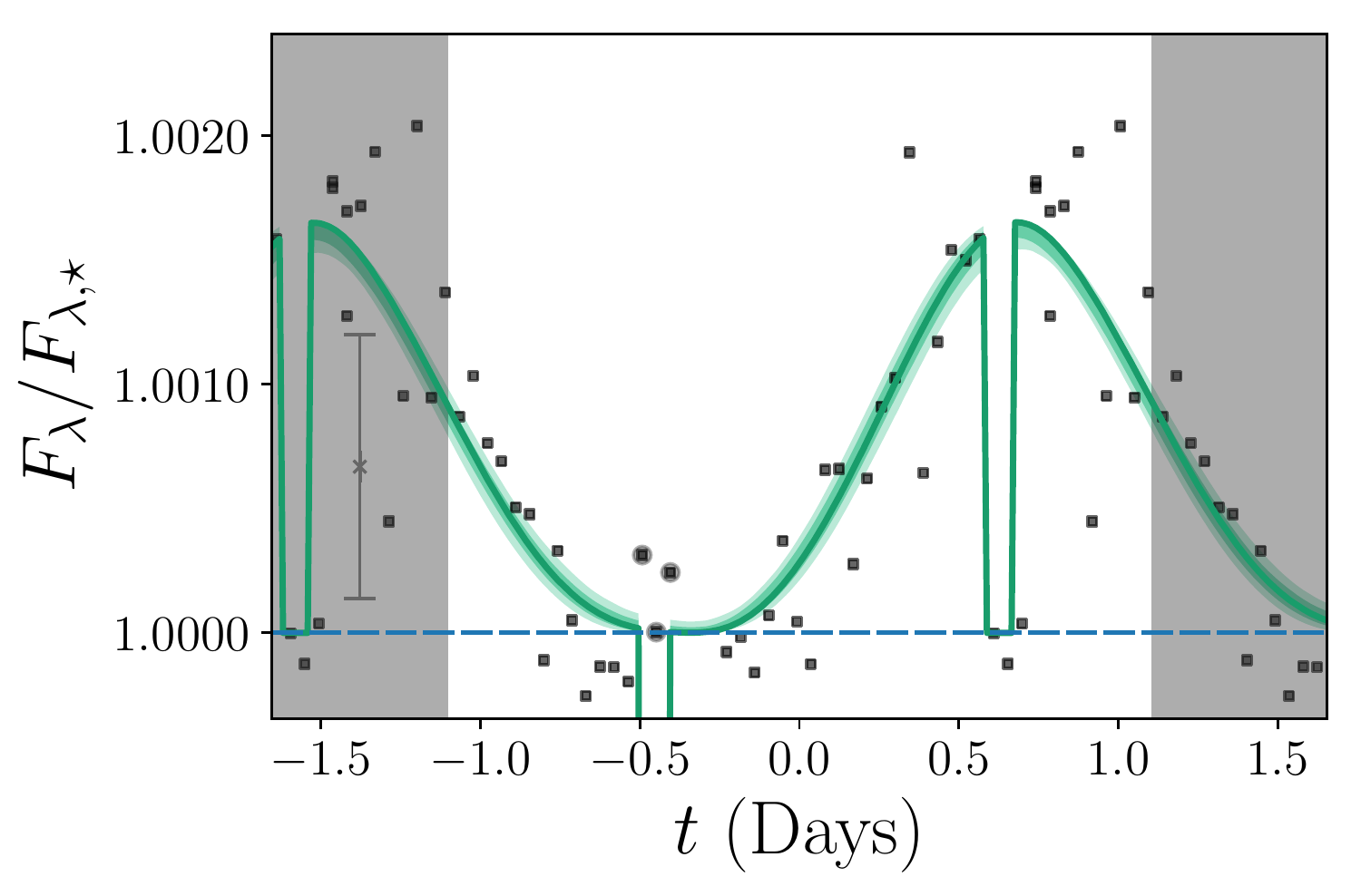} &
\includegraphics[height=5cm]{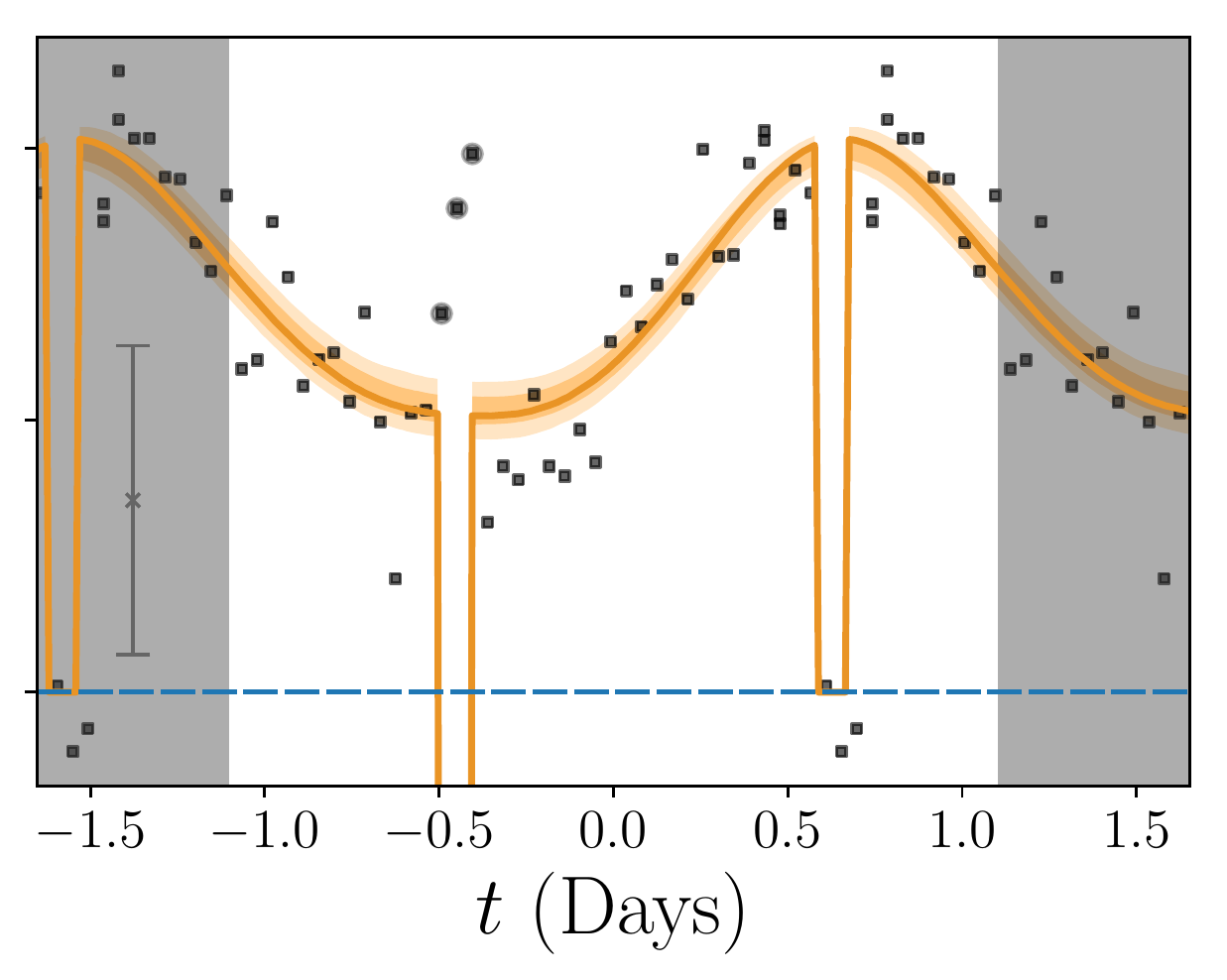}
\end{tabular}
\caption{If differences in the opacities between IRAC bands can be neglected, the transit depths in each band should be marginally consistent. We visualize this by adding back the geometric transit depth to the binned photometry within the observed transit. Here we present similar light curves as in Table \ref{fig:lightcurves}, but with the in-transit and in-eclipse shifted by the geometric depth (shown as slightly enlarged points). The resulting points are marginally consistent for 3.6 $\mu$m, but noticeably higher for 4.5 $\mu$m, indicating a clear difference in the transit depths by wavelength.}
\label{fig:HATP7b_delta}
\end{center}
\end{figure*}

We construct a check on the consistency of the photometry among bands of a single planet, which we highlight using HAT-P-7 b (Figure \ref{fig:HATP7b_delta}). We take the sampled photometry which lie within the transit and shift them up by the geometric transit depth. Assuming the differences in opacities of the atmosphere at each wavelength do not contribute significantly to the observed transit depth, we expect the resulting flux levels to be at least marginally consistent.

\subsubsection{HD 149026 b}\label{sec:results:circular:HD149026b}
Only the night-side temperatures for the 3.6 and 4.5 $\mu$m models agree within $1\sigma$. While the light curves both have muted amplitudes relative to the instantaneously and completely re-radiating case, the models prefer different means of recreating the shallow variations. In the 3.6 $\mu$m model, a very long radiative timescale is preferred, with a very small albedo. The 4.5 $\mu$m model, in contrast, a short radiative timescale is paired with a quite high albedo, absorbing only half of the incident stellar radiation. This suggests at least a modest degeneracy between these parameters in this case; it is not clear whether this degeneracy may be generalized.

Beyond these disagreements, the light curves also disagree visibly with respect to the observed phase offsets. The 4.5 $\mu$m light curve exhibits a slight positive phase offset, which we model with a rotation rate faster than synchronous. However, the 3.6 $\mu$m data appear to have a flux minimum which follows, rather than precedes, the transit. Therefore, the model returns a rotation rate slower than synchronous. This echoes the concerns raised in \citet{zha18}, where the authors point out that the positive phase offset disagrees with the negative offset in the 4.5 $\mu$m data.

\subsubsection{HD 189733 b}\label{sec:results:circular:HD189733b}
Both the 3.6 and 4.5 $\mu$m light curves exhibit minima in the flux prior to the transits, possibly indicating a significant phase offset. Accordingly, the rotation periods in each wavelength are consistent both with each other and with a planet whose relevant photospheres are super-rotating.\footnote{It should be noted that the $1\sigma$ range at 4.5 $\mu$m captures synchronous rotation; accordingly, the 4.5 $\mu$m photometry shows the weaker phase offset of the full-phase light curves.} All three night-side temperatures disagree within uncertainties. This is immediately evident in the 8.0 $\mu$m data which are elevated entirely above either of the other light curves, suggesting a significant discrepancy in the temperatures of the material responsible for emission at each band. The albedos in each wavelength are also both significantly nonzero and inconsistent with each other, pointing to further evidence of distinct photospheres with distinct properties such as temperature and opacity.

\subsubsection{HD 209458 b}\label{sec:results:circular:HD209458b}
\begin{figure}[htb!]
\begin{center}
\includegraphics[width=8.5cm]{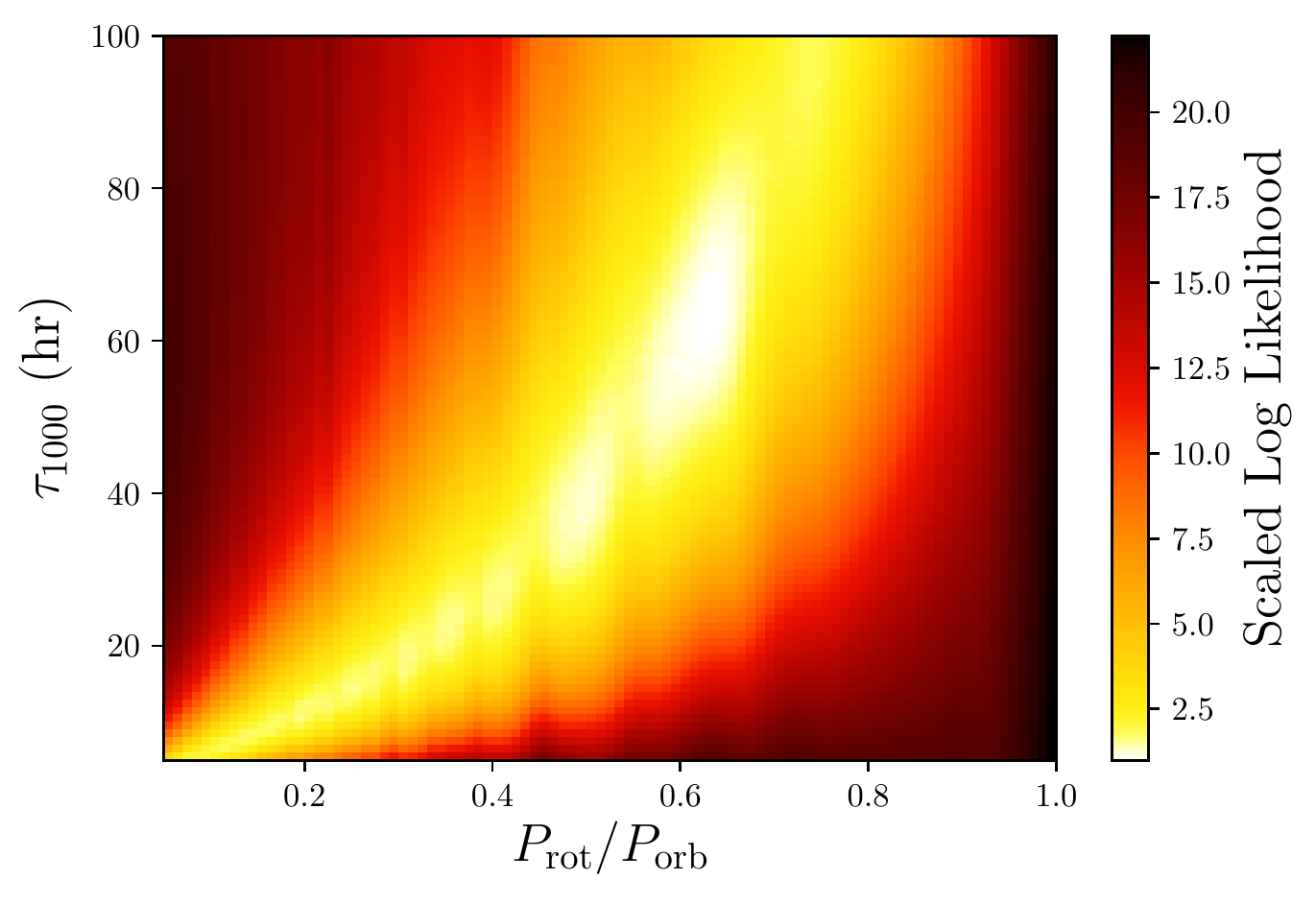}
\caption{The 2D projection of likelihood space for the model of HD 209458 b. The likelihoods are presented in logarithmic space, with unity fixed to the minimum likelihood. The most favorable likelihoods, shown in white, are not well-localized in one region but instead follow a track in the rotation period-radiative timescale plane. This is the primary contributor to the large $1\sigma$ uncertainties in both rotation period and radiative timescale.}
\label{fig:HD209458b_degen}
\end{center}
\end{figure}

In contrast with the fits of 3-D models to the data, our simple thermal model readily reproduces the flux minimum prior to transit, within uncertainties. This is accomplished by both relaxing the assumption of a synchronous rotation period and having a radiative time scale which is non-negligible compared with the rotation period. Here we find a best-fit rotation period that is $\sim\!0.46$ the orbital period, but with a significant upper range. The radiative timescale, which is of similar scale to the rotation period, has a similarly large upper range, suggesting there is considerable degeneracy between the two timescales. The MCMC chain used to determine uncertainties remains quite close to the best-fit values in minimum temperature and albedo, while exploring an extended region of favorable likelihoods in projected parameter space for both rotation period and radiative timescale (Figure \ref{fig:HD209458b_degen}).

\subsubsection{WASP-12 b}\label{sec:results:circular:WASP-12b}
Due to the nature of the WASP-12 system, we expect limited success in effectively capturing the behavior of phase variations with a simple spherical-planet model. Prior to eclipse, the data in both bands exhibit very hot day-side temperatures; just prior to transit, the observed temperatures drop precipitously to a level which is effectively consistent with zero flux. Our models converge on results which prefer to cut through the rough median of the high-amplitude observations. Super-synchronous rotation rates produce pronounced phase offsets in each band; beyond this, our models fail to capture the extreme features in the photometry.

It is possible that the phase curve is influenced by the tidal distortion of the planet. We consider the effects of relaxing the spherical shape assumption for our planets in \S \ref{appendix:tidal}. Even with an ellipsoidal model, with ellipticity determined by the known mass ratio of the system, the effect of including the distortion makes a negligible change in both the best-fit parameters and resulting model light curves (Figure \ref{fig:WASP12b_prolate}).

\subsubsection{WASP-14 b}\label{sec:results:circular:WASP-14b}
While our models do a reasonable job of fitting the broad phase variations in each band, the photometry in both bands indicates that the total amplitudes might not be fully captured by our simple thermal model. In particular, the models capture the maxima well but unable to quite reach the minima, even with short radiative timescales.

\subsubsection{WASP-18 b}\label{sec:results:circular:WASP-18b}
In contrast with the results for WASP-14 b, here we are now unable to capture the peak fluxes, especially at 4.5 $\mu$m, even with short radiative timescales, elevated minimum temperatures, and effectively complete absorption. Otherwise, the phase offsets in each band are minimal, leading to rotation periods roughly consistent with synchronous rotation. 

\subsubsection{WASP-19 b}\label{sec:results:circular:WASP-19b}
Both models prefer rotation rates faster than synchronous, relatively fast radiative timescales, elevated night-side temperatures, and modest albedos. Across all parameters the $1\sigma$ bounds agree between 3.6 and 4.5 $\mu$m, though the minima exhibit a similar case to the other WASP planets in remaining elevated with respect to the data minima.

\subsubsection{WASP-33 b}\label{sec:results:circular:WASP-33b}
WASP-33 b has the lowest signal-to-noise ratio of all the WASP planets considered here. Nevertheless, the variations are largely fit by short radiative timescales, very high night-side temperatures, and significant albedos. The model rotation periods are in disagreement, but the pronouncement of the phase offsets is debatable given the quality of the data. Both photometric light curves show a strong drop-off just prior to transit, which the models are unable to capture.

\subsubsection{WASP-43 b}\label{sec:results:circular:WASP-43b}
Both models prefer rotation rates slightly faster than synchronous. The models disagree on the night-side temperature but with ambiguous cause. One reason for the seemingly inaccurate low night-side temperature at 3.6 $\mu$m might be due to a strange feature in the photometry following the eclipse. The flux drops off briefly, before returning to a level consistent with the rest of the phase variations\footnote{Note that we have already excluded the data which were originally excluded in the original analysis presented in \citet{ste17}. The feature we discuss here is one which persists in the published data.}.

\subsection{Eccentric Hot Jupiters}\label{sec:results:eccentric}
\subsubsection{GJ 436 b}\label{sec:results:eccentric:GJ436b}
Our analyses are limited by the poor quality of the photometry, despite having more than one whole orbital period of observations. Our best results suggest the planet's emitting layer is rotating significantly faster than the eccentric pseudo-synchronous rate, which allows for a significant phase offset which is suggested by the observations prior to transit. The $1\sigma$ range in radiative timescales approach the rotation period at the upper limit, which limits the amplitude of flux variations, and suggests significant night-side temperatures.

\subsubsection{HAT-P-2 b}\label{sec:results:eccentric:HAT-P-2b}
The results for 4.5 and 8.0 $\mu$m are consistent with pseudo-synchronous rotation, while the 3.6 $\mu$m model favors a slightly faster rotation of that emitting layer. All radiative timescales suggest a response significantly faster than the expected rotation period, implying a small phase offset between peak flux and secondary eclipse, which is observed.

There is a strong disagreement in the minimum temperatures of the 3.6 and 8.0 $\mu$m models with that of the 4.5 $\mu$m model. This is not surprising given the data, which settles to a baseline flux ratio consistent with zero planetary contribution for the 3.6 $\mu$m data (and, arguably, the $8.0$ $\mu$m data prior to the peak), while remaining significantly higher than unity for the 4.5 $\mu$m data. Under the assumption that the Spitzer photometry systematic uncertainties have been well characterized, this would imply that the observations probe different atmospheric layers depending on wavelength: a warm layer visible in the 4.5 $\mu$m band, and a cooler layer in the others.

While the best-fitting night-side temperature at 8.0 $\mu$m is the lowest of all bands, we also at the same time fail to reach the peak observed flux in the band. Given that the available photometry only captures a single transit and eclipse, with the relatively short time separation between them, we are missing much of the orbital phases that constrain how the planet cools following periastron.

The observed eclipse depths are variable, particularly in 3.6 and 8.0 $\mu$m. In 3.6 $\mu$m, one eclipse depth is consistent with no planetary contribution to the flux, while the second is relatively elevated. For our analyses we phase-fold all photometry into a single orbit, thereby effectively folding these eclipses on top of one another. At 8.0 $\mu$m, while only one eclipse was observed in the partial-phase observation, the spread in the reduced data leads to binned median points below unity. This is an effect of the noise in the data.

This disagreement persists for albedo, with the 3.6 and 8.0 $\mu$m models favoring effectively zero albedo (both with only modest upper limits), while the 4.5 $\mu$m model favors a significant albedo of 0.42.

Moreover, the observed transit depth is considerably different in 3.6 and 8.0 $\mu$m from that in 4.5 $\mu$m. While wavelength-dependent opacity can contribute to differences in transit depths, from a scale height argument it is unlikely that a non-gray opacity would explain the entire difference. This is because HAT-P-2 b's high mass (9.09 $M_J$) should mitigate much of the differences in effective scale height.\footnote{See the discussion in \S 4.4 of \citet{lew13}.}

\subsubsection{HD 80606 b}\label{sec:results:eccentric:HD80606b}
HD 80606 b has an extremely eccentric orbit ($e=0.93366$) which prompts consideration of the role of tidal heating and dissipation (see also \S \ref{sec:model:rotation}). Since HD 80606 b is thought to be the only planet in its system, and the age of its host star (and hence the system as a whole) is inconsistent with a young system ($\sim 1$--10 Gyr) \citep{tak07}, the tidal $Q$ must be $\gg 1$; \citet{dew16} calculate $Q > 2.5 \times 10^6$. Therefore, the planet is expected to be extremely slow in both circularizing its orbit and slowing its rotation.

Radiative timescales considerably shorter than the theoretical pseudo-synchronous rotation period of 39.9 hours allow the time of peak flux to be close enough to periastron. The bands require distinct, significant albedos to fit the observed maximum fluxes. While the best fits for the rotation period disagree, the uncertainties are among the largest of all planets considered. This is almost certainly a consequence of the limited phase photometry from the long 111-day orbital period. Our results therefore are at least marginally consistent with the published results from \citet{dew16}, who find a rotation period longer than pseudo-synchronous.

\subsection{Discerning Trends in the Planet Sample}\label{sec:results:trends}
\begin{figure}[htb!]
\begin{center}
\includegraphics[width=8.5cm]{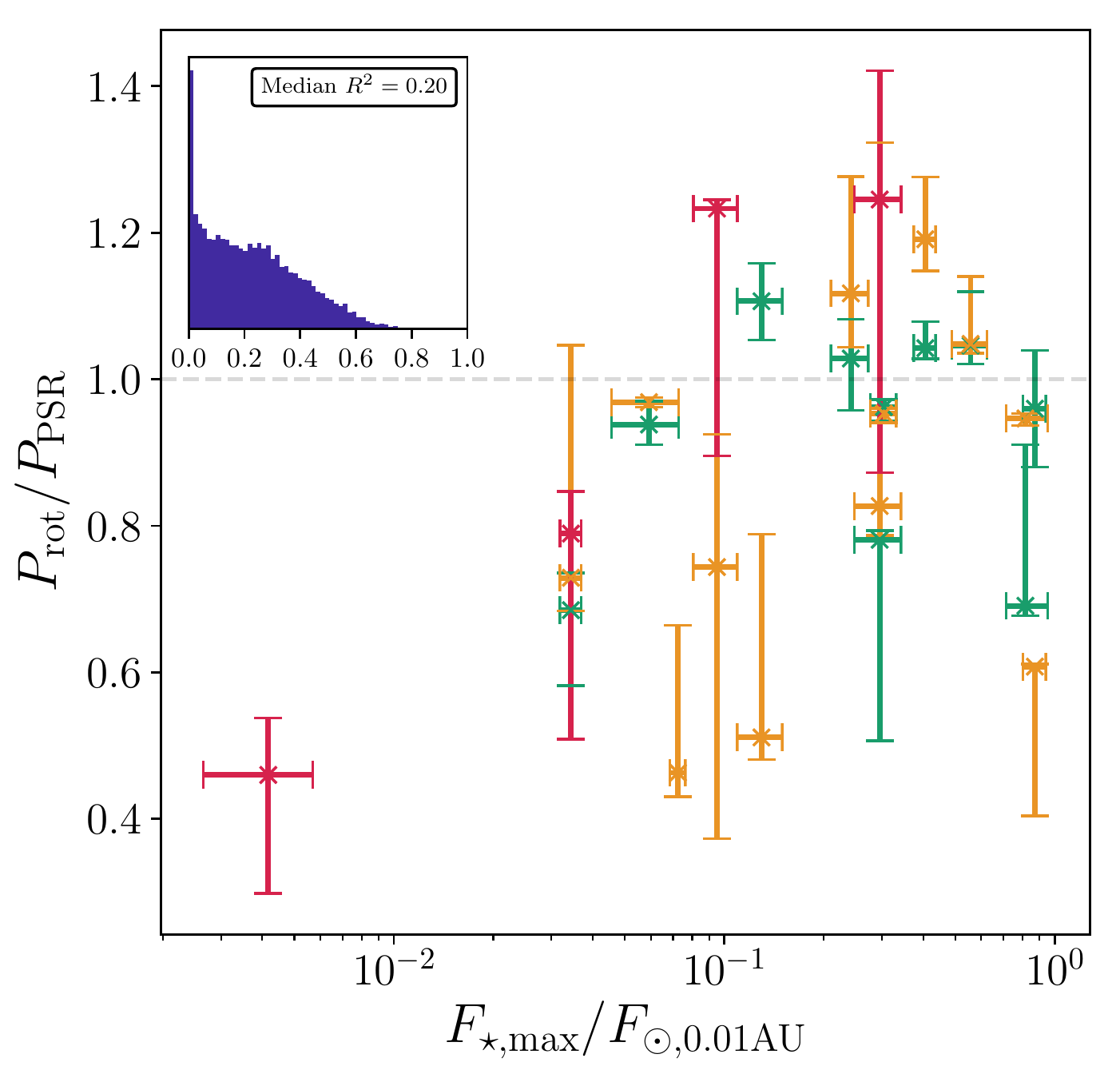}
\caption{Comparison of the peak instellation (relative to that of the Solar System at 0.01 AU) with the best-fit rotation period from our model. Each band of each planet from our light curve analysis is represented here, with errors in instellation propagated from system properties and errors in rotation period from the uncertainties in our model. The dashed line indicates a rotation period equal to the (pseudo-)synchronous period.}
\label{fig:instellation_rotper}
\end{center}
\end{figure}

Despite a sample of 13 planets with a combined 26 light curves spanning 3 Spitzer bands, comparison of the known and the inferred properties of the systems generates a frustratingly inclonclusive overall picture. Very few substantial correlations seem to exist among the fitted parameters and system properties. We do, however, note a potentially interesting trend by drawing inspiration from \citet{zha18}, who noticed a connection between the ``instellation temperature'' (equivalent to equilibrium temperature) of a planet and its phase offset. We construct an analogous plot in Figure \ref{fig:instellation_rotper}, comparing the peak instellation for each planet with the preferred rotation period from our model. In principle, the phase offset is dictated by the interplay between the rotation period and radiative timescale; a non-synchronous rotation coupled with a non-zero timescale for heating allows for a displacement of the peak thermal emission from the substellar point\footnote{In principle, the other two parameters (minimum temperature and albedo) contribute to the calculation of the phase offset. However, their effects are generally secondary.}. The diagram suggests a weak positive correlation, which we quantify through bootstrap resampling from both the data points and from their uncertainties. For each resampled realization, we then calculate a linear regression, and plot the distribution of the $R^2$ statistics. The median $R^2$ value is 0.20, with a significant peak in the lowest bin. This indicates a considerable sensitivity of the bootstrapping process to the selection of just one (or a few) data points; in this case, the selection of GJ 436 b appears to drive the correlation up markedly.

The strongest correlation appears in comparing the orbit-averaged (mean) instellation with the radiative timescale relative to the periastron equilibrium temperature. Despite a median $R^2$ of 0.36, the distribution is again quite sensitive to the selection of a single planet's data, in this case HD 80606 b. This, and the previous result, as well as many other tests for correlations, suggest that our understanding of trends in the physical properties may be fundamentally limited by the small numbers of planets with comparable light curves, especially those with appreciable eccentricity.

\begin{figure}[htb!]
\begin{center}
\includegraphics[width=8.5cm]{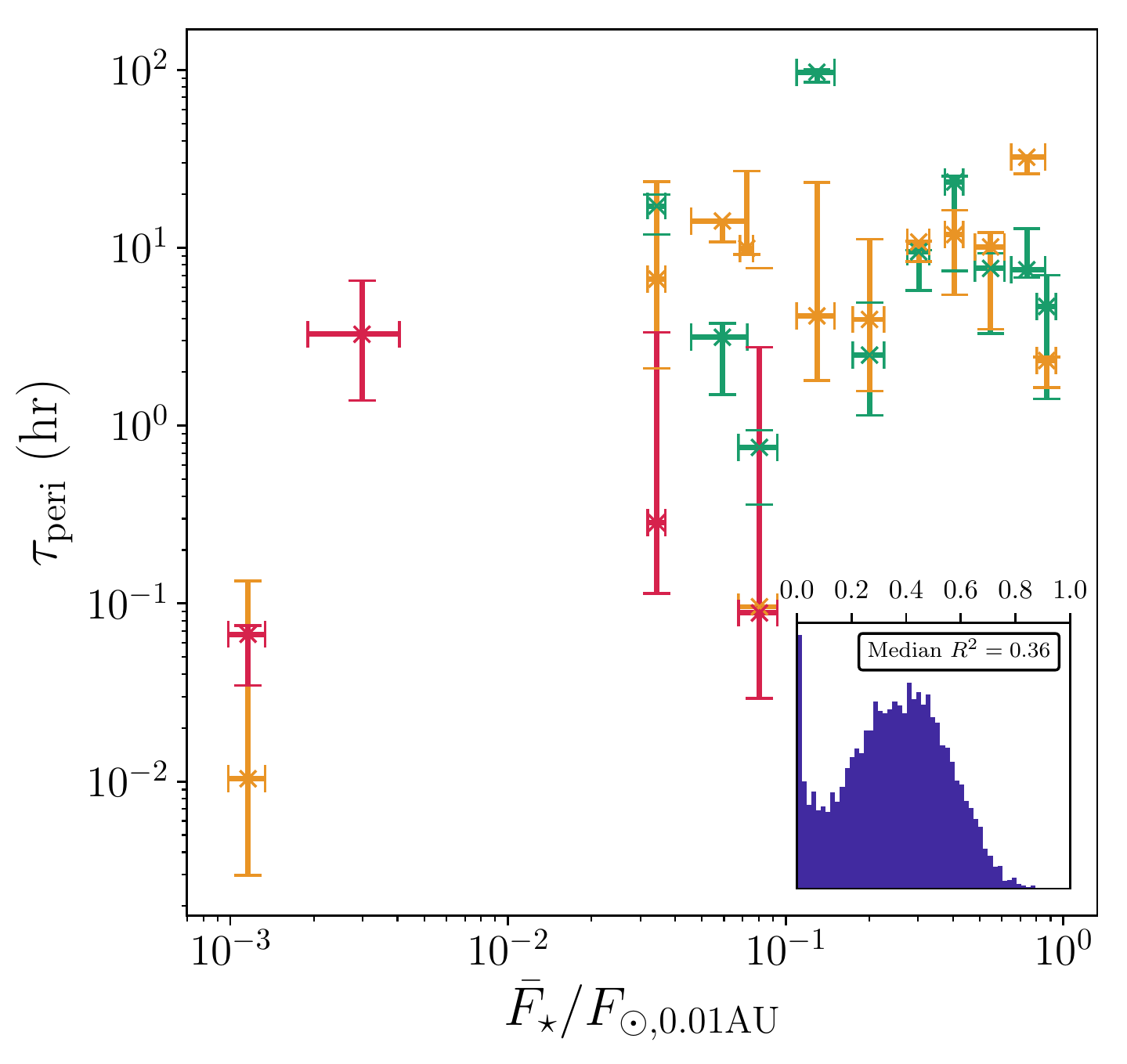}
\caption{Comparison of the time-averaged instellation (relative to that of the Solar System at 0.01 AU) with the radiative timescale expressed relative to the equilibrium temperature at periastron. Each band of each planet from our light curve analysis is represented here, with errors in instellation propagated from system properties and errors in radiative timescale from the uncertainties in our model.}
\label{fig:instellation_radtime}
\end{center}
\end{figure}

\section{Discussion}\label{sec:discussion}
The existing full-orbit photometry from Spitzer are readily amenable to re-analysis using the simplest physically interesting radiative model. Our thermal model reproduces effectively the large-scale features in much of the data, capturing both the timescale of heating/cooling, the time positions of minimum and maximum flux, and the depths of both the transits and the secondary eclipses. The existence of photometry in multiple wavelength bands allows us to compare the consistency of the model parameters. The most prevalent disagreements among bands for individual planets tend to be between night-side temperatures and albedos. Arguably these properties are often among the most robustly determined from the data, since they arise almost entirely from the minima and amplitudes of the phase variations. These disagreements also show up in HAT-P-2 b and HD 80606 b, the planets in our sample with the highest eccentricity. The simplest explanation for these differences is that the photospheres for each wavelength are distinct, each having their own characteristic temperature and opacity. However, the photospheric depths can vary not just by wavelength but also longitude, due to day-night differences in the IR absorbers (CO vs. CH$_4$, see the discussion in \citet{dob17}). There is certainly the possibility of cloud formation due to their time-dependent instellation, which would introduce non-gray opacities.

For the majority of the planets studied here, rotation periods are not well-constrained observationally. There is a wide range of rotation periods among the planets, from roughly half-synchronous to slightly super-synchronous. Moreover, among planets on highly eccentric orbits, there may not be a guarantee of zero obliquity. Acknowledging that many of our circular-orbit planets are expected to have their solid-body rotation be synchronous with their orbital periods, our results suggest a range of coherent atmospheric motions. Given that eastward hot spots have been demonstrated for many Hot Jupiters on circular orbits, there is observational evidence for bulk equatorial flow of the atmospheric photosphere(s). The expectations for the characteristics of bulk flow for planets on eccentric orbits are even less well-constrained. As we have shown, discerning any meaningful trends among atmospheric and system properties, for example between rotation/phase offsets and instellation, is difficult given both the quality and quantity of the photometry. The primary challenge is the scarcity of full-orbit light curves for eccentric Hot Jupiters; of the known planets, HAT-P-2 b is currently the only planet that meets this criterion. HAT-P-2 b will therefore be an interesting target for further observation to refine the constraints on its rotation and bulk atmospheric dynamics. For the most eccentric planets, despite the near-impossible prospects of obtaining full orbits, having much stronger signal-to-noise data encompassing periastron and occultations will be crucial to more confidently constraining the unique atmospheric conditions of the flash heating during periastron.

Beyond the possible astrophysical causes of variation among the model parameters, we also express caution in over-interpreting the data currently available. There is a very real possibility that some number of observed features in the photometry are at least partially the result of uncharacterized systematic uncertainties \citep[e.g.][]{cow12,max13,zel14,won16,ste17,zha18}. In the case of HAT-P-2 b, for example, there are two separate eclipse measurements whose fit depths have a slight overlap at $1\sigma$ but nevertheless disagree by roughly 20\%. For HAT-P-2 b and HAT-P-7 b in particular, the observed differences in transit depths are likely not entirely explained by predicted differences in photospheric scale heights.

It is important to note that in many of the cases where 3-D circulation models (GCMs) have been applied to partial and full phase photometry, they have been unable to fully explain the observed shape of the light curves. This is potentially due to physical phenomena which have not been incorporated into the radiative transfer calculations. \citet{zel14} speculate that the discrepancy in minimum night-side fluxes between the models of \citet{sho09} and the observed light curve could, for example, be due to not modeling disequilibrium chemistry. In particular, a mechanism that could elevate CH$_4$ abundances in the atmosphere would provide a cooling source that could bring the theoretical night-side flux in 4.5 $\mu$m emission closer to the observed flux. More recently, \citet{dru18} have re-analyzed the 4.5 $\mu$m and conclude disequilibrium chemistry cannot entirely explain the model-data discrepancy. At higher eccentricities, \citet{lew17} apply similar GCMs to the periastron-centered photometry of HD 80606 b in an attempt to explain the observed shape and rate of heating in the 4.5 and 8.0 $\mu$m light curves, but are not able to simultaneously fit both the amplitude and phase offset.

As atmospheric detection and characterization moves forward with the anticipated launch of the James Webb Space Telescope, comprehensive characterization of the instrumental response will be crucial; it is expected that JWST data will contain systematics of a similar nature to those seen in past space telescopes, especially on the time scales needed for phase photometry \citep[see][for a thorough and up-to-date discussion]{bea18}. The increased resolution alone will not be sufficient to disentangle the current ambiguities in exoplanet atmospheric data. Through careful target selection we hope to gather additional phase photometry that will aid in the determination of fundamental atmospheric properties, especially in the regime of high eccentricity.

\appendix

\section{Tidal Distortion: The Case of WASP-12 b}\label{appendix:tidal}
WASP-12 b is very likely tidally distorted due to its low density and proximity to its host star. Its radius (1.83 $R_\mathrm{J}$) is quite inflated given its mass (1.43 $M_\mathrm{J}$) \citep{sou12}, which suggests that the outermost layers of the planet's atmosphere are suspectible to significant mass loss. \citet{lis10} made the initial claim that WASP-12 b should be filling a significant part of its Roche lobe, a theoretical prediction that was expanded in subsequent works such as \citet{lai10}.

One can use the ratio of the stellar and planetary masses to estimate the extent to which the planet's shape should be distorted. The \emph{Roche radius} describes the distance between the center of the planet and the L1 Lagrange point of the star-planet system:

\begin{equation}\label{eq:roche}
    \frac{R_\mathrm{Roche}}{a} = \left(\frac{M_{\mathrm{P}}}{3 M_\star}\right)^{1/3}.
\end{equation}
For WASP-12 b, $R_\mathrm{Roche}/R_{\mathrm{P}} \approx 1.9$.

Given that WASP-12 b should be tidally distorted, one possible way to fit the observed light curves more accurately is to update the planet geometry in our model. To do this we assume the planet's distorted shape may be approximately modeled as a prolate spheroid, whose prolate axis is assumed to always lie along the star-planet line. We define the ``eccentricity'' $\eta$ of the elliptical cross-section of the planet along its prolate axis. That is, for semi-major and semi-minor axes $A_\mathrm{P}$ and $B_\mathrm{P}$,

\begin{equation}\label{eq:prolate_ecc}
    \eta = \sqrt{1 - \left(\frac{B_\mathrm{P}}{A_\mathrm{P}}\right)^2}.
\end{equation}

For planets whose radii are determined from transit depths, the measured radii $R_\mathrm{P}$ are equivalent to $B_\mathrm{P}$, the semi-minor axes. This sets the axis ratio

\begin{equation}
    \frac{A_\mathrm{P}}{R_\mathrm{P}} = \left( 1-\eta^2 \right)^{-1/2}.
\end{equation}

\citet{lis10} estimate that the difference in observed emission flux from the planet between transit, where the minimum cross-sectional area is seen, and quadrature, where we see the maximum area, is of order $\sim 10\%$. This corresponds to a tidal eccentricity on order $\eta \sim 0.42$. This result can be accurately reproduced by solving for the maximum extent of the planet along each Cartesian axis, assuming those extreme points lie on the same surface of constant Jacobi energy, which is defined by

\begin{align}\label{eq:constant_jacobi}
\begin{split}
\Phi \!\left(\vec{x}\right) = -\frac{GM_\star}{a} \Bigg\{&\left[\left(\frac{x}{a}+\frac{\xi}{1+\xi}\right)^2 + \left(\frac{y}{a}\right)^2 + \left(\frac{z}{a}\right)^2\right]^{-1/2} \\ + &\left[\left(\frac{x}{a}+\frac{1}{1+\xi}\right)^2 + \left(\frac{y}{a}\right)^2 + \left(\frac{z}{a}\right)^2\right]^{-1/2} \\ + &\frac{1+\xi}{2} \left(\frac{x^2+y^2}{a^2}\right)\Bigg\}
\end{split}
\end{align}
where $\xi$ is the planet-star mass ratio $M_P / M_\star$, and the coordinates are set up such that $x$ increases from star to planet, $y$ points in the direction of the planet's motion in its orbit, and $z$ is tangent to the orbital plane according to a right-handed coordinate system. The transit depth, calculated from observations, gives the product of the extent of the planet along the two dimensions normal to the star-planet axis ($y$ and $z$)\footnote{Note that solving this equation in general leads to different extents along $y$ and $z$. For WASP-12 b, the differences are minor enough that the corresponding differences in calculated ellipticity are $\lesssim$ 0.01.}.

The effect of the distortion on the area of each grid cell can be calculated using the first fundamental form of differential geometry. The surface of our planet can be parametrized in Cartesian coordinates as

\begin{equation}
    \boldsymbol{r} = r_x\hat{x} + r_y\hat{y} + r_z\hat{z}
\end{equation}
where

\begin{align*}
    r_x &= A_\mathrm{P} \sin \theta \cos \phi,\\
    r_y &= B_\mathrm{P} \sin \theta \sin \phi,\\
    r_z &= B_\mathrm{P} \cos \theta.
\end{align*}
The coordinates are defined as in Equation \ref{eq:constant_jacobi}, with $\hat{z}$ along the rotation axis and $\hat{x}$ along the prolate (star-planet) axis.

In general, the area of an element on a solid surface is given as
\begin{equation}
    A_{ij} = \int_{\phi_i}^{\phi_j}\!\int_{\theta_i}^{\theta_j}\!\sqrt{\left(\frac{d\boldsymbol{r}}{d\theta} \cdot \frac{d\boldsymbol{r}}{d\theta}\right) \left(\frac{d\boldsymbol{r}}{d\phi} \cdot \frac{d\boldsymbol{r}}{d\phi}\right) - \left(\frac{d\boldsymbol{r}}{d\theta} \cdot \frac{d\boldsymbol{r}}{d\phi}\right)^2} \,d\theta\,d\phi.
\end{equation}

In the case of a prolate spheroid, this integral simplifies to

\begin{equation}
    A_{ij} = \, \frac{R^2_\mathrm{P}}{1-\eta^2} \int_{\phi_i}^{\phi_j}\!\int_{\theta_i}^{\theta_j}\! \sin\theta \, \Big[ 1 - \eta^2 \left(2-\cos^2\theta\right) + \eta^4 \left(\sin^2\theta + \cos^4\theta\sin^2\phi\cos^2\phi\right) \Big]^{1/2} \,d\theta\,d\phi.
\end{equation}

\begin{figure*}[htb!]
\begin{center}
\textbf{WASP-12 b}\par\medskip
\begin{tabular}{cc}
\includegraphics[height=5cm]{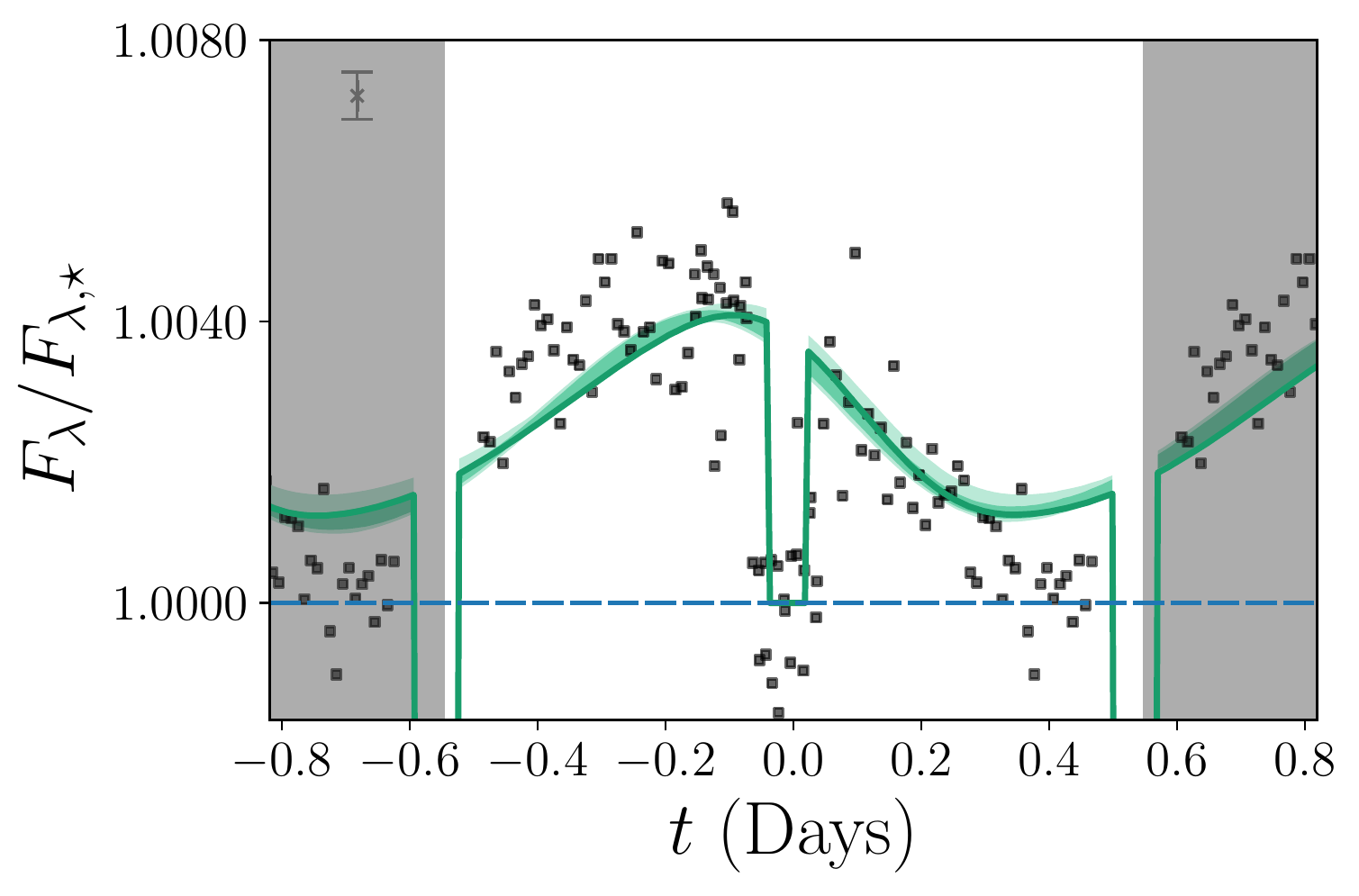} &
\includegraphics[height=5cm]{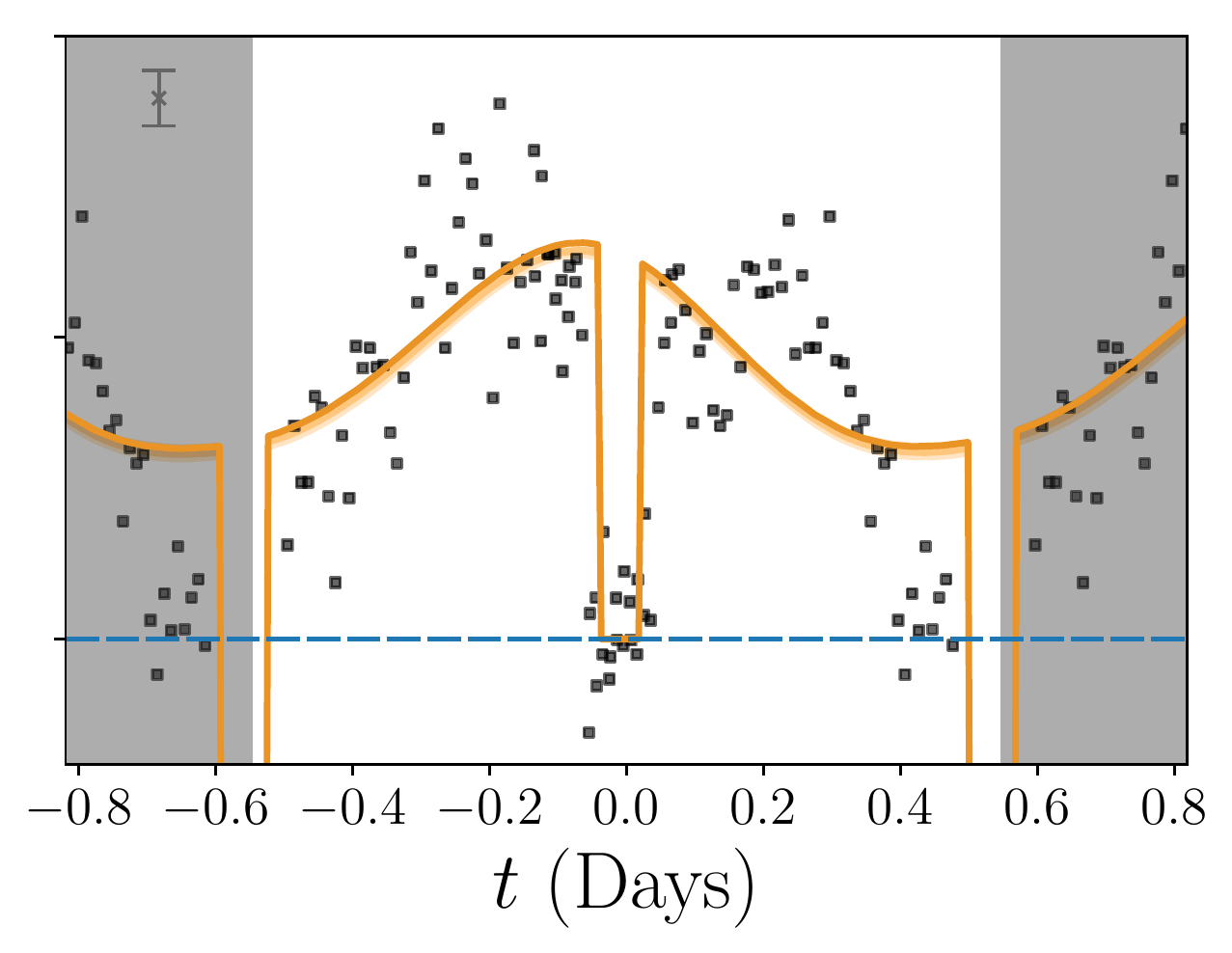}
\end{tabular}
\caption{With a prolate eccentricity $\eta \approx 0.42$ (as defined and calculated from Equations \ref{eq:roche}--\ref{eq:constant_jacobi}), we make little progress in how accurately our light curves reproduce the peak fluxes in the data.}
\label{fig:WASP12b_prolate}
\end{center}
\end{figure*}

\newcommand\tnH{\tablenotemark{H}}

\clearpage
\section{Tables of Eclipse and Transit Properties}

\startlongtable
\begin{deluxetable*}{ccccccc}
\centering
\tabletypesize{\scriptsize}
\tablecaption{Eclipse Properties for Exoplanets with Spitzer Light Curves}
\tablewidth{0pt}
\tablehead{
\colhead{Band} & \colhead{$E_{\textrm{mid}}$ $\left(\mathrm{BJD}-2450000\right)$} & \colhead{$E_{1,4}$ (days)} & \colhead{$E_{1,2}=E_{3,4}$ (days)} & Depth & Ref.}
\startdata
\cutinhead{GJ 436 b}
3.6 $\mu$m & $4496.4888\pm0.0010$ & $0.0508\pm0.0021$ & & $0.00041\pm0.00003$ & \citet{ste10} \\
4.5 $\mu$m & $4499.1330\pm0.0010$ & $0.0505\pm0.0021$ & & $<0.00010$ $\left(3\sigma\right)$ & \citet{ste10} \\
5.8 $\mu$m & $4501.778\pm0.005$ & $0.0505\pm0.0021$ & & $0.00033\pm0.00014$ & \citet{ste10} \\
8.0 $\mu$m & $4866.63444\pm0.00082$ & 0.04347 & 0.00700 & $0.000452\pm0.000027$ & \citet{knu11} \\
& $4282.3331\pm0.0016$ & $0.0492\pm0.0037$ & & $0.00054\pm0.00008$ & \citet{ste10} \\
& $4282.33\tnH\pm0.01$ & & & $0.00057\pm0.00008$ & \citet{dem07} \\
16 $\mu$m & $4477.981\pm0.003$ & $0.0505\pm0.0061$ & & $0.00140\pm0.00027$ & \citet{ste10} \\
24 $\mu$m & $4470.053\pm0.002$ & $0.0505\pm0.0061$ & & $0.00175\pm0.00041$ & \citet{ste10} \\
\cutinhead{HAT-P-2 b}
$z$ & $4388.546\pm0.011$ & $0.1650\pm0.0034$ & & $0.00522\pm0.00061$ & \citet{pal10} \\
3.6 $\mu$m & $5289.9302\pm0.0014$ & $0.1550\pm0.0027$ & $0.01090\pm0.00075$ & $0.000996\pm0.000072$ & \citet{lew13} \\
4.5 $\mu$m & $5757.5130\pm0.0011$ & $0.1651\pm0.0023$ & $0.01444\pm0.00083$ & $0.001031\pm0.000061$ & \citet{lew13} \\
5.8 $\mu$m & $4906.8561^{+0.0076}_{-0.0062}$ &  &  & $0.00071^{+0.00029}_{-0.00013}$ & \citet{lew13} \\
8.0 $\mu$m & $4354.7757\pm0.0022$ & $0.1610\pm0.0043$ & $0.0121\pm0.0016$ & $0.001392\pm0.000095$ & \citet{lew13} \\
\cutinhead{HAT-P-7 b}
$K_P$ &  &  &  & $6.930\times10^{-5}\pm6\times10^{-7}$ & \citet{ang15} \\
3.6 $\mu$m & $5418.4562^{+0.0021}_{-0.0019}$ & & & $0.00156\pm0.00009$ & \citet{won16} \\
4.5 $\mu$m & $5429.4780^{+0.0010}_{-0.0009}$ & & & $0.00190\pm0.00006$ & \citet{won16} \\
5.8 $\mu$m & $4768.05200\pm0.0035$ & & & $0.00245\pm0.00031$ & \citet{chr10} \\
8.0 $\mu$m & $4770.26413\pm0.0039$ & & & $0.00225\pm0.00052$ & \citet{chr10} \\
\cutinhead{HD 80606 b}
$i/R_C/r/z$ & $4424.736\tnH\pm0.004$ & $0.0762\pm0.0023$ & $0.007188\pm0.000263$ & & \citet{win09a} \\
4.5 $\mu$m & & & & $0.000651\pm0.000049$ & \citet{dew16} \\
8.0 $\mu$m & & & & $0.001053\pm0.000094$ & \citet{dew16} \\
& $4424.736\tnH\pm0.003$ & $0.075\pm0.010$ & $0.005\pm0.005$& $0.00136\pm0.00018$ & \citet{lau09} \\
\cutinhead{HD 149026 b}
3.6 $\mu$m & $4535.8764\pm0.0010$ & $0.274\pm0.002$ & $0.0192\pm0.0009$ & $0.00040\pm0.00003$ & \citet{ste12} \\
4.5 $\mu$m & $4596.2676\pm0.0019$ & $0.274\pm0.002$ & $0.0195\pm0.0010$ & $0.00034\pm0.00006$ & \citet{ste12} \\
5.8 $\mu$m & $4903.990\pm0.012$ & $0.274\pm0.002$ & $0.0195\pm0.0010$ & $0.00044\pm0.00010$ & \citet{ste12} \\
8.0 $\mu$m & $4912.614\pm0.002$ & $0.274\pm0.002$ & $0.0195\pm0.0010$ & $0.00052\pm0.00006$ & \citet{ste12} \\
\cutinhead{HD 189733 b}
3.6 $\mu$m & $5560.66515\tn\pm0.00046$ & & & $0.001466\pm0.000040$ & \citet{knu12} \\
& & & & $0.00256\pm0.00014$ & \citet{cha08} \\
4.5 $\mu$m & $5187.94447\tn\pm0.00040$ & & & $0.001787\pm0.000038$ & \citet{knu12} \\
& & & & $0.00214\pm0.00020$ & \citet{cha08} \\
\tn & $2455560.66451\pm0.00025$ & & & & \citet{knu12} \\
5.8 $\mu$m & & & & $0.00310\pm0.00034$ & \citet{cha08} \\
8.0 $\mu$m & & & & $0.00391\pm0.00022$ & \citet{cha08} \\
16 $\mu$m & & & & $0.00519\pm0.00020$ & \citet{cha08} \\
& & & & $0.00551\pm0.00030$ & \citet{dem06} \\
24 $\mu$m & & & & $0.00598\pm0.00038$ & \citet{cha08} \\
\cutinhead{HD 209458 b}
3.6 $\mu$m & $3702.5244\tnH\pm0.0024$ & & & $0.00094\pm0.00009$ & \citet{knu08} \\
4.5 $\mu$m & $3702.5198\tnH\pm0.0024$ & & & $0.00213\pm0.00015$ & \citet{knu08} \\
5.8 $\mu$m & $3702.5251\tnH\pm0.0026$ & & & $0.00301\pm0.00043$ & \citet{knu08} \\
8.0 $\mu$m & $3702.5299\tnH\pm0.0022$ & & & $0.00240\pm0.00026$ & \citet{knu08} \\
24 $\mu$m & $3346.5278\tnH\pm0.0049$ & & & $0.00260\pm0.00046$ & \citet{dem05} \\
\cutinhead{WASP-12 b}
$J$ & & & & $0.00131^{+0.00027}_{-0.00029}$ & \citet{cro11} \\
$H$ & & & & $0.00176^{+0.00016}_{-0.00051}$ & \citet{cro11} \\
$K_S$ & & & & $0.00309^{+0.00013}_{-0.00012}$ & \citet{cro11} \\
3.6 $\mu$m & & & & $0.00421\pm0.00011$ & \citet{ste14} \\
 & & & & $0.0038\pm0.0004$ & \citet{cow12} \\
 & $4773.6481\pm0.0006$ & $0.1229\pm0.0012$ & 0.0147 & $0.00379\pm0.00013$ & \citet{cam11} \\
4.5 $\mu$m & & & & $0.00428\pm0.00012$ & \citet{ste14} \\
 & & & & $0.0039\pm0.0003$ & \citet{cow12} \\
 & $4769.2819\pm0.0008$ & $0.1244\pm0.0017$ & 0.0147 & $0.00382\pm0.00019$ & \citet{cam11} \\
5.8 $\mu$m & & & & $0.00696\pm0.0006$ & \citet{ste14} \\
 & $4773.6481\pm0.0006$ & $0.1229\pm0.0012$ & 0.0147 & $0.00629\pm0.00052$ & \citet{cam11} \\
8.0 $\mu$m & & & & $0.00696\pm0.00096$ & \citet{ste14} \\
 & $4769.2819\pm0.0008$ & $0.1244\pm0.0017$ & 0.0147 & $0.00636\pm0.00067$ & \citet{cam11} \\
\cutinhead{WASP-14 b}
3.6 $\mu$m & $5274.6617\pm0.0006$ & $0.1079\pm0.0013$ & $0.0121\pm0.0003$ & $0.00187\pm0.00007$ & \citet{ble13} \\
4.5 $\mu$m & $4908.9298\pm0.0011$ & $0.1079\pm0.0013$ & $0.0121\pm0.0003$ & $0.00224\pm0.00018$ & \citet{ble13} \\
8.0 $\mu$m & $4908.9298\pm0.0011$ & $0.1079\pm0.0013$ & $0.0121\pm0.0003$ & $0.00181\pm0.00022$ & \citet{ble13} \\
\cutinhead{WASP-18 b}
3.6 $\mu$m & $4820.7160\pm0.0006$ & $0.0944\pm0.0007$ & 0.0099 & $0.0030\pm0.0002$ & \citet{nym11} \\
4.5 $\mu$m & $4824.4809\pm0.0005$ & $0.0944\pm0.0007$ & 0.0099 & $0.0039\pm0.0002$ & \citet{nym11} \\
5.8 $\mu$m & $4820.7160\pm0.0006$ & $0.0944\pm0.0007$ & 0.0099 & $0.0037\pm0.0003$ & \citet{nym11} \\
8.0 $\mu$m & $4824.4809\pm0.0005$ & $0.0944\pm0.0007$ & 0.0099 & $0.0041\pm0.0002$ & \citet{nym11} \\
\cutinhead{WASP-19 b}
1.6--8.0 $\mu$m & $5183.56158\tnH^{+0.00042}_{-0.00035}$ & $0.06564^{+0.00043}_{-0.00040}$ & $0.01358\pm0.00054$ & & \citet{and13} \\
2.09 $\mu$m & & & & $0.00366\pm0.00067$ & \citet{and13} \\
3.6 $\mu$m & $5776.77019^{+0.00082}_{-0.00083}$ & & & $0.00485\pm0.00024$ & \citet{won16} \\
 & & & & $0.00483\pm0.00025$ & \citet{and13} \\
4.5 $\mu$m & $5787.02381^{+0.00078}_{-0.00076}$ & & & $0.00584\pm0.00029$ & \citet{won16} \\
 & & & & $0.00572\pm0.00030$ & \citet{and13} \\
5.8 $\mu$m & & & & $0.0065\pm0.0011$ & \citet{and13} \\
8.0 $\mu$m & & & & $0.0073\pm0.0012$ & \citet{and13} \\
\cutinhead{WASP-33 b}
0.91 $\mu$m & & $0.11224\pm0.00084$ & $0.01149^{+0.00097}_{-0.00084}$ & $0.00109\pm0.00030$ & \citet{smi11} \\
$K_s$ & & & & $0.00244^{+0.00027}_{-0.00020}$ & \citet{dem13} \\
 & $5844.8156\pm0.0040$ & & & $0.0027\pm0.0004$ & \citet{dem12} \\
3.6 $\mu$m & $5647.1978\pm0.0001$ & & & $0.0026\pm0.0005$ & \citet{dem12} \\
4.5 $\mu$m & $5650.8584\pm0.0005$ & & & $0.0041\pm0.0002$ & \citet{dem12} \\
\cutinhead{WASP-43 b}
1.19, 2.09 $\mu$m & $5726.95069^{+0.00084}_{-0.00078}$ & $0.05037^{+0.00023}_{-0.00021}$ & & $0.02542^{+0.00024}_{-0.00025}$ & \citet{gil12} \\
3.6 $\mu$m & $5773.3179\pm0.0003$ & $0.05208\pm0.00083$ & $0.01117\pm0.00075$ & $0.00323\pm0.00006$ & \shortstack{\citet{ble14},\\\citet{ste17}} \\
4.5 $\mu$m & $5772.5045\pm0.0003$ & $0.05208\pm0.00083$ & $0.01117\pm0.00075$ & $0.00383\pm0.00008$ & \shortstack{\citet{ble14},\\\citet{ste17}} \\
\enddata
\tablenotetext{H}{Times are in HJD$-2450000$.}
\tablenotetext{*}{\citet{knu12} report a combined secondary eclipse ephemeris.}
\tablecomments{Times are in BJD$-2450000$, unless otherwise noted.}

\label{table:eclipse_properties}
\end{deluxetable*}

\startlongtable
\begin{deluxetable*}{cccccc}
\centering
\tabletypesize{\scriptsize}
\tablecaption{Transit Properties for Exoplanets with Spitzer Light Curves}
\tablewidth{0pt}
\tablehead{
\colhead{Band} & \colhead{$T_{\textrm{mid}}$ $\left(\mathrm{BJD}-2450000\right)$} & \colhead{$T_{1,4}$ (days)} & \colhead{$T_{1,2}=T_{3,4}$ (days)} & Depth & Ref.}
\startdata
\cutinhead{GJ 436 b}
$V$ & & & & $0.00696\pm0.000117$ & \citet{tor08} \\
NICMOS & $4449.99141\tnH\pm0.00008$ & $0.0317\pm0.0004$ & & $0.006906\pm0.000083$ & \citet{pon09} \\
$H$ & $4534.59584\tnH\pm0.00015$ & & & $0.00707\pm0.00019$ & \citet{alo08} \\
$K_s$ & $4238.47898\tnH\pm0.00046$ & & & $0.0064\pm0.0003$ & \citet{cac09} \\
3.6 $\mu$m & & & & $0.006694\pm0.000061$ & \citet{knu11} \\
4.5 $\mu$m & & & & $0.006865\pm0.000078$ & \citet{knu11} \\
8.0 $\mu$m & & & & $0.006831\pm0.000083$ & \citet{knu11} \\
& $4280.78149\tnH\pm0.00016$ & & & & \citet{dem07} \\
\tn & $4865.083208\pm0.000042$ & $0.04227\pm0.00016$ & $0.01044\pm0.00014$ & $0.006907\pm0.000043$ & \citet{knu11} \\
\cutinhead{HAT-P-2 b}
$z$ & $4387.43975\pm0.00074$ & $0.1787\pm0.0013$ & $0.0141^{+0.0015}_{-0.0012}$ & $0.00522\pm0.00061$ & \citet{pal10} \\
3.6 $\mu$m & $5288.84988\pm0.00060$ & $0.1770\pm0.0011$ & $0.0128\pm0.0010$ & $0.004653\pm0.000102$ & \citet{lew13} \\
4.5 $\mu$m & $5756.42520\pm0.00067$ & $0.1813\pm0.0013$ & $0.0177\pm0.0012$ & $0.004958\pm0.000084$ & \citet{lew13} \\
8.0 $\mu$m & $4353.6911\pm0.0012$ & $0.1789\pm0.0023$ & $0.0144\pm0.0021$ & $0.004462\pm0.000213$ & \citet{lew13} \\
\cutinhead{HAT-P-7 b}
$i$ & $4731.67929\tnH\pm0.00043$ & $0.1669\pm0.0027$ & $0.0198^{+0.0025}_{-0.0039}$ & $0.0060202\pm0.0000047$ & \citet{win09b} \\
3.6 $\mu$m & $5419.55818^{+0.00054}_{-0.00070}$ & & & $0.00629^{+0.00024}_{-0.00019}$ & \citet{won16} \\
4.5 $\mu$m & $5430.58278^{+0.00040}_{-0.00047}$ & & & $0.00604\pm0.00012$ & \citet{won16} \\
\cutinhead{HD 80606 b}
$i/R_C/r/z$ & $4987.7842\tnH\pm0.0049$ & $0.4850\pm0.0104$ & $0.1083\pm0.0075$ & $0.01067\pm0.00023$ & \citet{win09a} \\
$R$ & $4876.344\tnH\pm0.011$ & & & $0.01117\pm0.00038$ & \citet{fos09} \\
\cutinhead{HD 149026 b}
NICMOS & $4456.78761\tnH\pm0.00014$ & $0.270^{+0.012}_{-0.013}$ & $0.020\pm0.001$ & $0.002933^{+0.000099}_{-0.000076}$ & \citet{car09} \\
8.0 $\mu$m & $4597.70713\pm0.00016$ & $0.2738\pm0.0016$ & $0.0195\pm0.0010$ & $0.00268\pm0.00006$ & \citet{ste12} \\
& & $0.270^{+0.012}_{-0.013}$ & $0.019\pm0.001$ & $0.002692^{+0.000087}_{-0.000089}$ & \citet{car09} \\
\cutinhead{HD 189733 b}
3.6 $\mu$m & $5559.554550\pm0.000035$ & & & $0.024059\pm0.000062$ & \citet{knu12} \\
4.5 $\mu$m & $5189.052491\pm0.000032$ & & & $0.024274\pm0.000059$ & \citet{knu12} \\
8.0 $\mu$m & $4279.436714\pm0.000015$ & $0.1789\pm0.0023$ & $0.0144\pm0.0021$ & $0.004462\pm0.000213$ & \citet{ago10} \\
\cutinhead{HD 209458 b}
290--1030 nm & $2826.628521\tnH\pm0.000087$ & & & $0.014607\pm0.000024$ & \citet{knu07a} \\
\ttn & & $0.13971\pm0.00074$ & $0.10027\pm0.00068$ & & \shortstack{\citet{sea03},\\\citet{tor08}} \\
\cutinhead{WASP-12 b}
$B/z'$ & $4508.9761\pm0.0002$ & $0.122\pm0.001$ & & $0.0138\pm0.0002$ & \citet{heb09} \\
$V$ & $4508.97605\pm0.00028$ & $0.1666\pm0.0015$ & $0.0135\pm0.0008$ & $0.01252\pm0.00045$ & \citet{cha12} \\
$R$ & $4508.97682\pm0.00020$ & $0.1220\pm0.0010$ & & $0.01380\pm0.00016$ & \citet{mac11} \\
\cutinhead{WASP-14 b}
$V+R, R, I$ & $4463.57583\tnH\pm0.00053$ & $0.1275^{+0.0028}_{-0.0031}$ & & $0.0102^{+0.0002}_{-0.0003}$ & \citet{jos09} \\
\cutinhead{WASP-18 b}
HARPS & $4664.90531^{+0.00016}_{-0.00017}$ & $0.09089^{+0.00080}_{-0.00061}$ & & $0.00916^{+0.00020}_{-0.00012}$ & \citet{tri10} \\
\cutinhead{WASP-19 b}
$r, z$ & $5183.16711\tnH\pm0.000068$ & $0.06549\pm0.00035$ & $0.01346\pm0.00052$ & $0.02050\pm0.00024$ & \shortstack{\citet{and13},\\\citet{heb10},\\\citet{hel11a}} \\
3.6 $\mu$m & $5777.16364^{+0.00022}_{-0.00021}$ & & & $0.01957^{+0.00039}_{-0.00050}$ & \citet{won16} \\
4.5 $\mu$m & $5787.41861^{+0.00023}_{-0.00022}$ & & & $0.02036^{+0.00049}_{-0.00071}$ & \citet{won16} \\
\cutinhead{WASP-33 b}
$R$+$I$, RV & $4163.22373\tnH\pm0.00026$ & & & $0.01136\pm0.00019$ & \citet{col10} \\
$U$ & $2984.82964\pm0.00030$ & & & $0.01179^{+0.00048}_{-0.00015}$ & \citet{tur16} \\
0.91 $\mu$m & $4590.17948\tnH\pm0.00028$ & $0.11224\pm0.00084$ & $0.01149^{+0.00097}_{-0.00084}$ & $0.01041^{+0.00023}_{-0.00021}$ & \citet{smi11} \\
\cutinhead{WASP-43 b}
$I+z$, $r'$ & $5726.54336\pm0.00012$ & $0.05037^{+0.00023}_{-0.00021}$ & & $0.02542^{+0.00024}_{-0.00025}$ & \citet{gil12} \\
3.6 $\mu$m & $7089.11181\pm0.00007$ & & & $0.02496\pm0.00010$ & \citet{ste17} \\
4.5 $\mu$m & $6897.13195\pm0.00007$ & & & $0.02525\pm0.00016$ & \citet{ste17} \\
\enddata
\tablenotetext{H}{Times are in HJD$-2450000$.}
\tablenotetext{*}{Transit properties are from combined wavelength fits.}
\tablenotetext{**}{Transit duration and ingress/egress times are calculated from the geometric relation described in \citet{sea03}, using the transit properties in \citet{tor08}.}
\tablecomments{Times are in BJD$-2450000$, unless otherwise noted. Where the source authors specify times with respect to the UTC and TDB systems \citep{eas10}, we choose to quote the TDB time.}

\label{table:transit_properties}
\end{deluxetable*}

\clearpage
\acknowledgements
This material is based upon work supported by the National Aeronautics and Space Administration through the NASA Astrobiology Institute under Cooperative Agreement Notice NNH13ZDA017C issued through the Science Mission Directorate. We acknowledge support from the NASA Astrobiology Institute through a cooperative agreement between NASA Ames Research Center and Yale University.

We would like to thank Drs. Peter Gao and Mark Marley for their help in interpreting physical processes in the data.

This research has made use of the Exoplanet Orbit Database
and the Exoplanet Data Explorer at \url{exoplanets.org}.

\facility{Spitzer (IRAC), Kepler}

\software{Astropy \citep{ast13}, Colorcet \citep{kov15}, Jupyter \citep{klu16}, Matplotlib \citep{hun07}, Numpy \citep{van11}, Paletton \citep{pal18}, Scipy \citep{jon01}, WebPlotDigitizer \citep{ank17}}

\bibliographystyle{apj}
\bibliography{apj-jour,paper}

\end{document}